\documentclass[aps, prb, reprint, superscriptaddress, nofootinbib]{revtex4-2} 
\usepackage{amsmath} 
\usepackage{amssymb} 
\usepackage{bm} 
\usepackage{color} 
\usepackage{comment} 
\usepackage{float} 
\usepackage{graphicx} 
\usepackage{hyperref} 
\usepackage[utf8]{inputenc} 
\usepackage{placeins} 
\usepackage{relsize} 

\newcommand{\Eqref}[1]{Eq.~\eqref{#1}}

\newcommand{\Eqsref}[1]{Eqs.~\eqref{#1}} 
\newcommand{\Figref}[1]{Fig.~\ref{#1}} 
\newcommand{\Figsref}[1]{Figs.~\ref{#1}} 
\newcommand{\Refref}[1]{Ref.~\onlinecite{#1}} 
\newcommand{\Refsref}[1]{Refs.~\onlinecite{#1}} 
\newcommand{\Secref}[1]{\textbf{Section~\ref{#1}}} 
\newcommand{\SIref}[2][]{\SI~\Sec~\textbf{#2}\textit{#1}} 

\newcommand{\Appendix}{\textbf{Appendix}} 

\newcommand{\bbR}{\mathbb R}

\newcommand{\bfJ}{\mathbf J} 
\newcommand{\bfM}{\mathbf M} 
 
\newcommand{\bfN}{\mathbf N} 
\newcommand{\bfv}{\mathbf v} 
 
\newcommand{\bfw}{\mathbf w} 
\newcommand{\bfx}{\mathbf x}

\newcommand{\bsell}{\boldsymbol\ell} 
\newcommand{\bsomega}{\boldsymbol\omega}

\newcommand{\calA}{\mathcal A} 
\newcommand{\calE}{\mathcal E} 
\newcommand{\calN}{\mathcal N} 

\newcommand{\const}{\mathrm{const.}} 

\newcommand{\defo}     {\mathrm{defo}} 
 
\newcommand{\isoDeform}{\delta_\iso{\bfx}} 
\newcommand{\SPDeform} {\delta_\SP {\bfx}} 
\newcommand{\dependence}{ \left(u^1, u^2\right) } 

\newcommand{\df}[1]{\rmd{#1}} 
\newcommand{\eq}{{\mathrm{eq}}} 
\newcommand{\frakd}{ \mathlarger{\mathfrak d} } 

\newcommand{\FvK}{{F\"{o}ppl-von K\'{a}rm\'{a}n}} 
\newcommand{\fvk}{\mathrm{FvK}} 

\newcommand{\h}{\\[0.25 em]}

\newcommand{\inn}{{\mathrm{in }}} 
\newcommand{\out}{{\mathrm{out}}} 

\newcommand{\inProd}[2]{\left\langle{#1, #2}\right\rangle} 
\newcommand{\iso}{\mathrm{iso}} 
\newcommand{\SP}{{\mathrm{SP}}} 
 
\newcommand{\norm}[1]{\left\lVert{#1}\right\rVert} 
 
\newcommand{\one}{\mathbf 1} 
\newcommand{\rmb}{\mathrm b} 
\newcommand{\rmC}{\mathrm C} 
\newcommand{\rmd}{\mathrm d} 
\newcommand{\rmm}{\mathrm m} 
\renewcommand{\rmp}{\mathrm p} 
\newcommand{\rms}{\mathrm s} 
\newcommand{\Sec}{\textbf{Sec.}} 
\newcommand{\SI}{\textbf{SI}} 
 
\newcommand{\textOr} {\mbox{\quad{or} \quad}} 
 
\newcommand{\unitvec}[1]{\hat{\mathbf#1}} 
\newcommand{\zero}{\mathbf 0} 

\numberwithin{equation}{section} 
\renewcommand{\theequation}{\arabic{section}.\arabic{equation}} 
 
\makeatletter 
\newcommand*\rel@kern[1]{\kern#1\dimexpr\macc@kerna} 
\newcommand*\widebar [1]{%
\begingroup 
\def\mathaccent##1##2{%
\rel@kern{+0.8}%
\overline{\rel@kern{-0.8}\macc@nucleus\rel@kern{+0.2}}%
\rel@kern{-0.2}%
}%
\macc@depth\@ne 
\let\math@bgroup\@empty\let\math@egroup\macc@set@skewchar 
\mathsurround\z@\frozen@everymath{\mathgroup\macc@group\relax}%
\macc@set@skewchar\relax 
\let\mathaccentV\macc@nested@a 
\macc@nested@a\relax111{#1}%
\endgroup 
} 
\makeatother 
\makeatletter 
\newcommand*\wt[1]{\mathpalette\wthelper{#1}} 
\newcommand*\wthelper[2]{%
\hbox{\dimen@\accentfontxheight#1%
\accentfontxheight#11.15\dimen@$\m@th#1\widetilde{#2}$%
\accentfontxheight#1\dimen@ 
}%
} 

\newcommand*\accentfontxheight[1]{%
\fontdimen5\ifx#1\displaystyle 
\textfont 
\else\ifx#1\textstyle 
\textfont 
\else\ifx#1\scriptstyle 
\scriptfont 
\else 
\scriptscriptfont 
\fi\fi\fi3 
} 
\makeatother 
\begin{document} 
\begin{abstract} 
Thin surfaces are ubiquitous in nature, from leaves to cell membranes, and in technology, from paper to corrugated containers. Structural thinness imbues them with flexibility, the ability to easily bend under light loads, even as their much higher stretching stiffness can bear substantial stresses. When surfaces have periodic patterns of either smooth hills and valleys or sharp origami-like creases this can substantially modify their mechanical response. We show that for any such surface, there is a duality between the surface rotations of an isometric deformation and the in-plane stresses of a force-balanced configuration. 
This duality means that of the six possible combinations of global in-plane strain and out-of-plane bending, exactly three must be isometries. We show further that stressed configurations can be expressed in terms of both the applied deformation and the isometric deformation that is dual to the pattern of stress that arises. 
We identify constraints rooted in symplectic geometry on the three isometries that a single surface can generate. 
This framework sheds new light on the fundamental limits of the mechanical response of thin periodic surfaces, while also highlighting the role that continuum differential geometry plays in even sharply creased origami surfaces. 
\end{abstract} 
\author{Wenqian Sun} 
\affiliation{School of Physics, Georgia Institute of Technology, Atlanta, GA 30332, USA} 
\author{Yanxin Feng} 
\affiliation{School of Physics, Georgia Institute of Technology, Atlanta, GA 30332, USA} 
\author{Christian D. Santangelo} 
\affiliation{Department of Physics, Syracuse University, Syracuse, New York 13244, USA} 
\author{D. Zeb Rocklin} 
\affiliation{School of Physics, Georgia Institute of Technology, Atlanta, GA 30332, USA} 
\title{Geometric Mechanics of Thin Periodic Surfaces} 
\date{\today} 

\maketitle 
\section{Introduction} \label{sec: introduction} 
Thin sheets occur widely in nature, from macroscale leaves and petals to cell membranes and viral capsids. Thin sheets are also used widely in engineering, from cutting-edge origami metamaterials~\cite{Lv2014_Origami-Based_Metamaterials, Kamrava2017_Origami-Based_Cellular_Metamaterials, Schenk2011_Origami_Folding} to nanotechnology such as graphene kirigami~\cite{Blees2015_Graphene_Kirigami, Yong2020_Kirigami-Inspired_Sensors, Mortazavi2017_Transport_Properties_of_Graphene_Kirigami} to ubiquitous corrugated structures such as cardboard and shipping containers~\cite{Twede2014_Handbook_on_Corrugated_Boxes, Pathare2014_Review_on_Design_of_Corrugated_Boxes, Frank2014_Review_on_Compression_of_Corrugated_Boxes}. In all cases, thinness means that the systems require little volume or mass while maintaining a combination of strength and flexibility that is determined largely by their geometry. 

Recent work on origami and smooth surfaces has hinted at deep mathematical relationships between different modes of deformation. In particular, even though the in-plane and out-of-plane Poisson's ratios of thin elastic slabs are necessarily equal, for an ever-widening classes of periodically patterned surfaces, such as those shown in \Figref{fig: periodic surfaces}, the ratios are instead exactly opposite~\cite{Wei2013_Miura-Ori, Schenk2013_Miura-Ori, Nassar2017_Eggbox, Pratapa2019_Morph, Nassar2024_Number_of_Modes}. The structure of these linear, infinitesimal properties endures even as the ratios themselves change under nonlinear deformation~\cite{McInerney2022_Discrete_Symmetries}, and may inform more complex nonlinear phenomena such as buckling, wrinkling and defect formation. 

In this paper, we reveal that the vertex duality between folding modes and equilibrium stresses previously explored in triangulated origami~\cite{McInerney2020_Hidden_Symmetries, Maxwell2011_Seminal_Work, Calladine1978_Seminal_Work, Gluck1975_Rigid_Surfaces, Crapo1982_Applied_Projective_Geometry, Tachi2012_Origami_Design, Tachi2016_Rigid_Folding, Rocklin2022_Topologically_Protected_Deformations_bookChapter} can be extended to smooth surfaces. We introduce the concept of shape-periodic deformations that combine uniform strain and bending of periodically patterned surfaces and show that any such sheet has six such deformations, exactly three of which must be isometries. No matter how a sheet is patterned, the set of isometries is not arbitrary and instead corresponds to a Lagrangian subspace~\cite{Arnold1991_Mechanics, da_Silva2008_Symplectic_Geometry, Jeffs2022_Symplectic_Geometry}, meaning that the form of each isometry limits which others are possible. 

The rest of the paper is structured as follows. 
In \Secref{sec: duality} we  review the notions of equilibrium stresses and modes of linear isometry. 
We then present the duality between them, which holds for smooth, simply-connected surfaces that can deform isometrically. 
In \Secref{sec: rigidity and flexibility of smooth periodic surfaces}, we apply this duality to doubly periodic surfaces to show that each such surface has three shape-periodic isometries and that these isometries form a Lagrangian subspace. 
In \Secref{sec: verification of the results} our results are extended to creased surfaces such as origami and verified numerically. 
Finally, in \Secref{sec: conclusion} we relate our results to the literature and suggest directions for future research investigations. 
\section{Thin sheets' isometries, equilibrium stresses and their duality} \label{sec: duality} 
In this work, following convention in the literature~\cite{Ventsel2001_Thin-Walled_Structures}, we work with the two-dimensional elasticity of the shell's mid-surface, assuming no strain along the thickness direction. 
Consequently, going forward, we disregard the shell's thickness direction, and all physical quantities introduced below are defined with respect to the mid-surface---for example, stress components have units of force per unit length.

\begin{figure*} 
\centering 
\includegraphics[width = \textwidth]{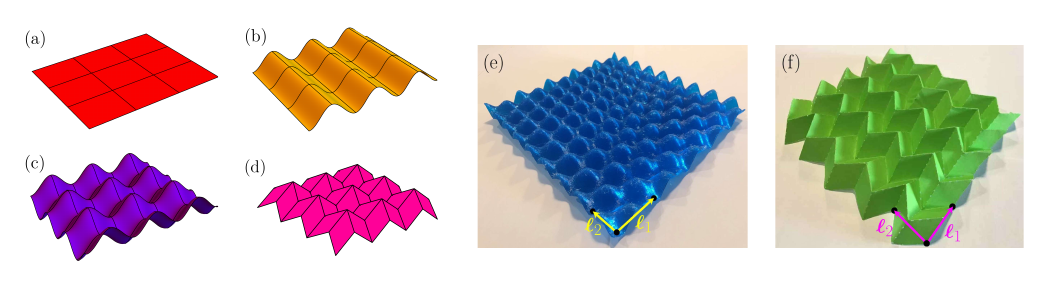} 
\caption{
We consider thin sheets that are either flat (a), singly periodic (b), doubly periodic (c). In addition to such smooth sheets, our results apply to creased sheets such as origami (d). Prototypical doubly periodic sheets are readily realized through common fabrication techniques (e) and (f). 
} 
\label{fig: periodic surfaces} 
\end{figure*}

As shown in \Figref{fig: local geometry of a surface}~(a), we consider a thin shell whose mid-surface is described by $\bfx\left(u^1, u^2\right)$, with the arguments representing local coordinates that have the dimension of length. 
The surface, which is embedded in the conventional three-dimensional Euclidean space, in general has a nonzero local Gaussian curvature and cannot be developed from a flat sheet. 
We assume the surface to be smoothly differentiable almost everywhere, though our results also apply to sheets like origami, which feature sharp creases along certain lines [see \Figsref{fig: periodic surfaces}~(d) and (f)]. 
\subsection{Linear isometries} \label{subsec: duality, isometric deformations} 
\begin{figure*} 
\centering 
\includegraphics[width = 0.3\linewidth]{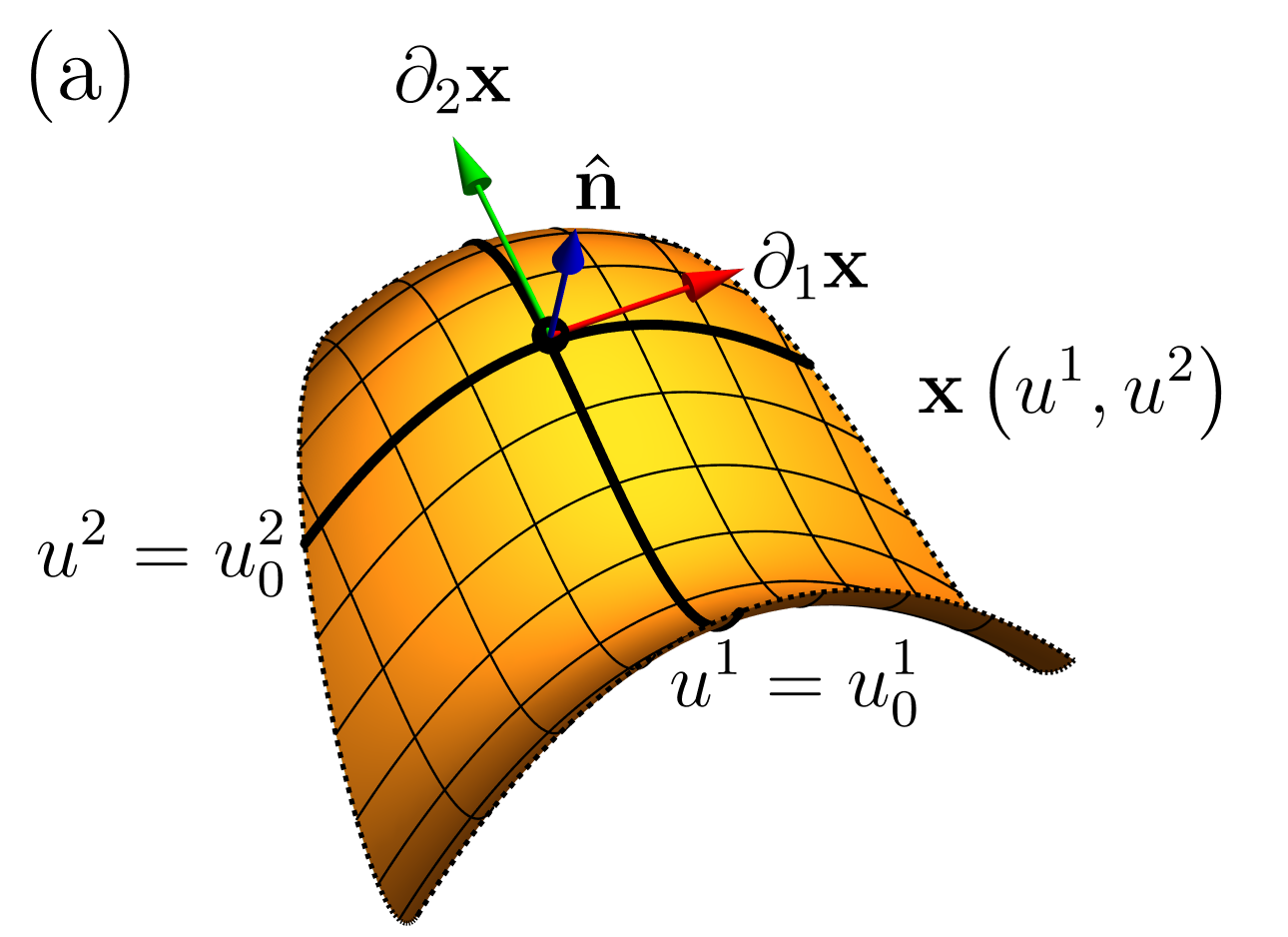} \quad 
\includegraphics[width = 0.3\linewidth]{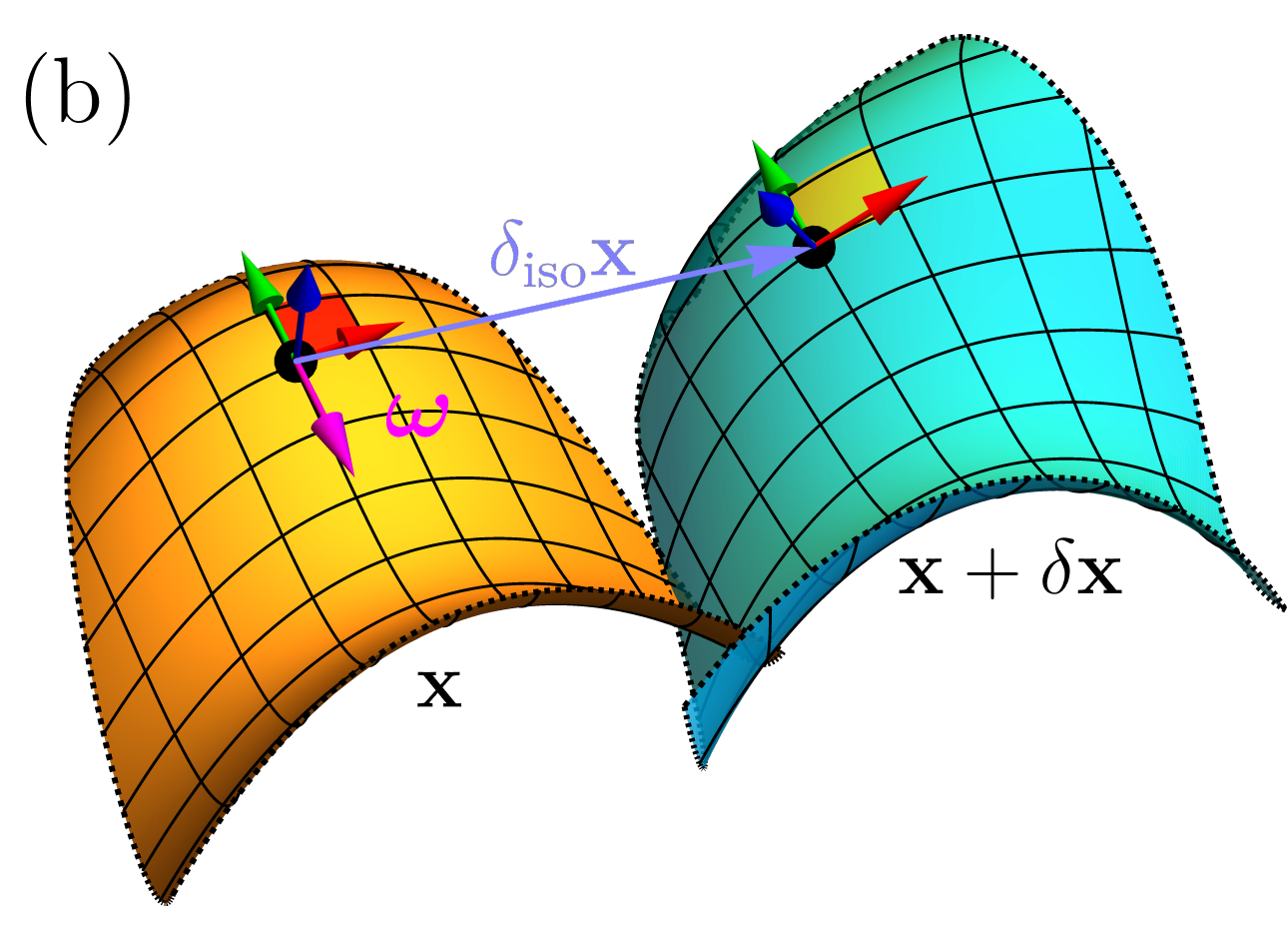}          \quad 
\includegraphics[width = 0.3\linewidth]{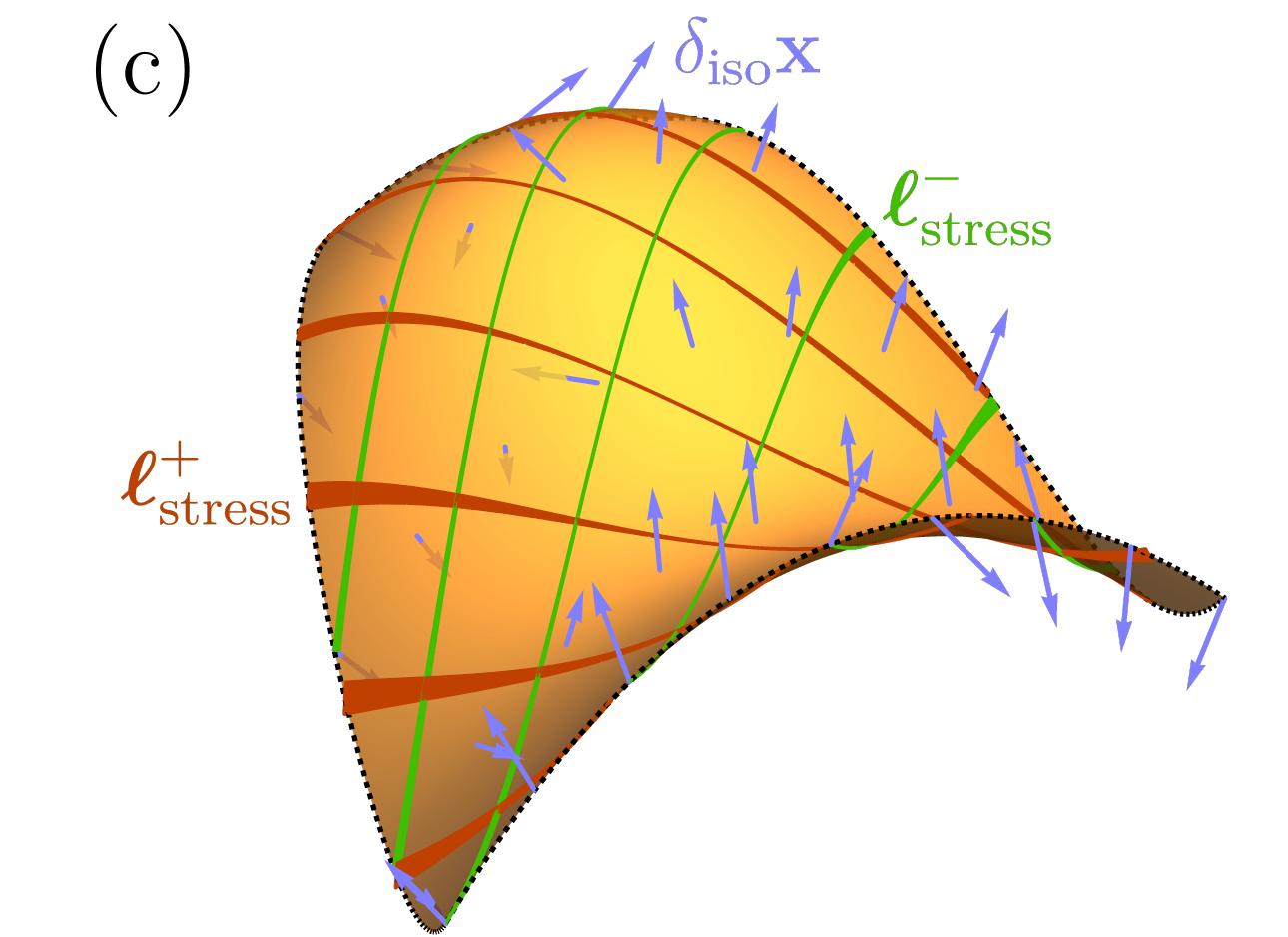} 
\caption{
(a) Any local portion of a smooth simply-connected surface can be represented by a vector field $\bfx\left(u^1, u^2\right)$, where $u^1$ and $u^2$ are the local coordinates. 
The grid lines depicted in the figure are the corresponding coordinate lines, i.e., curves of constant $u^1$ or constant $u^2$. 
For every point on the surface there is a local frame composed of three vectors, including the unit surface normal vector ($\unitvec{n}$) and the vectors tangent to the two coordinate lines, $\partial_1{\bfx}$ and $\partial_2{\bfx}$. 
In the figure we use the red-green-blue (RGB) triad of vectors to represent the local frame $\left\{\partial_1{\bfx}, \partial_2{\bfx}, \unitvec{n}\right\}$ at a point. 
(b) When the local portion is deformed isometrically each point on the surface is displaced by a displacement vector ($\delta_\iso{\bfx}$) in such a way that infinitesimal patches of the surface are locally rotated with respect to a so-called angular velocity field ($\bsomega$) without being stretched. 
In the figure the RGB triads of vectors still represent the local frames, but we rescale the lengths of the surface tangent vectors in the frames, so that their rescaled lengths are approximately equal to the side lengths of the corresponding infinitesimal patches. 
The local rotations of the infinitesimal patches can hence be understood as rotations of the corresponding local frames with respect to the angular velocity field. 
(c) While an isometric deformation---such as the one represented by the light-blue arrows in the figure---does not stress the surface it is mathematically dual to a stress field, as described in \Secref{subsec: duality, the duality}. 
The corresponding lines of principal stress are depicted in the figure as the green and the red curves, respectively, indicating the directions along which the surface is most (the red curves) and least (the green curves) stressed. 
The thickness of each curve is proportional to the corresponding principal stress. 
} 
\label{fig: local geometry of a surface} 
\end{figure*}

Isometric deformations do not locally stretch a material. 
For thin shells, they often correspond to low-energy deformation modes, as stretching is energetically costlier than bending~\cite{LL1986_Elasticity, Audoly2019_Elasticity_and_Geometry}. 
The relative importance of stretching compared to bending is captured by the two-dimensional \FvK{} number associated with the shell's mid-surface~\cite{ProfP2013_Soft_Spots}: 
\begin{align} 
        \fvk 
 \equiv \frac 
        {\mathrm{Stretching\ stiffness} \times L_\defo^2} 
        {\mathrm{Bending   \ stiffness}                 } 
\approx 10\left(\frac{L_\defo}{t}\right)^2, 
\label{eqn: FvK number} 
\end{align} 
which is much greater than unity for thin surfaces undergoing large-scale deformations compared to their thickness (i.e., $L_\defo \gg t$). 
Such an isometric deformation field, denoted $\isoDeform\dependence$, thus consists of local infinitesimal rotations of the reference state, so that: 
\begin{align} 
       \partial_\alpha{\isoDeform} 
     = \bsomega 
\times \partial_\alpha{\bfx      }, 
\label{eqn: defn. of the angular velocity} 
\end{align} 
where $\bsomega$ is a dimensionless angular velocity field~\cite{Spivak1999_Differential_Geometry} (the nomenclature does not imply dynamics), as shown in~\Figref{fig: local geometry of a surface}~(b). 
In fact, the compatibility of an isometry can depend only on gradients of this angular velocity, and such gradients are compatible if and only if evolving the isometry over a closed curve $\gamma$ on the surface does not change the position or orientation: 
\begin{align} 
\oint_\gamma \df{ \left(\isoDeform \right) } = \zero \textOr 
\oint_\gamma \df{       \bsomega           } = \zero. 
\label{eqn: closure conditions} 
\end{align} 
\par 
The angular acceleration vectors, which can be expressed in terms of the tangent vectors and the normal vectors as 
$
                          \partial_\alpha{\bsomega} 
\equiv {a_\alpha}^\beta\, \partial_\beta {\bfx    } 
     + {a_\alpha}      \, \unitvec{n} 
$, describe the variation of the angular velocity along the surface. 
As shown in \SIref{II~A}, the closure conditions [\Eqref{eqn: closure conditions}] imply: 
\refstepcounter{equation} \label{eqn: compatibility equations} 
\begin{gather} 
a_\alpha = 0, 
\label{eqn: compatibility equation 1: vanishing in-plane components} 
\tag{\theequation, a} 
\h 
{a_\alpha}^\alpha = 0, 
\label{eqn: compatibility equation 2: traceless equation} 
\tag{\theequation, b} 
\h 
\partial_\alpha\left(\sqrt{g}\, \calE^{\alpha\gamma}\, {a_\gamma}^\beta\, \partial_\beta{\bfx}\right) = \zero, 
\label{eqn: compatibility equation 3: isometry equations} 
\tag{\theequation, c} 
\end{gather} 
where $\calE^{\alpha\beta} \equiv \epsilon^{\alpha\beta} / \sqrt{g}$ denotes the contravariant Levi-Civita tensor, with $\epsilon^{\alpha\beta}$ the Levi-Civita symbol (whose components are $\epsilon^{1 2} = -\epsilon^{2 1} = 1$ and $\epsilon^{1 1} = \epsilon^{2 2} = 0$) and $g \equiv \det\left(g_{\alpha\beta}\right)$ the metric determinant~\cite{David2004_Geometry_and_Field_Theory}. 
As a consequence of the traceless condition \Eqref{eqn: compatibility equation 2: traceless equation}, $\calE^{\alpha\gamma}\, {a_\gamma}^\beta$ must be symmetric. Because of its relation to (spatial) derivatives of the angular velocity vector, we refer to ${a_\alpha}^\beta$ as the traceless \emph{angular acceleration} tensor. Its use automatically incorporates the translational and rotational invariance of the metric while introducing the above nontrivial compatibility requirements. As derived in \Appendix~\textbf{B}~\textit{3} of the \SI, it is closely related to the changes in the second fundamental form induced by the isometry.

\subsection{Equilibrium stresses} \label{subsec: duality, equilibrium stresses} 
These conditions on the angular acceleration tensor resemble the equilibrium conditions for thin sheets. 
The locally generated internal forces within a thin sheet can be described in terms of a symmetric stress tensor $\sigma^{\alpha\beta}$~\cite{LL1986_Elasticity, Audoly2019_Elasticity_and_Geometry}. 
In the case of thin sheets (membranes) whose forces are exerted in-plane (i.e., with negligible bending moments), the equilibrium conditions are given by (see \SIref{II~B}): 
\begin{align} 
\partial_\alpha\left(\sqrt{g}\, \sigma^{\alpha\beta}\, \partial_\beta{\bfx}\right) = \zero. 
\label{eqn: equilibrium equations} 
\end{align} 
\subsection{The duality between isometries and equilibrium stresses} \label{subsec: duality, the duality} 
By comparing \Eqsref{eqn: compatibility equations} and \eqref{eqn: equilibrium equations}, we arrive at the duality, which is a key result of this paper. 
For any given infinitesimal isometry with specified angular acceleration tensor, an equilibrium stress may be generated via: 
\begin{align} 
                     \sigma^{\alpha\beta } 
= \left(\const\right) \calE^{\alpha\gamma}\, {a_\gamma}^\beta, 
\label{eqn: relation between equilibrium stress and isometry} 
\end{align} 
where the constant must have units of force. 
The equation is easily inverted, so that an equilibrium stress corresponds to a unique isometry. 
Note that although each isometry can be mathematically mapped onto an equilibrium stress, deforming a sheet isometrically generates no stress. 
This correspondence mirrors that of the vertex duality of triangulated origami~\cite{McInerney2020_Hidden_Symmetries, Maxwell2011_Seminal_Work, Calladine1978_Seminal_Work, Gluck1975_Rigid_Surfaces, Crapo1982_Applied_Projective_Geometry, Tachi2012_Origami_Design, Tachi2016_Rigid_Folding, Rocklin2022_Topologically_Protected_Deformations_bookChapter}. 
\par 
This duality is depicted in \Figref{fig: local geometry of a surface}~(c). 
The red and green lines in the figure respectively depict the local maximum and minimum stresses of a force equilibrium state of the surface, with line thickness proportional to the stress magnitudes. 
According to the isometry-stress duality [\Eqref{eqn: relation between equilibrium stress and isometry}], the corresponding equilibrium stress can be mapped onto an angular acceleration tensor, which can then be integrated once over the surface to obtain the angular velocity field, and integrated a second time to yield the isometric displacement field $\isoDeform$, shown as the blue arrows. 
In this way, once an equilibrium stress is known, an isometry can be generated, and vice versa. This stress/isometry duality appears analogous to those developed by Calladine for thin shells~\cite{Calladine1977_Static-Geometric_Analogy} and subsequently extended to more general surfaces (see, e.g., \Refref{Niordson1985_Shell_Theory}). 
\section{The relation between rigidity and flexibility of smooth periodic surfaces} \label{sec: rigidity and flexibility of smooth periodic surfaces}

\begin{figure*} 
\centering 
\includegraphics[width = \textwidth]{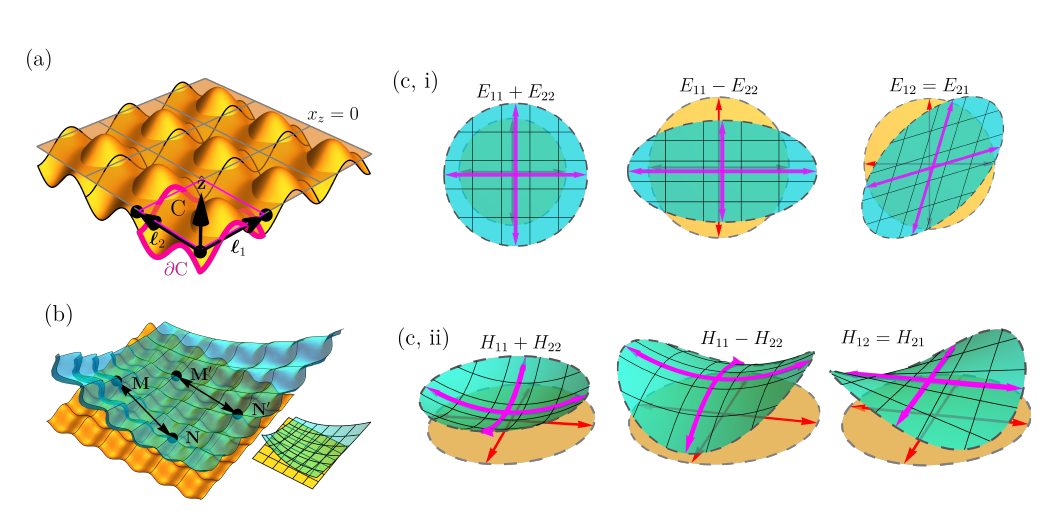} 
\caption{Geometry of doubly periodic surfaces and their shape-periodic deformations.  
(a) A doubly periodic surface is generated by repeatedly translating the unit cell (denoted by $\rmC$), which is enclosed by the magenta curve $\partial{\rmC}$ in the figure, along the directions of $\bsell_1$ and $\bsell_2$. 
The coarse-grained geometry of the doubly periodic surface is flat. 
The corresponding flat plane is spanned by $\bsell_1$ and $\bsell_2$ and has the normal vector $\unitvec{z}$. 
We choose as the zero-height level the mid-height plane of the periodic surface, which coincides with the coarse-grained flat plane. 
(b) Under a shape-periodic deformation, the Euclidean distance between any pair of points on the deformed surface remains invariant if both points are shifted by the same number of unit cells; e.g., 
$
  \norm{\bfM  - \bfN } 
= \norm{\bfM' - \bfN'} 
$ in the figure. 
The inset illustrates the corresponding deformation of the coarse-grained flat plane. 
(c) The deformation of the coarse-grained flat plane can be decomposed into six deformation modes: three in-plane modes and three out-of-plane modes, as illustrated by $\mathrm{ (c, i) }$ and $\mathrm{ (c, ii) }$, respectively. 
} 
\label{fig: smooth periodic surface and its deformations} 
\end{figure*}

\subsection{Smooth periodic surfaces} \label{subsec: rigidity and flexibility, periodic surfaces} 
This duality has important implications for the response of surfaces that are doubly periodic; i.e., those invariant under two linearly independent translations. 
Let $\bsell_{ 1 (2) } \equiv \ell^{ 1 (2) }\, \unitvec{\bsell}_{ 1 (2) }$ denote the two translation vectors. 
We can thus always generate a doubly periodic surface by first choosing the geometry of its unit cell (denoted by $\rmC$) and then repeatedly translating the unit cell by $\bsell_1$ and $\bsell_2$. [See \Figsref{fig: periodic surfaces} and \ref{fig: smooth periodic surface and its deformations}~(a)]. 
The geometry of the unit cell can be characterized by some doubly periodic vector-valued function $\bfx^\rmp\left(u^1, u^2\right)$, satisfying the periodic conditions: 
\refstepcounter{equation} \label{eqn: function periodic conditions} 
\begin{align} 
\bfx^\rmp\left(u^1, u^2\right) & = \bfx^\rmp\left(u^1 + \ell^1, u^2         \right), \tag{\theequation, a} \h 
\bfx^\rmp\left(u^1, u^2\right) & = \bfx^\rmp\left(u^1         , u^2 + \ell^2\right). \tag{\theequation, b} 
\end{align} 
And the corresponding doubly periodic surface can be accordingly parameterized as: 
\begin{align} 
                                            \bfx     \left(u^1, u^2\right) 
\equiv u^\alpha\, \unitvec{\bsell}_\alpha + \bfx^\rmp\left(u^1, u^2\right). 
\label{eqn: parametrization of doubly periodic surfaces} 
\end{align} 
\par
In the expression above, the surface can be thought of as a (macroscopic) smooth plane and a (microscopic) periodic unit cell, the latter of which we take to average to zero.
Consequently, since the  surface macroscopically resembles a plate---specifically, the one spanned by its translation vectors---the macroscopic deformations can be described using the concepts of plate deformations. 
The following notations are used to characterize the geometry of the macroscopic plate corresponding to a periodic surface: 
$\phi$ denotes the angle between the translation vectors; 
$\hat{g}_{\alpha\beta} \equiv \unitvec{\bsell}_\alpha \cdot \unitvec{\bsell}_\beta$ denotes the planar metric components with determinant $\hat{g} \equiv \det\left(\hat{g}_{\alpha\beta}\right) = \sin^2{\phi}$; 
$\hat{\calE}^{\alpha\beta} \equiv \epsilon^{\alpha\beta} / \sqrt{\hat g}$ denotes the planar contravariant Levi-Civita tensor; 
and $\unitvec{z} \equiv \unitvec{\bsell}_1 \times \unitvec{\bsell}_2 / \sqrt{\hat g}$ denotes the unit normal vector of the plate. 
\subsection{Shape-Periodic deformations} \label{subsec: rigidity and flexibility, shape-periodic deformations} 
We focus on deformations that induce the same shape in every unit cell, which we thus refer to as \emph{shape-periodic}. 
In a shape-periodic deformed state of a periodic surface, the \emph{Euclidean distance} between any pair of points is invariant if we shift both of them by an arbitrary number of unit cells, just like in the undeformed state. [See \Figref{fig: smooth periodic surface and its deformations}~(b).] 
Mathematically, let $(s, t)$ and $(S, T)$ denote the local coordinates of a pair of points on the periodic surface $\bfx$, and we write a deformed state of $\bfx$ as 
$
             {\bfx'}\left(u^1, u^2\right) 
\equiv       {\bfx }\left(u^1, u^2\right) 
     + \delta{\bfx }\left(u^1, u^2\right) 
$. 
By our definition, a shape-periodic deformation thus has to satisfy: 
\begin{align} 
\begin{split} 
&        \norm{ \bfx'\left(S            , T            \right) - \bfx'\left(s            , t            \right) } \h 
& \quad= \norm{ \bfx'\left(S + n_1\ell^1, T + n_2\ell^2\right) - \bfx'\left(s + n_1\ell^1, t + n_2\ell^2\right) } 
\end{split} 
\label{eqn: defn. of shape-periodic deformations} 
\end{align} 
for any integer tuple $(n_1, n_2)$. Rigid-body displacements (translations and rotations) are trivially shape-periodic but do not deform the surface. 
\par 
In simple terms, a shape-periodic deformation causes a periodic surface to deform uniformly, with each unit cell of the surface experiencing the same deformation. 
Based on the notion of separation of length scales, such a deformation can therefore be microscopically characterized by a doubly periodic displacement field $\SPDeform^\rmp\dependence$, which assumes the same value in every unit cell. 
The macroscopic part of the shape-periodic deformation corresponds to the uniform deformation modes of the macroscopic plate associated with the periodic surface. 
According to plate theory, these uniform deformations typically consist of uniform in-plane stretching and shearing, as well as out-of-plane bending and twisting. 
Consequently, a shape-periodic deformation can be macroscopically described using two \emph{constant} symmetric matrices, $\left(E^{\alpha\beta}\right)$ and $\left(H^{\alpha\beta}\right)$, which quantify the constant in-plane strains and out-of-plane curvature changes of the macroscopic plate, respectively. [See \Figref{fig: smooth periodic surface and its deformations}~(c).] 
\par 
Let 
$x_\alpha \equiv \bfx \cdot \unitvec{\bsell}_\alpha$ and 
$x_z      \equiv \bfx \cdot \unitvec{z     }       $ denote the macroscopic in-plane and out-of-plane components of the periodic surface $\bfx$, respectively. 
Based on the expressions for the uniform deformations of a plate~\cite{Audoly2019_Elasticity_and_Geometry, LL1986_Elasticity}, a generic shape-periodic displacement field may therefore be expressed as: 
\begin{align} 
\begin{split} 
         \SPDeform     \dependence 
& \equiv               E^{\alpha\beta}x_\alpha\,           \unitvec{\bsell}_\beta \h 
& \quad+ \frac{1}{2}\, H^{\alpha\beta}x_\alpha\, x_\beta\, \unitvec{z     } 
       -               H^{\alpha\beta}x_\alpha\, x_z    \, \unitvec{\bsell}_\beta \h 
& \quad\quad+ 
         \SPDeform^\rmp\dependence. 
\end{split} 
\label{eqn: ansatz for shape-periodic deformations} 
\end{align} 
Given the form of a shape-periodic mode $\SPDeform\dependence$, the corresponding changes in the macroscopic (i.e., unit-cell-averaged) first and second fundamental forms, $E_{\alpha\beta}$ and $H_{\alpha\beta}$, can be obtained using: 
\refstepcounter{equation} \label{eqn: expressions for the macroscopic strains and curvature changes} 
\begin{align*} 
H_{\alpha\beta} & =                  \inProd{ \frakd_\alpha{ \frakd_\beta{\SPDeform} } }{\unitvec{z     }       },       \tag{\theequation, a} \h 
E_{\alpha\beta} & = \frac{1}{2}\left(\inProd{ \frakd_\alpha{             {\SPDeform} } }{\unitvec{\bsell}_\beta } 
                  +                  \inProd{              { \frakd_\beta{\SPDeform} } }{\unitvec{\bsell}_\alpha}\right) 
                  + 
H_{\alpha\beta}\, 
x_{z          },                                                     \label{eqn: expression for the macroscopic strains} \tag{\theequation, b} 
\end{align*} 
where the lattice derivative of a function $f\dependence$ is defined as 
$
       \frakd_1{f} 
\equiv \left[  {f}\left(u^1 + \ell^1, u^2\right) 
     -         {f}\left(u^1         , u^2\right) 
       \right]              / \ell^1 
$ (and similarly for $\frakd_2{f}$). 
We note that \Eqsref{eqn: expressions for the macroscopic strains and curvature changes} exactly mirror the usual expressions used to compute the microscopic infinitesimal strains and curvature changes. [See \SIref{VI~B} for a discussion of the gauge-invariance properties of the macroscopic strains $E_{\alpha\beta}$, which give rise to the term $H_{\alpha\beta}\, x_z$ in \Eqref{eqn: expression for the macroscopic strains}.] 
In \SIref[~4]{III~B}, we further discuss the physical meaning of each term in \Eqref{eqn: ansatz for shape-periodic deformations} and explicitly show that the equation satisfies the formal definition of shape-periodic deformations [\Eqref{eqn: defn. of shape-periodic deformations}] to linear order. 
\par 
If a shape-periodic deformation is also microscopically isometric, the concept of an angular velocity field, as discussed in \Secref{subsec: duality, isometric deformations}, can be used to describe it, alongside \Eqref{eqn: ansatz for shape-periodic deformations}. 
Due to the separation of length scales, the angular velocity field corresponding to an isometric shape-periodic deformation, $\bsomega_\SP\dependence$, naturally consists of both microscopic and macroscopic parts. 
As with the surface embedding itself [see \Eqref{eqn: parametrization of doubly periodic surfaces}], the microscopic part, describing the identical local rotations of the infinitesimal area elements within each unit cell, can be captured by a doubly periodic function $\bsomega_\SP^\rmp\dependence$. 
The macroscopic part of $\bsomega_\SP$ characterizes the relative rotations between each unit cell and its neighboring unit cells. 
By shape periodicity, these relative rotations must also be identical for every unit cell and can thus be captured by a pair of \emph{constant} angular acceleration vectors, $\bfw_1$ and $\bfw_2$, which characterize the constant variations across unit cells along the two translation directions. 
Based on these facts, the angular velocity field associated with an isometric shape-periodic deformation can be parameterized as follows: 
\begin{align} 
                                \bsomega_\SP     \dependence 
\equiv u^\alpha\, \bfw_\alpha + \bsomega_\SP^\rmp\dependence. 
\label{eqn: parametrization of the angular velocity} 
\end{align} 
\par 
This form also follows from the assumption of a periodic angular velocity tensor. 
In \SIref[~1]{IV~B}, we show that the constant angular acceleration vectors $\bfw_\alpha$ and the curvature changes $H^{\alpha\beta}$, both describing the macroscopic out-of-plane isometric deformations of a periodic surface, are related as follows: 
\begin{align} 
                                       {\bfw  }_\alpha 
= \hat{\calE}^{\beta \gamma}\, 
      {H    }_{\gamma\alpha}\, \unitvec{\bsell}_\beta, 
\end{align} 
where $H_{\gamma\alpha} = \hat{g}_{\gamma\mu}\, \hat{g}_{\alpha\nu}\, H^{\mu\nu}$, with $\hat{g}_{\alpha\beta}$ denoting the planar metric components. 
\par 
To summarize, any shape-periodic deformation mode can be characterized by the six macroscopic quantities $E_{\alpha\beta}$ and $H_{\alpha\beta}$, together with a microscopic periodic function $\delta{\bfx_\SP^\rmp}$. It is useful to represent the former as a single Voigt-type~\cite{Helnwein2001_Vector_Representation_of_Tensors} vector $\bfv$, whose exact form will become convenient later: 
\begin{align} 
       \bfv^\intercal 
\equiv \left( 
       \frac{1}{ \sqrt{\hat g} }H_{22}, 
       \frac{1}{ \sqrt{\hat g} }H_{11}, 
     - \frac{1}{ \sqrt{\hat g} }H_{12}, 
                                E_{11}, 
                                E_{22}, 
            {2}                 E_{12} 
       \right), 
\label{eqn: vector representation of a MP mode} 
\end{align} 
where $\hat{g}$ denotes the determinant of the planar metric tensor. 
The vector space formed by all such six-dimensional vectors is called the \emph{deformation phase space}. 
\par 
The linearity of these deformations, combined with the isometry-stress duality [\Eqref{eqn: relation between equilibrium stress and isometry}] has important consequences for the number and type of isometries that a doubly periodic sheet can possess. Consider a large sheet whose boundaries are held at positions and orientations so as to enforce an overall stretching of the unit cell give by $E^{\alpha\beta}$ and bending given by $H^{\alpha\beta}$ corresponding to some $\bfv$. The system is then allowed to relax to an energetic minimum, which implies that any stress that is present must be given by the equilibrium equation [\Eqref{eqn: equilibrium equations}]. If such a stress is present, the isometry-stress duality [\Eqref{eqn: relation between equilibrium stress and isometry}] can be used to map the original deformation $\bfv$ to some new deformation $\wt{\bfv}$ that corresponds to an isometry or is a zero vector that corresponds to no deformation at all. We can denote this map, which will in general depend on material quantities such as bulk and shear moduli as well as the surface geometry, as $\wt{\bfv} = \bfM\, \bfv$. The number of isometries is thus both the dimension of the right nullspace and the dimension of the column space of the six-dimensional linear operator $\bfM$. From the rank-nullity theorem, the number of isometries must be exactly three. More loosely, we can say that because the isometries and stresses are in one-to-one correspondence, they must form linearly independent three-dimensional subspaces within the six-dimensional deformation space. 

The existence of exactly three linear isometries is surprising. For example, the flat surface [\Figref{fig: periodic surfaces}~(a)] has an infinite number of linear isometries corresponding to small local displacements in the transverse direction. However, any such purely local deformation induces no global strain or curvature. Instead, the three global isometries of the surface consist of the three independent components of $H^{\alpha\beta}$, with the sheet curling up like a cylinder or a dome. 

Thus, contrary to the common perception that adding corrugation to cardboard or curving a pizza slice so it does not bend downward adds strength to the structure, we find that any (periodic) sheet will always have three global, shape-periodic isometries and three shape-periodic deformations that are not isometries. We refer to this result as \emph{conservation of flexibility}. 
\subsection{Deformation energy and boundary work} \label{subsec: rigidity and flexibility, energy and work} 
Having characterized the deformation modes of interest, we now investigate how they couple, leading to the relation between the rigidity and flexibility of a periodic surface. 
We begin by formulating the energies incurred by an energy-minimizing deformation. 
\par 
Recall that we consider small elastic deformations in the limit of large \FvK{} number [\Eqref{eqn: FvK number}], where stretching dominates over bending. 
In this regime, the corresponding energy is typically captured by the following quadratic membrane energy functional~\cite{Niordson1985_Shell_Theory}: 
\begin{align} 
  E_\rmm 
= \frac{1}{2}\, \iint_\rmC \df{\calA}\      C^{\alpha\beta\gamma\rho}\, \varepsilon_{\gamma\rho}\, \varepsilon_{\alpha\beta} 
= \frac{1}{2}\, \iint_\rmC \df{\calA}\ \sigma^{\alpha\beta          }\,                            \varepsilon_{\alpha\beta}. 
\label{eqn: expression for membrane energy} 
\end{align} 
Here, the integral is taken over a unit cell ($\rmC$) of a periodic surface, and $\df{\calA}$ representing its area element. 
The symbol $\varepsilon_{\alpha\beta}$ denotes the strain tensor, which quantifies the change in the surface metric induced by the deformation. 
In the elastic regime, the strain tensor is linearly related to the stress tensor via the generalized Hooke's law: $\sigma^{\alpha\beta} = C^{\alpha\beta\gamma\rho}\, \varepsilon_{\gamma\rho}$, where $C^{\alpha\beta\gamma\rho}$ denotes the components of the rank-four two-dimensional (proportional to the thickness of the sheet) stiffness tensor, which encodes the elastic properties of the surface material, such as Young's modulus and Poisson's ratio. 
\par 
At equilibrium, the internal stresses in the surface bulk balance out, satisfying the equilibrium equations [\Eqref{eqn: equilibrium equations}]. 
As a result, by the principle of energy conservation, the deformation energy stored in the bulk must equal the net work done by the deformation on the boundary. 
As shown in \SIref{II~B}, the stored energy can be expressed as: 
\begin{align} 
  E_\rmm^\eq = W 
= \frac{1}{2}\oint_{\partial\rmC} \df{u^\gamma}\, \inProd{ 
   \calE_{\alpha\gamma}\, 
  \sigma^{\alpha\beta }\, \partial_\beta {\bfx} }{ 
                                   \delta{\bfx} }, 
\label{eqn: membrane energy and boundary work, stress version} 
\end{align} 
where $\partial{\rmC}$ denotes the boundary of the unit cell $\rmC$, $\calE_{\alpha\gamma}$ is the inverse of the contravariant Levi-Civita tensor, and $\delta{\bfx}$ is the displacement field associated with the deformation. 
\par 
The stress tensor $\sigma^{\alpha\beta}$ in \Eqref{eqn: membrane energy and boundary work, stress version} corresponds to an energy-minimizing deformation. 
For a periodic surface, the isometry-stress duality [\Eqref{eqn: relation between equilibrium stress and isometry}] enables us to express the stored energy in terms of the angular velocity field associated with an isometric shape-periodic mode of the surface, as follows: 
\begin{align} 
  E_\rmm^\eq = W 
= \frac{1}{2}\left(\const\right)\oint_{\partial\rmC} \df{u^\alpha}\, \inProd{ 
  \partial_\alpha{\wt{\bsomega}_\SP} }{ 
           \delta{   {\bfx    }_\SP} }, 
\label{eqn: membrane energy and boundary work, isometry version} 
\end{align} 
where the tilde notation is henceforth used to denote quantities associated with the virtual isometric mode mapped from an energy-minimizing deformation. This represents the actual physical energy as a purely geometric quantity, given in terms of an isometric deformation.
\par 
As a special case of \Eqref{eqn: membrane energy and boundary work, isometry version}, when there is a prestress and the imposed deformation is itself isometric---thus incurring no energy cost---the equation implies (see \SIref[~2]{IV~C} for further discussion): 
\begin{align} 
  \oint_{\partial\rmC} \df{u^\alpha}\inProd{ 
  \partial_\alpha{\bsomega_\SP^a} }{ 
    \delta_\iso  {\bfx    _\SP^b} } 
= 0, 
\label{eqn: membrane energy and boundary work, special isometry version} 
\end{align} 
where the indices $a$ and $b$ label two arbitrary isometric shape-periodic modes of the periodic surface. 
\subsection{The relation between rigidity and flexibility} \label{subsec: rigidity and flexibility, the relation} 
The coupling between an isometric mode and its corresponding energy-minimizing mode, as well as with a second isometric mode, is described by \Eqsref{eqn: membrane energy and boundary work, isometry version} and \eqref{eqn: membrane energy and boundary work, special isometry version}, respectively. 
\par 
Substituting the expressions for the shape-periodic displacement field [\Eqref{eqn: ansatz for shape-periodic deformations}] and the angular velocity field [\Eqref{eqn: parametrization of the angular velocity}] allows us to express the couplings in terms of the corresponding macroscopic strains and curvature changes. 
After considerable algebra, detailed in \SIref{V~B}, the energy can be expressed solely in terms of the actual changes in the macroscopic fundamental forms associated with the imposed deformation ($E_{\alpha\beta}$ and $H_{\alpha\beta}$), and the virtual changes in the forms corresponding to the mapped isometry ($\wt{E}_{\alpha\beta}$ and $\wt{H}_{\alpha\beta}$): 
\begin{align} 
\begin{split} 
    E_\rmm^\eq = W 
& = \frac{1}{2}\left(\const\right)\norm{\bsell_1 \times \bsell_2} 
    \h 
& \quad\times 
    \hat{\calE}^{\alpha\mu}\, 
    \hat{\calE}^{\beta \nu}\left( 
    \wt{H}_{\mu\nu}\,    {E}_{\alpha\beta} 
  -    {H}_{\mu\nu}\, \wt{E}_{\alpha\beta} 
    \right). 
\end{split} 
\label{eqn: relation between rigidity and flexibility, general version} 
\end{align} 
where, intriguingly, all the microscopic periodic components are integrated out. 
The fact that the couplings depend solely on the macroscopic quantities implies that all doubly periodic surfaces can be treated macroscopically as plates, with their mechanical properties modified by the geometries of their unit cells---a notion also established using homogenization techniques (see, e.g., \Refsref{Xu2024_Effective_Plate_Theory_for_Origami} and \onlinecite{Li2025_Homogenizing_Non-rigid_Origami}). 
\par 
Of exceptional interest is the case, mentioned previously, in which the deformation is itself isometric. In that case, the energy vanishes and, upon dividing out constant terms, we have one of our main results, which we term the \emph{surface mode compatibility condition}. For any two isometric modes of a single periodic surface $a, b$: 
\begin{align} 
  \epsilon^{\alpha\mu}\, 
  \epsilon^{\beta \nu}\left( 
  H^a_{\mu\nu}\, E^b_{\alpha\beta} 
- H^b_{\mu\nu}\, E^a_{\alpha\beta} 
  \right) 
= 0. 
\label{eqn: relation between rigidity and flexibility, special version} 
\end{align} 
In the vector representation introduced previously [\Eqref{eqn: vector representation of a MP mode}], the couplings assume the following form: 
\begin{align} 
E_\rmm^\eq = W & \propto \wt{\bfv}  ^\intercal\, \bfJ\, \bfv                      , \label{eqn: relation between rigidity and flexibility, general matrix version} \h 
             0 &       =    {\bfv}_a^\intercal\, \bfJ\, \bfv_b^{\phantom\intercal}, \label{eqn: relation between rigidity and flexibility, special matrix version} 
\end{align} 
where $\bfJ$ denotes the canonical symplectic matrix~\cite{Morrison1998_Hamiltonian_Fluid}: 
\begin{align} 
       \bfJ 
\equiv \begin{pmatrix} 
       \phantom{-}\zero_{3 \times 3} &  \one_{3 \times 3} \h 
               {-} \one_{3 \times 3} & \zero_{3 \times 3} 
       \end{pmatrix}. 
\end{align} 
\par 
The fact that the couplings between shape-periodic modes can be expressed as symplectic inner products, as shown in \Eqsref{eqn: relation between rigidity and flexibility, general matrix version} and \eqref{eqn: relation between rigidity and flexibility, special matrix version}, means that the deformation phase space possesses a natural symplectic structure, akin to the phase space of classical Hamiltonian mechanics. 
Specifically, the curious combination $\left(H_{22}, H_{11}, -H_{12}\right) / \sqrt{\hat g}$ in \Eqref{eqn: vector representation of a MP mode}---more compactly expressed as $\Sigma^{\alpha\beta} \equiv \hat{\calE}^{\alpha\mu}\, \hat{\calE}^{\beta\nu}\, H_{\mu\nu}$---plays the role of generalized coordinates in classical mechanics. 
The macroscopic strain components $E_{\alpha\beta}$, with lowered indices, are conjugate to $\Sigma^{\alpha\beta}$, thereby serving as the conjugate momenta associated with the generalized coordinates. 
\par 
In the symplectic language, as shown in \Eqref{eqn: relation between rigidity and flexibility, general matrix version}, the energy cost incurred by an energy-minimizing shape-periodic mode can be interpreted geometrically as the phase-space area of the parallelogram formed by its associated vector $\bfv$ and that of the corresponding isometric mode, $\wt{\bfv}$. 
More remarkably, the mode compatibility condition \Eqref{eqn: relation between rigidity and flexibility, special matrix version} implies that the subspace formed by the three isometric shape-periodic modes is \emph{Lagrangian}, with the property that the symplectic inner product between any pair of its elements vanishes~\cite{Arnold1991_Mechanics, da_Silva2008_Symplectic_Geometry, Jeffs2022_Symplectic_Geometry}. 
\par 
The mode compatibility condition constrains the possible forms of isometries for a given surface, thereby restricting its intrinsic responses to stretching and bending---particularly when some of its isometry data are known. 
Specifically, if a given vector $\bfv$ corresponds to an isometric mode, then, based on \Eqref{eqn: relation between rigidity and flexibility, special matrix version}, the mode associated with the vector $\bfJ\, \bfv$ will inevitably incur an energy cost. 
\par 
For concreteness, consider singly corrugated surfaces as an example [see \Figref{fig: periodic surfaces}~(b)]. 
Uniformly stretching such surfaces along the corrugation direction (and flattening the corrugations) does not incur an energy cost and is therefore an isometric mode. 
By aligning the first translation vector $\bsell_1$ with the corrugation direction, we can represent this isometry as $\bfv_\rms^\intercal \equiv \left(0, 0, 0, E_{11}, 0, 0\right)$. 
Given this information, \Eqref{eqn: relation between rigidity and flexibility, special matrix version} implies that bending the surface along the second translation vector $\bsell_2$, represented by $\bfv_\rmb^\intercal \equiv \left(H_{22}, 0, 0, 0, 0, 0\right)$, cannot be isometric, as consistent with everyday experience (e.g., bending a corrugated piece of paper). 
More generally, \Eqref{eqn: relation between rigidity and flexibility, special matrix version} goes beyond this simple scenario, showing that \emph{no} periodic sheet---regardless of its unit cell geometry---can simultaneously exhibit the uniaxial strain mode of the corrugated sheet in this example and the transverse uniaxial bending mode of a flat sheet. 
\par 
In the special case where the isometric modes of a periodic surface involve only pure stretching or pure bending, the mode compatibility condition [\Eqref{eqn: relation between rigidity and flexibility, special version}] reduces to: 
\begin{align} 
\begin{split} 
  0 
= \epsilon^{\alpha\mu}\, 
  \epsilon^{\beta \nu}\, E^a_{\alpha\beta}\, H^b_{\mu\nu} 
=                        E^a_{1     1    }\, H^b_{2  2  } 
+                        E^a_{2     2    }\, H^b_{1  1  } 
-                    2\, E^a_{1     2    }\, H^b_{1  2  }, 
\end{split} 
\label{eqn: relation between rigidity and flexibility, Nassar's version} 
\end{align} 
which recovers the result reported in \Refref{Nassar2024_Rigidity_and_Flexibility}, obtained via homogenization techniques and the method of averaging~\cite{Strogatz1994_Dynamics}. 
Consequently, as established in \Refref{Nassar2024_Rigidity_and_Flexibility}, a coordinate transformation that sets either $E^a_{1 2}$ or $H^b_{1 2}$ to zero causes the mode compatibility condition to imply that the surface possesses opposite macroscopic in-plane and out-of-plane Poisson's ratios: 
\begin{align} 
 \hat{\upsilon}_\inn \equiv  \frac{ E^a_{2' 2'} }{ E^a_{1' 1'} } 
                          = -\frac{ H^b_{2' 2'} }{ H^b_{1' 1'} } 
                     \equiv 
-\hat{\upsilon}_\out, 
\end{align} 
where the primed indices refer to the transformed coordinates. 
\section{Origami and triangulation of smooth surfaces} \label{sec: verification of the results}

\begin{figure*} 
\centering 
\includegraphics[width = \linewidth]{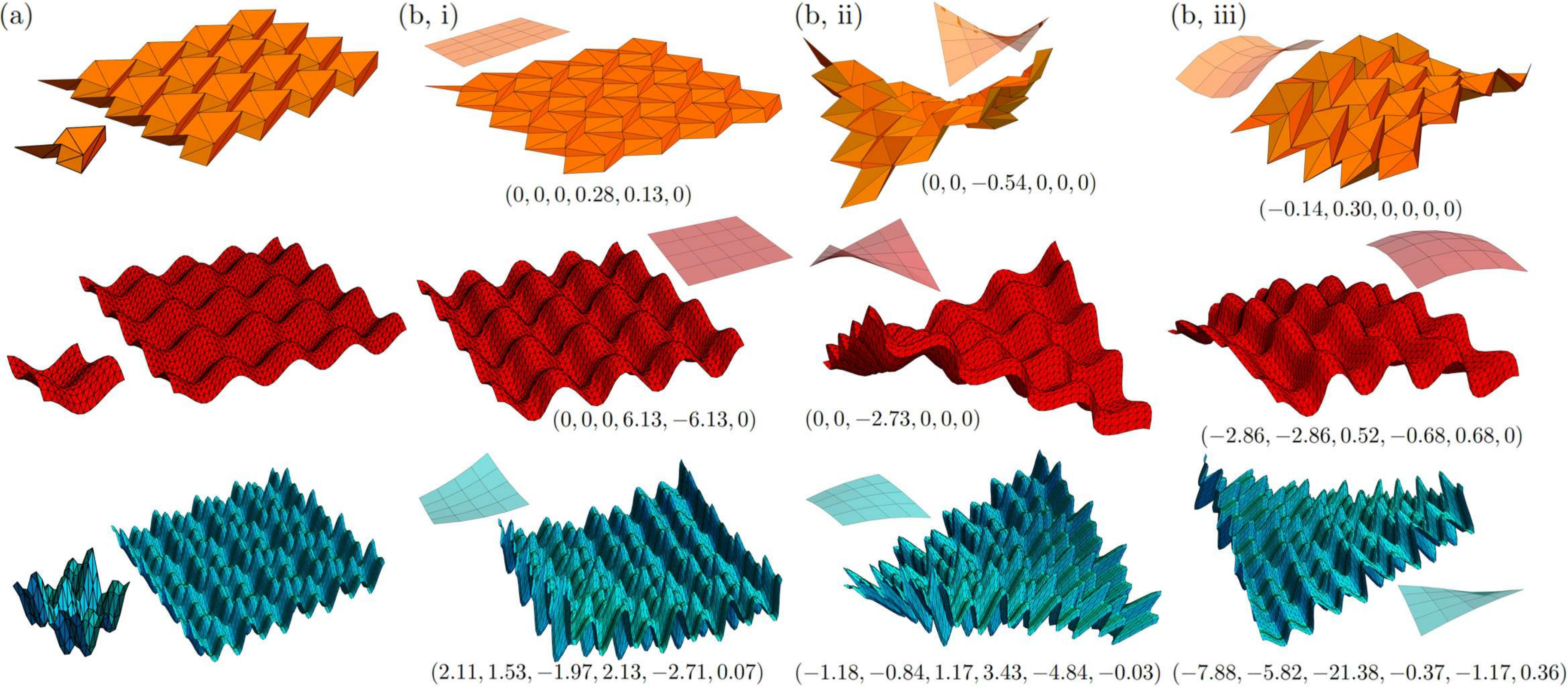} 
\caption{
Numerically computed isometries of  surfaces. 
(a) The  surfaces and their triangulations used in the simulations: a Miura-ori (first row), a graph of translation (second row) and a perturbed graph of translation (third row). 
Explicit parametrizations of the latter two surfaces are provided in Fig.~2 of the \SI. 
Unit cell geometries are shown in the corresponding insets. 
The unit cell of the triangulated Miura-ori comprises eight triangular faces, whereas each of the other two triangulated surfaces has $2\calN^2 = 288$ triangular faces in its unit cell, where $\calN$ denotes the number of grid squares per row (and per column) and serves as the parameter controlling the triangulation density. 
(b) The three numerically computed linear isometries are shown in the three columns, along with their vector representations [of the form in \Eqref{eqn: vector representation of a MP mode}] and the corresponding macroscopic plate deformations, displayed in the insets. 
The macroscopic strains and curvature changes associated with each mode are computed using \Eqsref{eqn: expressions for the macroscopic strains and curvature changes}, with the surface average heights set to zero, as shown for example in \Figref{fig: smooth periodic surface and its deformations}~(a). (See \SIref[~3]{VII~B} for more details.) 
} 
\label{fig: triangulated surfaces and their isometries} 
\end{figure*}

Although our analysis has assumed smooth surfaces, the analytical results for periodic surfaces---\Eqsref{eqn: relation between rigidity and flexibility, general version} and \eqref{eqn: relation between rigidity and flexibility, special version}---also apply to piecewise smooth geometries, such as origami tessellations and triangulated surfaces that are smooth except at sharply creased interfaces [\Figref{fig: triangulated surfaces and their isometries}~(a)]. 
As discussed in \SIref{VII~A}, this result can be derived by decomposing the line integral around the unit cell boundary into integrals along the creases, rendering the relevant quantities smooth within the regions bounded by the creases, following the approach of \Refref{Nassar2024_Rigidity_and_Flexibility}. 
\par 
Origami is a rich and vibrant field~\cite{Misseroni2024_Origami_Review, Meloni2021_Origami_Review}, and triangulations are commonly used to model the isometries of origami tessellations, including those with non-triangular faces, to capture face bending~\cite{Tachi2009_Simulating_Origami, Tachi2010_Simulating_Origami, Zhu2022_Simulating_Origami_Review, McInerney2020_Hidden_Symmetries}. 
Origami and triangulated surfaces thus serve as a natural testbed for our results: the presence of three isometric modes that obey the mode compatibility condition [\Eqsref{eqn: relation between rigidity and flexibility, special version} or \eqref{eqn: relation between rigidity and flexibility, special matrix version}]. 
As detailed in \SIref[~2]{VII~B}, the isometries of a triangulated surface can be characterized by changes in the dihedral angles between adjoining triangular faces, subject to the requirement that the total change around any closed loop satisfies the closure conditions in \Eqref{eqn: closure conditions}. 
In this formulation, the scalar dihedral angle changes serve the role of the angular acceleration tensor introduced around \Eqsref{eqn: compatibility equations}. 
\par 
In our simulations, after triangulating the selected surfaces, we obtain their isometry data---the dihedral angle changes and, subsequently, the isometric displacements---by computing the zero modes of their associated compatibility matrices (see \SIref[~2]{VII~B} for details). 
With the isometry data in hand, \Eqsref{eqn: expressions for the macroscopic strains and curvature changes} are used to extract the macroscopic strains and curvature changes corresponding to the isometries. 
A caveat applies when we report the numbers: Because the macroscopic strains are gauge-dependent (see \SIref{VI~B} for details), the average heights of the selected surfaces are set to zero to produce physically intuitive results; in particular, in this height gauge, bending a surface leaves its mid-height plane strainless. 
\par 
As shown in the first row of \Figref{fig: triangulated surfaces and their isometries}, a triangulated Miura-ori indeed generates three isometries that lead to global strains and curvatures whose forms are consistent with those reported in the literature~\cite{Schenk2013_Miura-Ori, Wei2013_Miura-Ori}. 
With the six-dimensional vectors [of the form given in \Eqref{eqn: vector representation of a MP mode}] corresponding to these isometries, as shown in the insets, one can verify that the mode compatibility condition is satisfied within numerical precision for each pair of isometries. 
This confirmation not only validates our theory but also demonstrates that the assumption that bending is concentrated at the creases is exact. 
In this way, we show that our results apply not only to strictly smooth surfaces but also to origami tessellations. 
\par
\begin{figure}[tbh] 
\centering 
\includegraphics[width = \linewidth]{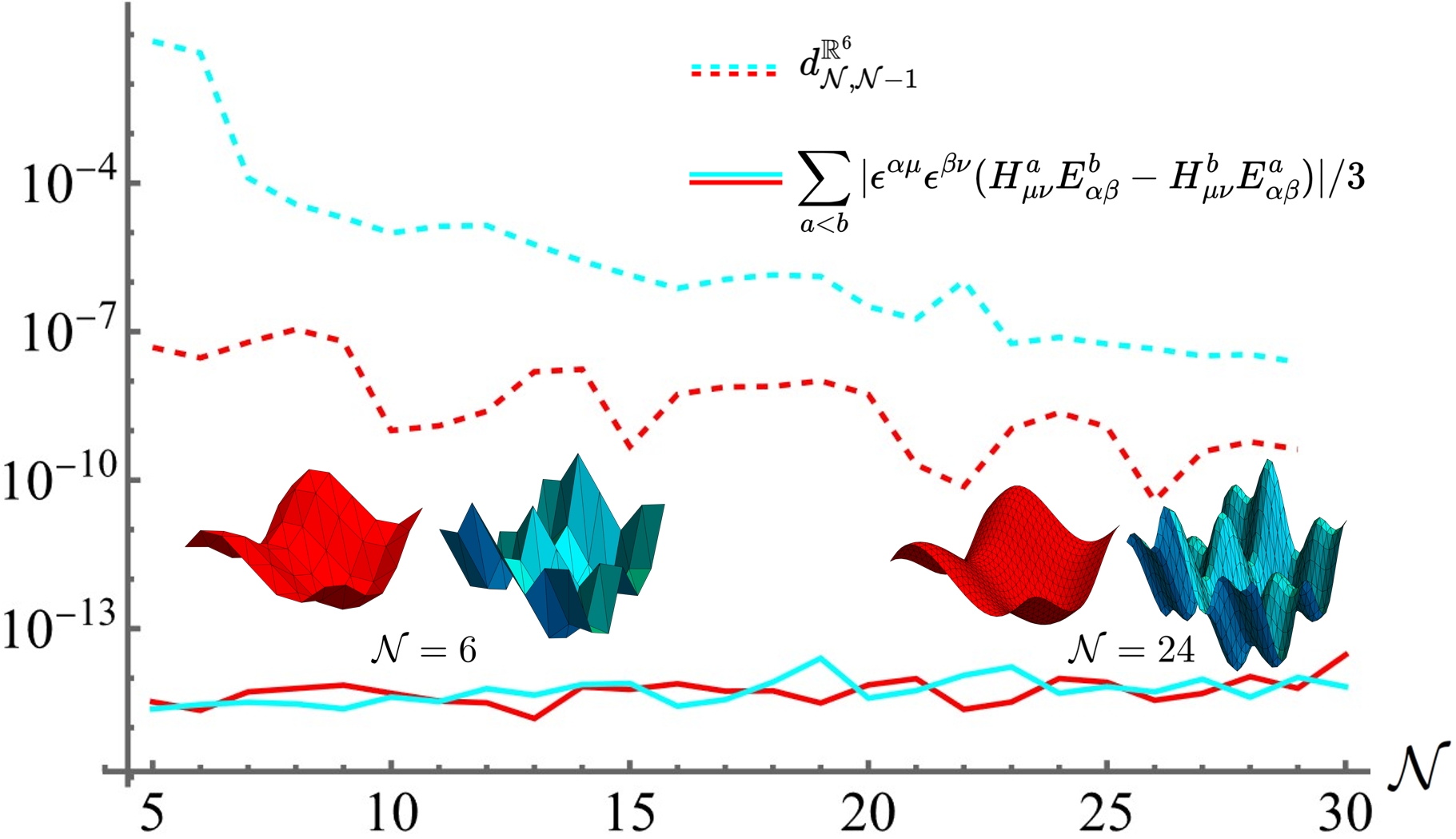} 
\caption{
Convergence of isometric subspaces and validation of the mode compatibility condition. 
The red and cyan curves correspond to the red and cyan surfaces in \Figref{fig: triangulated surfaces and their isometries}, respectively. 
For each triangulated surface, $\calN$ denotes the number of grid squares per row and per column, so that the unit cell contains $2\calN^2$ triangular faces, as seen in the insets. 
The solid curves show the values obtained by substituting the isometry data into the mode compatibility condition [\Eqsref{eqn: relation between rigidity and flexibility, special version} or \eqref{eqn: relation between rigidity and flexibility, special matrix version}], averaged over all non-redundant isometry pairs. 
The dashed curves depict the convergence of isometric subspaces for the triangulated surfaces. 
Each data point measures the ``distance'' between subspaces corresponding to $\calN$ and $\calN - 1$ grid squares per row and column, quantified by $d^{\bbR^6}_{\calN,\, \calN - 1}$, a Euclidean-based metric for comparing equal-dimensional subspaces in $\bbR^6$ (see \SIref{VII~C} for details and for convergence behavior using other types of metrics). 
} 
\label{fig: convergence plot} 
\end{figure}

 
The second row of \Figref{fig: triangulated surfaces and their isometries} corresponds to triangulations of a graph of translation~\cite{Struik1988_Classical_DG}, for which two of the three isometric modes are known analytically~\cite{Nassar2023_Isometries_of_Translation_Surfaces}. 
In contrast to the case of triangulated Miura-ori, we now use a finite number of triangles to approximate an ideal curved surface. 
In \Figref{fig: convergence plot}, we observe that, as the number of triangles used increases, the isometries always obey the mode compatibility condition and gradually converge to a fixed subspace, indicating that the isometries of the graph of translation are faithfully reproduced. 
Indeed, two of the three isometries that can be determined analytically lie within the subspace. 
Finally, the last row of \Figref{fig: triangulated surfaces and their isometries} depicts a perturbed graph of translation whose isometries are not known analytically and involve both macroscopic stretching and bending. 
As illustrated in \Figref{fig: convergence plot}, our results hold for the surface. 
This intrinsic coupling between stretching and bending arises for generic surfaces, in contrast with the modes considered in~\Refref{Nassar2024_Rigidity_and_Flexibility}, which were either pure bending or pure stretching. 
\par 
While such triangulations can, in principle, approximate any smooth surface, numerical limitations may hinder the convergence of isometric subspaces for arbitrary geometries, possibly due to rigidifying curves~\cite{Mosleh2017_Surfaces_with_Rigidifying_Curves} that arise near regions of vanishing Gaussian curvature. 
Such regions must exist in any of the periodic surfaces considered here, as the Gauss-Bonnet theorem implies that the average Gaussian curvature over a unit cell vanishes---even under isometric deformations, since Gaussian curvature is intrinsic.

\section{Discussion} \label{sec: conclusion} 
We have uncovered a general duality associated with any piecewise smooth thin membrane surface: Any isometric deformation can be mapped to an equilibrium stress (and thus energy-minimizing deformation), and vice-versa. 
For periodic surfaces, the duality establishes that, within the six-dimensional space of combined macroscopic in-plane (coarse-grained strain) and out-of-plane (coarse-grained bending) deformation modes, exactly three modes (and their linear combinations) must correspond to microscopic isometries, thereby being energetically favored. 
Moreover, we find that the three isometric modes couple symplectically with each other, satisfying the surface mode compatibility condition [\Eqsref{eqn: relation between rigidity and flexibility, special version} or \eqref{eqn: relation between rigidity and flexibility, special matrix version}]. 
In the special case for which the isometric modes consist of pure macroscopic strains or curvature changes, the mode compatibility condition reduces to the exciting result recently established in the literature~\cite{Nassar2024_Rigidity_and_Flexibility}, thus providing a mathematical foundation for the equal and opposite macroscopic in-plane and out-of-plane Poisson's ratios possessed by several types of origami tessellations---including the Miura-ori~\cite{Schenk2013_Miura-Ori, Wei2013_Miura-Ori}, the eggbox~\cite{Nassar2017_Eggbox} and the morph pattern~\cite{Pratapa2019_Morph}. 
\par 
Our main results for periodic surfaces---namely, the symplectic energy-cost formula [\Eqsref{eqn: relation between rigidity and flexibility, general version} or \eqref{eqn: relation between rigidity and flexibility, general matrix version}] and the mode compatibility condition---are universal in the sense that neither expression depends on the microscopic details of the unit cells of a periodic surface. 
Despite this, it is important to emphasize that, unlike other existing theories for origami tessellations~\cite{Czajkowski2023_Orisometry}, the theory developed in this work is an exact microscopic theory that does not rely on any coarse-graining or homogenization techniques. 
\par 
There are a few puzzles in our work that remain to be solved. 
First, as mentioned in \Secref{sec: verification of the results}, it remains unclear why the isometric Lagrangian subspace associated with a general triangulated periodic surface does not converge as the number of triangles increases, even though the mode compatibility condition is consistently obeyed. 
Second, the revealed symplectic structure of the deformation phase space and the Lagrangian nature of its isometric subspace point to deeper underlying principles that are not yet fully understood. 
It would also be interesting to investigate whether these structures, as well as the isometry-stress duality, persist when the assumptions of membrane states and small deformations are relaxed. 
\par 
The theoretical framework developed in this work can be extended in the following directions. 
First, similar to the direction pursued for origami tessellations~\cite{McInerney2020_Hidden_Symmetries}, one could incorporate the constraints imposed by geometric nonlinearities into the theory to investigate which linear isometric modes of a periodic surface extend to the nonlinear regime, where microscopic deformations are no longer small. 
Another promising direction would be to introduce extra microscopic structures, such as symmetries, fluctuations or disorder, into the unit cells of a periodic surface and examine their effects on the surface's macroscopic behavior. 
Finally, perhaps of greater practical interest is the inverse problem: How to design unit cells that realize a desired set of isometries permitted by the mode compatibility condition. 
\par 
In conclusion, our work reveals the intrinsic mechanical behaviors of general periodic surfaces, and we anticipate that our results, along with future investigations, will prove valuable for the industrial design of corrugated structures, with specific applications such as cardboard, shipping containers and other related materials. 
\section{ACKNOWLEDGMENTS} The authors acknowledge financial support from the Army Research Office through the MURI program \#W911NF2210219 (DZR) and through the National Science Foundation CAREER program \#2338492 (DZR), NSF CMMI-
2247095 (CDS) and through NSF PHY-2309135 to the Kavli Institute for Theoretical Physics (KITP) (DZR). 
We would also like to thank Evan Dickson for fabricating the 3D-printed surface and folding the Miura-ori shown in \Figref{fig: periodic surfaces}. 
\bibliographystyle{unsrt} 
\bibliography{Refs_MT} 
\end{document}


\count\footins = 1000 
\author{Wenqian Sun} 
\affiliation{School of Physics, Georgia Institute of Technology, Atlanta, GA 30332, USA} 
\author{Yanxin Feng} 
\affiliation{School of Physics, Georgia Institute of Technology, Atlanta, GA 30332, USA} 
\author{Christian D. Santangelo} 
\affiliation{Department of Physics, Syracuse University, Syracuse, New York 13244, USA} 
\author{D. Zeb Rocklin} 
\affiliation{School of Physics, Georgia Institute of Technology, Atlanta, GA 30332, USA} 
\title{Supplementary Information for Geometric Mechanics of Thin Periodic Surfaces} 
\date{\today} 

\maketitle 

\tableofcontents 
\newpage 
\section{A brief review of the differential geometry of surfaces} 
In this section, we provide a brief review of the differential geometry of curved surfaces and establish the notation used in our derivations. 
In this work, we primarily study smooth, simply connected, two-dimensional surfaces embedded in three-dimensional Euclidean space $\bbR^3$. 
It is worth noting that the geometry of the surfaces considered is generally non-Euclidean, in the sense that they do not correspond to a bent configuration of a flat sheet, even though the embedding space is Euclidean. 
Such a surface can always be represented locally by a three-dimensional Euclidean vector field $\bfx\left(u^1, u^2\right)$, where $u^1$ and $u^2$ denote the corresponding local coordinates~\cite{do_Carmo2016_Classical_Differential_Geometry}. 
In our formulation, the two local coordinates have the dimension of length. 
Throughout this work, we use bold symbols to represent vectors in $\bbR^3$. 
\subsection{Tangent vectors and one-forms} \label{subsec: DG, tangent vectors and one-forms} 
The vector fields $\partial_\alpha{\bfx} \equiv \partial{\bfx} / \partial{u^\alpha}$ ($\alpha \in \{1, 2\}$) are tangent to the surface coordinate lines, 
$\bfx\left(u^1         , u^2 = \const\right)$ and 
$\bfx\left(u^1 = \const, u^2         \right)$, 
and hence span the tangent planes of a surface. 
For this reason, they are often called the coordinate basis vector fields. 
Any arbitrary tangent vector field $\bfv\left(u^1, u^2\right)$ on the surface can be expressed as a linear combination of the coordinate basis vector fields: 
\begin{align} 
                                 {\bfv} 
\equiv v^\alpha\, \partial_\alpha{\bfx}, 
\end{align} 
where $v^\alpha\left(u^1, u^2\right)$ denotes the \emph{contravariant} components of $\bfv$. 
The Einstein summation convention is implied over repeated indices. 
\par 
One-forms (or, equivalently, covectors) are linear functionals that take a surface tangent vector as their argument and output a real number. 
They form the cotangent planes of the surface, which are isomorphic to the surface tangent planes. 
The basis one-forms $\df{u^\alpha}$, which span the surface cotangent planes, are defined with respect to the coordinate basis vectors: 
\begin{align} 
\df{u^\alpha}\left(\partial_\beta{\bfx}\right) \equiv \delta^\alpha_\beta, 
\label{eqn: defn. of the basis one-forms} 
\end{align} 
where $\delta^\alpha_\beta$ is the Kronecker delta symbol. 
As with the tangent vector fields, any arbitrary one-form field $\bsomega\left(u^1, u^2\right)$ can be expressed as a linear combination of the basis one-forms: 
\begin{align} 
       \bsomega 
\equiv   \omega_\alpha\, \df{u^\alpha}, 
\end{align} 
where $\omega_\alpha\left(u^1, u^2\right)$ denotes the \emph{covariant} components of $\bsomega$. 
\par 
With the one-forms introduced, tangent vectors can also be viewed as linear functionals which map a one-form to a real number. 
Therefore, the relation between the basis one-forms and the coordinate basis vectors [\Eqref{eqn: defn. of the basis one-forms}] can be written equivalently as: 
\begin{align} 
\partial_\beta{\bfx}\left(\df{u^\alpha}\right) \equiv \delta^\alpha_\beta. 
\end{align} 
\subsubsection{Basis tensors} 
Like the tangent vectors and one-forms, higher-rank tensors can also be decomposed into a linear combination of basis tensors, with the corresponding tensor components. 
The basis tensors are constructed by taking tensor products of the tangent vectors and one-forms. 
Take as examples the following rank-two tensors, which are of primary interest in this work: 
\begin{itemize} 
\item The stress tensor $\bssigma$ (discussed in more detail in \Secref{subsec: the isometry-stress duality, equilibrium stress}), as a $\binom{2}{0}$ tensor, can be expressed in       component form as: 
      \begin{align} 
             \bssigma 
      \equiv   \sigma^{\alpha\beta}\, \partial_\alpha{\bfx} \otimes 
                                      \partial_\beta {\bfx}, 
      \label{eqn: introducing the stress tensor} 
      \end{align} 
      where $\sigma^{\alpha\beta}$ denotes the stress components and $\otimes$ denotes the tensor product. 
\item The shape operator $\shape$ (discussed further in \Secref{subsec: DG, the second fundamental form}) is a $\binom{1}{1}$ tensor and can be expressed in component form as: 
      \begin{align} 
             \shape 
      \equiv {b^\alpha}_\beta\, \df{u^\beta} \otimes \partial_\alpha{\bfx}, 
      \label{eqn: introducing the shape operator} 
      \end{align} 
      where the component matrix $\left( {b^\alpha}_\beta \right)$ encodes information about the surface curvatures. 
\item The metric tensor $\bfg$ (discussed further in \Secref{subsec: DG, the first fundamental form}) is a $\binom{0}{2}$ tensor and can be expressed in component form as: 
      \begin{align} 
             \bfg 
      \equiv    g_{\alpha\beta}\, \df{u^\alpha} \otimes 
                                  \df{u^\beta }, 
      \label{eqn: introducing the metric tensor} 
      \end{align} 
      where the components $g_{\alpha\beta}$ characterize the infinitesimal distance between neighboring points on the surface. 
\end{itemize} 
\subsubsection{The wedge product and exterior derivative} 
The wedge product of two one-forms is the antisymmetrized tensor product of them. 
For example, for the basis one-forms, it is given by: 
\begin{align} 
       \df{u^1} \wedge  \df{u^2} 
\equiv \df{u^1} \otimes \df{u^2} 
     - \df{u^2} \otimes \df{u^1}. 
\end{align} 
By this definition, it follows that: 
\begin{align} 
   \df{u^1} \wedge \df{u^1} 
=  \df{u^2} \wedge \df{u^2} 
=  0, 
   \mbox{\quad{and}\quad} 
   \df{u^1} \wedge \df{u^2} 
= -\df{u^2} \wedge \df{u^1}. 
\label{eqn: the skew-commutative property of the wedge product} 
\end{align} 
The object $\df{u^1} \wedge \df{u^2}$ is a two-form, i.e., a totally antisymmetric $\binom{0}{2}$ tensor. 
In fact, it is the only basis two-form associated with the two-dimensional surface. 
\par 
The exterior derivative maps a one-form to a two-form. 
Let $\bsomega \equiv \omega_\alpha\, \df{u^\alpha}$ be a one-form. 
Taking its exterior derivative gives: 
\begin{align} 
                                          \df{\bsomega} 
\equiv \partial_\beta{\, \omega_\alpha}\, \df{u^\beta } \wedge 
                                          \df{u^\alpha} 
     = \left( 
       \partial_1{\omega_2} 
     - \partial_2{\omega_1} 
       \right)\, \df{u^1} \wedge 
                 \df{u^2}. 
\end{align} 
For a two-dimensional surface $S$, Stokes' theorem~\cite{Schutz1999_Geometrical_Methods} states that the line integral of a one-form along the boundary $\partial{S}$ equals the surface integral of its exterior derivative: 
\begin{align} 
  \oint_{\partial S}    {\bsomega} 
=  \int_{         S} \df{\bsomega}. 
\label{eqn: Stokes' theorem} 
\end{align} 
\subsection{The first fundamental form of surfaces} \label{subsec: DG, the first fundamental form} 
As discussed earlier [see \Eqref{eqn: introducing the metric tensor}], the metric tensor, also known as the first fundamental form, is a $\binom{0}{2}$ tensor that takes two tangent vectors as its arguments and yields a real number. 
Its components, which describe both the lengths of the coordinate basis vectors and the angle between them, are given by: 
\begin{align} 
       \bfg\left( 
       { \partial_\alpha{\bfx} }, 
       { \partial_\beta {\bfx} } 
       \right) 
     = g_{\alpha\beta} 
\equiv \inProd 
       { \partial_\alpha{\bfx} } 
       { \partial_\beta {\bfx} }, 
\label{eqn: defn. of the metric components} 
\end{align} 
where the inner product $\inProd{\cdot}{\cdot}$ denotes the usual Euclidean dot product. 
While the inner product in \Eqref{eqn: defn. of the metric components} can be used to calculate the metric components from an embedding $\bfx\left(u^1, u^2\right)$ in $\bbR^3$, it is also possible to directly define a metric tensor without recourse to an embedding, i.e., without utilizing the Euclidean metric in $\bbR^3$. 
It follows from the former case, or can be safely assumed in the latter, that the metric components are symmetric: 
\begin{align} 
  g_{\alpha\beta } 
= g_{\beta \alpha}. 
\end{align} 
With the introduction of the metric components, the squared length of a tangent vector $\bfv \equiv v^\alpha\, \partial_\alpha{\bfx}$ can be expressed in terms of its components as follows: 
\begin{align} 
         \norm      {\bfv}^2 
\equiv \inProd{\bfv}{\bfv} 
\equiv \inProd 
       { v^\alpha\, \partial_\alpha{\bfx} } 
       { v^\beta \, \partial_\beta {\bfx} } 
     = \inProd 
       { \partial_\alpha{\bfx} } 
       { \partial_\beta {\bfx} }v^\alpha\, v^\beta 
\equiv g_{\alpha\beta}\,        v^\alpha\, v^\beta. 
\end{align} 
\par 
The inverse metric tensor is a $\binom{2}{0}$ tensor, with components given by the matrix inverse of the metric components: 
\begin{align} 
       \bfg^{-1} 
\equiv    g^{\alpha\beta}\, \partial_\alpha{\bfx} \otimes 
                            \partial_\beta {\bfx}, 
\end{align} 
where $g^{\alpha\beta}\, g_{\beta\gamma} \equiv \delta^\alpha_\gamma$. 
Since the inverse of a symmetric matrix is symmetric, the inverse metric components are symmetric: 
\begin{align} 
  g^{\alpha\beta } 
= g^{\beta \alpha}. 
\end{align} 
\par 
The metric tensor and the inverse metric tensor together establish a one-to-one mapping between tangent vectors and one-forms. 
More specifically, the metric tensor maps a tangent vector $\bfv \equiv v^\alpha\, \partial_\alpha{\bfx}$ to its associated one-form, which is given by: 
\begin{align} 
       \bfg\left(\bfv, \cdot\right) 
     =    g_{\beta \gamma}\, \df{u^\beta}\left(v^\alpha\, \partial_\alpha{\bfx}\right)\, \df{u^\gamma} 
     =    g_{\gamma\beta }\,                   v^\beta                                \, \df{u^\gamma} 
\equiv                                         v_\gamma                               \, \df{u^\gamma}, 
\end{align} 
where $v_\gamma \equiv g_{\gamma\beta}\, v^\beta$. 
In a similar manner, the inverse metric tensor maps a one-form $\bsomega \equiv \omega_\alpha\, \df{u^\alpha}$ to its associated tangent vector: 
\begin{align} 
       \bfg^{-1}\left(\bsomega, \cdot\right) 
     =    g^{\beta \gamma}\, \partial_\beta{\bfx}\left(\omega_\alpha\, \df{u^\alpha}\right)\, \partial_\gamma{\bfx} 
     =    g^{\gamma\beta }\,                           \omega_\beta                        \, \partial_\gamma{\bfx} 
\equiv                                                 \omega^\gamma                       \, \partial_\gamma{\bfx}, 
\end{align} 
where $\omega^\gamma \equiv g^{\gamma\beta}\, \omega_\beta$. 
Therefore, with the introduction of the metric tensor and the inverse metric tensor, the distinction between tangent vectors and one-forms becomes obscured. 
It is common to write: 
\begin{align} 
\bfv     \sim      v^\alpha\, \partial_\alpha{\bfx} 
         \sim      v_\alpha\,    \df{u^\alpha}      \mbox{\quad{and}\quad} 
\bsomega \sim \omega_\alpha\,    \df{u^\alpha} 
         \sim \omega^\alpha\, \partial_\alpha{\bfx} 
\end{align} 
and to refer to $v_\alpha$ as the \emph{covariant} components of $\bfv$ and $\omega^\alpha$ as the \emph{contravariant} components of $\bsomega$. 
More generally, the metric components $g_{\alpha\beta}$ and the inverse metric components $g^{\alpha\beta}$ are respectively used to lower and raise indices, mapping different types of tensors of the same rank into one another. 
\subsubsection{The area two-form of surfaces} 
The area two-form is a totally antisymmetric $\binom{0}{2}$ tensor that takes two tangent vectors as its arguments and returns the \emph{signed} area of the parallelogram they form. 
Let $g \equiv \det\left(g_{\alpha\beta}\right)$ denote the determinant of the metric components. 
For two-dimensional surfaces, it is strictly positive. 
In fact, the square root of the metric determinant equals the unsigned area of the parallelogram formed by the coordinate basis vectors~\cite{do_Carmo2016_Classical_Differential_Geometry, Shifrin2021_Classical_Differential_Geometry}: 
\begin{align} 
  \sqrt{g} 
= \norm{ 
  \partial_1{\bfx} \times 
  \partial_2{\bfx} 
  }. 
\label{eqn: geometric meaning of the metric determinant} 
\end{align} 
In terms of the metric determinant, the area two-form is given by: 
\begin{align} 
       d{\calA} 
\equiv            \sqrt{g}\,                          \df{u^1     } \wedge \df{u^2    } 
     = \frac{1}{2}\sqrt{g}\, \epsilon_{\alpha\beta}\, \df{u^\alpha} \wedge \df{u^\beta} 
\equiv \frac{1}{2}        \,    \calE_{\alpha\beta}\, \df{u^\alpha} \wedge \df{u^\beta}, 
\label{eqn: expression for area two-form} 
\end{align} 
where $\epsilon_{\alpha\beta}$ is the Levi-Civita symbol, and $\calE_{\alpha\beta} \equiv \sqrt{g}\, \epsilon_{\alpha\beta}$ is sometimes referred to as the Levi-Civita tensor~\cite{David2004_Geometry_and_Field_Theory}. 
For later reference, using the inverse metric components $g^{\alpha\beta}$ to raise both indices of $\calE_{\alpha\beta}$ yields the contravariant components of the area two-form, up to a factor of one-half: 
\begin{align} 
     \calE^{\gamma\rho  } 
=        g^{\gamma\alpha}\, 
         g^{\rho  \beta }\, 
     \calE_{\alpha\beta } 
= \frac{1}{ \sqrt{g} }\, 
  \epsilon^{\gamma\rho  }, 
\label{eqn: contravariant Levi-Civita tensor} 
\end{align} 
where $\epsilon^{\gamma\rho}$ is again the Levi-Civita symbol, for which the index positions do not affect its values. 
One quick way to verify that the second equality in \Eqref{eqn: contravariant Levi-Civita tensor} holds is to apply the following formula for the determinant of a two-by-two matrix $\bfM \equiv \left(M^{\alpha\beta}\right)$~\cite{Schutz1999_Geometrical_Methods}: 
\begin{align} 
  \det{\bfM} 
= \frac{1}{2}\, 
  \epsilon_{\alpha\beta }\, 
  \epsilon_{\gamma\rho  }\, 
         M^{\alpha\gamma}\, 
         M^{\beta \rho  }. 
\end{align} 
We leave the verification as an exercise for the interested reader. 
\subsection{The second fundamental form of surfaces} \label{subsec: DG, the second fundamental form} 
In terms of the contravariant components of the area two-form, the surface unit normal vector can be expressed as: 
\begin{align} 
       \unitvec{n} 
\equiv                                     \frac{      { \partial_1     {\bfx} \times \partial_2    {\bfx} } } 
                                                { \norm{ \partial_1     {\bfx} \times \partial_2    {\bfx} } } 
     = \frac{1}{2}\, \calE^{\alpha\beta}\,      {      { \partial_\alpha{\bfx} \times \partial_\beta{\bfx} } }. 
\label{eqn: defn. of the surface unit normal vector} 
\end{align} 
Like the metric tensor, the second fundamental form of a surface is also a $\binom{0}{2}$ tensor. 
It describes the changes in the local normal direction along the coordinate lines, thus quantifying how the surface curves at a point. 
Its components are defined as follows: 
\begin{align} 
       b_{\alpha\beta} 
\equiv -               { \inProd{ \partial_\alpha{\unitvec n} } {                { \partial_\beta{\bfx} } } } 
     = -\partial_\alpha{ \inProd{                {\unitvec n} } {                { \partial_\beta{\bfx} } } } 
     +                 { \inProd{                {\unitvec n} } { \partial_\alpha{ \partial_\beta{\bfx} } } } 
     =                 { \inProd{                {\unitvec n} } { \partial_\alpha{ \partial_\beta{\bfx} } } }. 
\end{align} 
By this definition, the components of the second fundamental form are symmetric: 
\begin{align} 
  b_{\alpha\beta } 
= b_{\beta \alpha}. 
\end{align} 
\par 
The curvatures of the surface can be obtained from the shape operator $\shape$, which is also known as the extrinsic curvature tensor. 
As mentioned earlier [see \Eqref{eqn: introducing the shape operator}], the shape operator is a $\binom{1}{1}$ tensor, and its components are related to those of the second fundamental form as follows: 
\begin{align} 
  {b^ \alpha}_\beta 
=  g^{\alpha  \gamma}\, 
   b_{\gamma  \beta }. 
\end{align} 
Among the useful properties of the shape operator, taking the trace and the determinant of the component matrix $\left( {b^\alpha}_\beta \right)$ yields the mean and the Gaussian curvature of the surface, respectively: 
\begin{align} 
H \equiv  \tr{ \left( {b^\alpha}_\beta \right) } = {b^\alpha}_\alpha                    \mbox{\quad{and}\quad} 
K \equiv \det{ \left( {b^\alpha}_\beta \right) } = \frac{1}{2}\, \calE^{\alpha\gamma}\, 
                                                                 \calE^{\beta \rho  }\, 
                                                                     b_{\alpha\beta }\, 
                                                                     b_{\gamma\rho  }. 
\label{eqn: mean curvature and Gaussian curvature} 
\end{align} 
\par 
It is worth pointing out that while the components of the shape operator ${b^\alpha}_\beta$ are generally not symmetric, the shape operator itself---mapping a tangent vector to another tangent vector---is symmetric in the following sense: 
\begin{align} 
  \inProd{ \shape\left(\partial_\alpha{\bfx}\right) }{ \partial_\beta {\bfx} } 
= b_{\alpha\beta } 
= b_{\beta \alpha} 
= \inProd{ \shape\left(\partial_\beta {\bfx}\right) }{ \partial_\alpha{\bfx} }. 
\end{align} 
\section{The duality between isometric deformations and equilibrium stresses} 
In this section, we introduce the physical quantities used to characterize isometric deformations, which are stress-free, and the equilibrium (i.e., force-balanced) stress patterns associated with energy-minimizing deformations for a given smooth surface. 
We then derive the partial differential equations these deformation modes must satisfy. 
Based on the derived governing equations, we reveal a duality between the two types of modes: Any isometric deformation can be mapped to an energy-minimizing deformation, and vice versa. 
\subsection{Isometric deformations} \label{subsec: the isometry-stress duality, isometry} 
Under a deformation, each point on a surface [denoted by $\bfx\left(u^1, u^2\right)$] is displaced to a new position. 
Let $\delta{\bfx}\left(u^1, u^2\right)$ denote the displacement field associated with the deformation. 
The corresponding deformed surface can therefore be expressed as: 
\begin{align} 
             {\bfx}'\left(u^1, u^2\right) 
\equiv       {\bfx} \left(u^1, u^2\right) 
     + \delta{\bfx} \left(u^1, u^2\right). 
\end{align} 
The metric of the deformed surface has components: 
\begin{align} 
\begin{split} 
         g'_{\alpha\beta} 
  \equiv \inProd 
         { \partial_\alpha{\bfx'} } 
         { \partial_\beta {\bfx'} } 
& \equiv \inProd 
         { \partial_\alpha\left(\bfx + \delta{\bfx}\right) } 
         { \partial_\beta \left(\bfx + \delta{\bfx}\right) } 
         \h 
&      = \inProd{ \partial_\alpha{      \bfx} }{ \partial_\beta{      \bfx} } 
       + \inProd{ \partial_\alpha{      \bfx} }{ \partial_\beta{\delta\bfx} } 
       + \inProd{ \partial_\alpha{\delta\bfx} }{ \partial_\beta{      \bfx} } 
       + \inProd{ \partial_\alpha{\delta\bfx} }{ \partial_\beta{\delta\bfx} } 
         \h 
& \equiv            g_{\alpha\beta} 
       + 2\varepsilon_{\alpha\beta}, 
\end{split} 
\label{eqn: defn. of the deformed metric components} 
\end{align} 
where 
$
       g_{\alpha\beta} 
\equiv \inProd 
       { \partial_\alpha{\bfx} } 
       { \partial_\beta {\bfx} } 
$ denotes the metric components of the original undeformed surface [see \Eqref{eqn: defn. of the metric components}], and $\varepsilon_{\alpha\beta}$ the components of the strain tensor, defined as half the difference between the deformed metric and the original metric~\cite{Niordson1985_Shell_Theory}. 
If the deformation considered is small, the nonlinear strain term in \Eqref{eqn: defn. of the deformed metric components}, 
$
\inProd 
{ \partial_\alpha{\delta\bfx} } 
{ \partial_\beta {\delta\bfx} } 
$, can be neglected without losing accuracy. 
\par 
Isometric deformations, as their name suggests, preserve the metric of a surface, so that the infinitesimal distance between neighboring points remains unchanged under such a deformation. 
In other words, an isometric deformation does not strain a surface, and the displacement field associated with a \emph{linear} isometric deformation hence satisfies: 
\begin{align} 
        0 
      = 2\varepsilon_{\alpha\beta} 
\approx \inProd{ \partial_\alpha{\bfx} }{ \partial_\beta {\delta\bfx} } 
      + \inProd{ \partial_\beta {\bfx} }{ \partial_\alpha{\delta\bfx} }. 
\label{eqn: linear isometry condition} 
\end{align} 
\par 
One way to solve the linear isometry condition [\Eqref{eqn: linear isometry condition}] is to introduce a dimensionless auxiliary vector field $\bsomega\left(u^1, u^2\right)$, such that: 
\begin{align} 
       \partial_\alpha{\delta\bfx} 
\equiv \bsomega 
\times \partial_\alpha{      \bfx}; 
\label{eqn: defn. of the angular velocity} 
\end{align} 
as can be verified: 
\begin{align} 
        \inProd{ \partial_\alpha{\bfx} }{                 \partial_\beta {\delta\bfx} } 
\equiv  \inProd{ \partial_\alpha{\bfx} }{ \bsomega \times \partial_\beta {      \bfx} } 
     = -\inProd{ \partial_\beta {\bfx} }{ \bsomega \times \partial_\alpha{      \bfx} } 
\equiv -\inProd{ \partial_\beta {\bfx} }{                 \partial_\alpha{\delta\bfx} }, 
\end{align} 
where the middle equality is due to the cyclic property of the scalar triple product. 
Geometrically, the vector $\bsomega$ describes a local rotation of an infinitesimal area element of the surface by a small angle equal to the magnitude of $\bsomega$, $\norm{\bsomega}$, about an axis pointing along the direction of $\bsomega$, $\unitvec{\bsomega} \equiv \bsomega / \norm{\bsomega}$. (See Rodrigues' rotation formula for justification.) 
The auxiliary field $\bsomega$ is often referred to as the angular velocity field, though it does not imply any dynamic behavior. 
\par 
For the local rotations of the infinitesimal area elements to be compatible, the following compatibility conditions must hold: 
\begin{itemize} 
\item position    compatibility, meaning the original surface must not tear under the corresponding isometric deformation; 
\item orientation compatibility, meaning the angular velocity field $\bsomega$ must be single-valued and vary smoothly across the original surface. 
\end{itemize} 
These local compatibility conditions imply the following position and orientation closure conditions when integrated along a simply closed curve $\gamma$ on the original surface: 
\refstepcounter{equation} \label{eqn: closure conditions} 
\begin{gather} 
\oint_\gamma \df{ \left(\delta{\bfx}\right) } = \zero, \label{eqn: position closure condition}    \tag{\theequation, pos.}    \h 
\oint_\gamma \df{       \bsomega            } = \zero. \label{eqn: orientation closure condition} \tag{\theequation, orient.} 
\end{gather} 
\par 
In the remainder of this subsection, we first show  the implications of the closure conditions [\Eqsref{eqn: closure conditions}], i.e., the derivation of the mathematical expressions for the local compatibility conditions, which are previously described only in words. 
By combining the resulting expressions, we obtain a compatibility equation that the derivatives of the angular velocity field, $\partial_\alpha{\bsomega}$, must satisfy. 
The vectors $\partial_\alpha{\bsomega}$ are hereafter referred to as the angular acceleration vectors, characterizing the variation of the angular velocity field across the surface. 
\subsubsection{Local implications of the position closure condition} \label{subsubsec: local position-compatibility condition} 
Let $S$ denote an arbitrary simply connected region of a surface and $\partial{S}$ its boundary. 
The main idea in deriving the local compatibility conditions is to express the closure conditions over the boundary curve $\partial{S}$ as surface integrals over $S$ using Stokes' theorem [\Eqref{eqn: Stokes' theorem}]; equating the integrands of the resulting surface integrals to zero then yields the desired local expressions. 
\par 
We begin by expressing the position closure condition [\Eqref{eqn: position closure condition}] as follows: 
\begin{align} 
       \zero 
     = \oint_{\partial S}                             \df{ \left(\delta{\bfx}\right) } 
     = \oint_{\partial S}                 \partial_\alpha{       \delta{\bfx}        }\, \df{u^\alpha} 
\equiv \oint_{\partial S} \bsomega \times \partial_\alpha{             {\bfx}        }\, \df{u^\alpha}, 
\label{eqn: an intermediate step toward obtaining the angular acceleration vectors, 0} 
\end{align} 
where the exterior derivative of the displacement field $\delta{\bfx}$ is taken to obtain the second equality, and the defining relation for the angular velocity field $\bsomega$ [\Eqref{eqn: defn. of the angular velocity}] is used for the third. 
Applying Stokes' theorem [\Eqref{eqn: Stokes' theorem}] yields: 
\begin{align} 
\begin{split} 
    \oint_{\partial S}               {       \bsomega \times \partial_\alpha{\bfx}\, \df{u^\alpha}        } 
& =  \int_{         S}            \df{ \left(\bsomega \times \partial_\alpha{\bfx}\, \df{u^\alpha}\right) } 
    \h 
& =  \int_{         S}            \df{ \left(\bsomega \times \partial_\alpha{\bfx}                \right) } \wedge \df{u^\alpha} 
  =  \int_{         S} \partial_\beta{ \left(\bsomega \times \partial_\alpha{\bfx}                \right) } 
                                                                                  \, \df{u^\beta }          \wedge \df{u^\alpha} 
    \h 
& =  \int_{         S} \partial_\beta{\bsomega} \times               { \partial_\alpha{\bfx} }\, \df{u^\beta} \wedge \df{u^\alpha} 
  +  \int_{         S}               {\bsomega} \times \partial_\beta{ \partial_\alpha{\bfx} }\, \df{u^\beta} \wedge \df{u^\alpha}. 
\end{split} 
\label{eqn: an intermediate step toward obtaining the angular acceleration vectors, 1} 
\end{align} 
The second surface integral in the last line of \Eqref{eqn: an intermediate step toward obtaining the angular acceleration vectors, 1} vanishes because the partial derivatives commute whereas the wedge product is skew-commutative [see \Eqref{eqn: the skew-commutative property of the wedge product}].

Thus, combining 
\Eqsref{eqn: an intermediate step toward obtaining the angular acceleration vectors, 0} and
 \eqref{eqn: an intermediate step toward obtaining the angular acceleration vectors, 1}  
\begin{align} 
\begin{split} 
    \zero 
  = \int_S                          \partial_\beta {\bsomega} \times \partial_\alpha{\bfx}       \, \df{u^\beta} \wedge \df{u^\alpha} 
& = \int_S \left(                   \partial_1     {\bsomega} \times \partial_2     {\bfx} 
  -                                 \partial_2     {\bsomega} \times \partial_1     {\bfx}\right)\, \df{u^1    } \wedge \df{u^2     } 
    \h 
& = \int_S \epsilon^{\alpha\beta}\, \partial_\alpha{\bsomega} \times \partial_\beta {\bfx}       \, \df{u^1    } \wedge \df{u^2     }, 
\end{split} 
\label{eqn: an intermediate step toward obtaining the angular acceleration vectors, 3} 
\end{align} 
where $\epsilon^{\alpha\beta}$ is the Levi-Civita symbol. 
Since the integration region $S$ is arbitrary, \Eqref{eqn: an intermediate step toward obtaining the angular acceleration vectors, 3} implies locally: 
\begin{align} 
  \epsilon^{\alpha\beta}\, \partial_\alpha{\bsomega} \times 
                           \partial_\beta {\bfx    } 
= \zero. 
\label{eqn: the final form of the position compatibility condition} 
\end{align} 
A priori, the angular acceleration vectors $\partial_\alpha{\bsomega}$ could  have both in-plane and out-of-plane components. 
However, as we will see below, \Eqref{eqn: the final form of the position compatibility condition} enforces that the out-of-plane components of the angular acceleration vectors must vanish. 
\par 
To demonstrate that the angular acceleration vectors $\partial_\alpha{\bsomega}$ are purely in-plane, we first decompose them in the moving frame $\left\{\partial_1{\bfx}, \partial_2{\bfx}, \unitvec{n}\right\}$, which spans the embedding space $\bbR^3$, as follows: 
\begin{align} 
                           \partial_\alpha{\bsomega} 
\equiv {a_\alpha}^\gamma\, \partial_\gamma{\bfx    } 
     + {a_\alpha}       \, \unitvec{n}. 
\label{eqn: defn. of the angular acceleration vectors} 
\end{align} 
Substituting \Eqref{eqn: defn. of the angular acceleration vectors} into \Eqref{eqn: the final form of the position compatibility condition} gives: 
\begin{align} 
  \zero 
= \epsilon^{\alpha\beta}\, {a_\alpha}^\gamma\, \partial_\gamma{\bfx} \times \partial_\beta{\bfx} 
+ \epsilon^{\alpha\beta}\, {a_\alpha}       \, \unitvec{n}           \times \partial_\beta{\bfx}. 
\label{eqn: an intermediate step toward obtaining the angular acceleration vectors, 4} 
\end{align} 
Recall that the square root of the metric determinant $g$ equals the magnitude of the surface normal vector [see \Eqref{eqn: geometric meaning of the metric determinant}]. 
From this, we have: 
\begin{align} 
                                      \partial_\gamma{\bfx} \times \partial_\beta{\bfx} 
=            \epsilon_{\gamma\beta}\, \partial_1     {\bfx} \times \partial_2    {\bfx} 
= \sqrt{g}\, \epsilon_{\gamma\beta}\, \unitvec{n}, 
\label{eqn: an identity involving the cross product of the coordinate basis vectors} 
\end{align} 
where $\epsilon_{\gamma\beta}$ is again the Levi-Civita symbol. 
Substituting \Eqref{eqn: an identity involving the cross product of the coordinate basis vectors} into \Eqref{eqn: an intermediate step toward obtaining the angular acceleration vectors, 4} leads to: 
\begin{align} 
\begin{split} 
    \zero 
& = \sqrt{g}\, \epsilon^{\alpha \beta }\, \epsilon_{\gamma\beta}\, {a_\alpha}^\gamma\, \unitvec{n} 
  +            \epsilon^{\alpha \beta }\,                          {a_\alpha}       \, \unitvec{n} \times \partial_\beta{\bfx} 
    \h 
& = \sqrt{g}\,   \delta^ \alpha_\gamma \,                          {a_\alpha}^\gamma\, \unitvec{n} 
  +            \epsilon^{\alpha \beta }\,                          {a_\alpha}       \, \unitvec{n} \times \partial_\beta{\bfx} 
    \h 
& = \sqrt{g}\,                                                     {a_\alpha}^\alpha\, \unitvec{n} 
  +            \epsilon^{\alpha \beta }\,                          {a_\alpha}       \, \unitvec{n} \times \partial_\beta{\bfx}. 
\label{eqn: an intermediate step toward obtaining the angular acceleration vectors, 5} 
\end{split} 
\end{align} 
\par 
Now, taking the inner product of both sides of \Eqref{eqn: an intermediate step toward obtaining the angular acceleration vectors, 5} with the coordinate basis vector $\partial_\gamma{\bfx}$, we obtain: 
\begin{align} 
\begin{split} 
    0 
& = \sqrt{g}\,                          {a_\alpha}^\alpha\inProd{ \partial_\gamma{\bfx} }{ \unitvec{n}                             } 
  +            \epsilon^{\alpha\beta}\, {a_\alpha}       \inProd{ \partial_\gamma{\bfx} }{ \unitvec{n} \times \partial_\beta{\bfx} } 
    \h 
& =            \epsilon^{\alpha\beta}\, {a_\alpha}       \inProd{ \partial_\gamma{\bfx} }{ \unitvec{n} \times \partial_\beta{\bfx} }. 
\end{split} 
\label{eqn: an intermediate step toward obtaining the angular acceleration vectors, 6} 
\end{align} 
Using the cyclic property of the scalar triple product and \Eqref{eqn: an identity involving the cross product of the coordinate basis vectors}, we simplify \Eqref{eqn: an intermediate step toward obtaining the angular acceleration vectors, 6} to: 
\begin{align} 
\begin{split} 
    0 
& = -           \epsilon^{\alpha \beta }\, a_\alpha\inProd{ \unitvec{n} }{ \partial_\gamma{\bfx} \times \partial_\beta{\bfx} } 
    \h 
& = -\sqrt{g}\, \epsilon^{\alpha \beta }\, 
                \epsilon_{\gamma \beta }\, a_\alpha\inProd{ \unitvec{n} }{ \unitvec{n}                                       } 
  = -\sqrt{g}\,   \delta^ \alpha_\gamma \, a_\alpha 
  = -\sqrt{g}\,                            a_\gamma. 
\end{split} 
\label{eqn: an intermediate step toward obtaining the angular acceleration vectors, 7} 
\end{align} 
Since the metric determinant $g$ is strictly positive, \Eqref{eqn: an intermediate step toward obtaining the angular acceleration vectors, 7} shows that the out-of-plane components of the angular acceleration vectors must vanish: 
\begin{align} 
  a_\gamma 
= 0. 
\end{align} 
\par 
By the same reasoning, taking the inner product of both sides of \Eqref{eqn: an intermediate step toward obtaining the angular acceleration vectors, 5} with the surface unit normal vector $\unitvec{n}$ shows that the component matrix $\left( {a_\alpha}^\gamma \right)$ is traceless: 
\begin{align} 
\begin{split} 
    0 
& = \sqrt{g}\,                          {a_\alpha}^\alpha\inProd{ \unitvec{n} }{ \unitvec{n}                             } 
  +            \epsilon^{\alpha\beta}\, {a_\alpha}       \inProd{ \unitvec{n} }{ \unitvec{n} \times \partial_\beta{\bfx} } 
    \h 
& = \sqrt{g}\,                          {a_\alpha}^\alpha, 
\end{split} 
\end{align} 
which implies that: 
\begin{align} 
  {a_\alpha}^\alpha 
= 0. 
\end{align} 
\par 
Thus, we have shown that, as a result of the position closure condition, the angular acceleration vectors lie in the tangent planes of the original surface and have traceless angular acceleration components. 
Consequently, \Eqref{eqn: defn. of the angular acceleration vectors} can now be expressed as: 
\begin{align} 
                           \partial_\alpha{\bsomega} 
\equiv {a_\alpha}^\gamma\, \partial_\gamma{\bfx    }. 
\label{eqn: defn. of the angular acceleration vectors, revised} 
\end{align} 
\subsubsection{Local implications of the orientation closure condition} \label{subsubsec: local orientation-compatibility condition} 
To derive the local orientation compatibility condition, we begin by rewriting the orientation closure condition [\Eqref{eqn: orientation closure condition}], following essentially the same steps as in the previous subsubsection [see \Eqsref{eqn: an intermediate step toward obtaining the angular acceleration vectors, 0}--\eqref{eqn: an intermediate step toward obtaining the angular acceleration vectors, 3}]: 
\begin{align} 
\begin{split} 
    \zero 
& = \oint_{\partial S}             \df{\bsomega} 
  = \oint_{\partial S} \partial_\alpha{\bsomega}\, \df{u^\alpha} 
    \h 
& = \int_S                                     \df{ \left(\partial_\alpha{\bsomega}\, \df{u^\alpha}\right) } 
  = \int_S                                     \df{ \left(\partial_\alpha{\bsomega}                \right) } \wedge \df{u^\alpha} 
  = \int_S                          \partial_\beta{       \partial_\alpha{\bsomega}                        } 
                                                                                   \, \df{u^\beta }          \wedge \df{u^\alpha} 
    \h 
& = \int_S \epsilon^{\beta\alpha}\, \partial_\beta{       \partial_\alpha{\bsomega}                        } 
                                                                                   \, \df{u^1     }          \wedge \df{u^2     }. 
\end{split} 
\label{eqn: an intermediate step toward deriving the compatibility equations, 0} 
\end{align} 
Again, since the integration region $S$ is arbitrary, \Eqref{eqn: an intermediate step toward deriving the compatibility equations, 0} implies: 
\begin{align} 
  \epsilon^{\beta  \alpha}\, 
  \partial_ \beta { 
  \partial_ \alpha{\bsomega} } 
= \zero. 
\label{eqn: the final form of the orientation compatibility condition} 
\end{align} 
\par 
Superficially, it seems that the local orientation compatibility condition [\Eqref{eqn: the final form of the orientation compatibility condition}] merely tells us that the partial derivatives of the angular velocity field commute, which is a trivial mathematical fact.\footnote{
In fact, the calculations we just performed reveal an important property of the exterior derivative: 
\begin{align*} 
\df{}^2 = 0. 
\end{align*} 
} 
However, for a locally position-compatible angular velocity field, whose derivatives must satisfy \Eqref{eqn: defn. of the angular acceleration vectors, revised}, the local orientation compatibility condition imposes a constraint on the variation of the associated angular acceleration components. 
Substituting \Eqref{eqn: defn. of the angular acceleration vectors, revised} into \Eqref{eqn: the final form of the orientation compatibility condition} and rewriting the resulting expression in terms of the contravariant components of the area two-form $\calE^{\alpha\beta}$ [see \Eqref{eqn: contravariant Levi-Civita tensor}] yields the constraint: 
\begin{align} 
\begin{split} 
    \zero 
  = \partial_\beta \left(                                        \epsilon^{\beta \alpha}\,                         \partial_\alpha{\bsomega}\right) 
& = \partial_\beta \left[\sqrt{g}\, \left(\frac{1}{ \sqrt{g} }\, \epsilon^{\beta \alpha}\right)                    \partial_\alpha{\bsomega}\right] 
    \h 
& \equiv 
    \partial_\beta \left(\sqrt{g}\,                                 \calE^{\beta \alpha}\,     {a_\alpha}^\gamma\, \partial_\gamma{\bfx    }\right) 
  = \partial_\alpha\left(\sqrt{g}\,                                 \calE^{\alpha\beta }\,     {a_\beta }^\gamma\, \partial_\gamma{\bfx    }\right), 
\end{split} 
\label{eqn: the final step in deriving the compatibility equations} 
\end{align} 
where the indices are relabeled in the final step. 
\par 
\Eqnref{eqn: the final step in deriving the compatibility equations}, henceforth referred to as the compatibility equations, provides the necessary and sufficient local conditions that any geometrically compatible linear isometry of a surface must satisfy. 
In the literature (see, e.g., \Refref{Audoly2019_Elasticity_and_Geometry}), the compatibility equations are more commonly expressed in terms of the components of the angular velocity field, rather than by introducing the angular acceleration components, as done in our work. 
However, it can be shown that our formulation is equivalent to the commonly used one in the literature. 
One advantage of our formulation is that our form of the compatibility equations closely resembles the equilibrium equations for thin membrane surfaces, which will be derived in the following subsection. 
\subsection{Equilibrium stresses} \label{subsec: the isometry-stress duality, equilibrium stress} 
Stresses arise when material points on a surface deviate from their energetically preferred positions, exerting forces on each other. 
The equilibrium stress characterizes the internal force distribution when the surface is in force equilibrium. 
As we will show in the following subsubsections, the equilibrium stress components $\sigma^{\alpha\beta}$ satisfy the equilibrium equations: 
\begin{align} 
  \partial_\alpha\left(\sqrt{g}\, \sigma^{\alpha\beta}\, \partial_\beta{\bfx}\right) 
= \zero, 
\label{eqn: equilibrium equations} 
\end{align} 
which take a similar form to the compatibility equations [\Eqref{eqn: the final step in deriving the compatibility equations}]. 
\subsubsection{Derivation of the equilibrium equations} 
To derive the equilibrium equations [\Eqref{eqn: equilibrium equations}], we follow the standard practice of first writing down the energy functional associated with an imposed deformation and then minimizing it by varying the corresponding displacement field. 
It is worth noting that, although unnecessary for the present analysis, stress can also be formulated purely in terms of forces and geometry, without assuming that the forces are conservative---that is, derivable from an energy functional. 
\paragraph{Expression for the membrane deformation energy functional.} For thin planar surfaces, it is well-established that bending deformations typically require much less energy than stretching deformations~\cite{Audoly2019_Elasticity_and_Geometry}. 
In the long-wavelength limit, the surfaces of interest in this work can be treated as flat since their local radii of curvature are much smaller than the overall length scale. 
Therefore, we assume they have a negligible bending stiffness and behave like membranes. 
\par 
Under small deformations of a surface, the net force exerted on a material point by its surrounding neighbors is directly proportional to its displacement, in accordance with Hooke's law. 
For area elements of the surface with finite spatial extent, this linearity is captured by the following relation between the stress and strain components, known as the generalized Hooke's law\footnote{
In this work, we adopt the standard Kirchhoff-Love hypothesis~\cite{Ventsel2001_Thin-Walled_Structures}, which, in essence, assumes uniform deformation along the membrane's thickness direction, so that the membrane's deformation is adequately captured by that of its mid-surface. 
Consequently, in the generalized Hooke's law presented below, both the stress and stiffness components pertain to the membrane's mid-surface, thereby having the dimension of force per unit length, rather than the more conventional force per unit area. 
Additionally, the resulting membrane deformation energy functional has the dimension of energy per unit length, with analogous dimensional considerations applying to related quantities. 
}: 
\begin{align} 
       \sigma^{\alpha\beta          } 
=           C^{\alpha\beta\gamma\rho}\, 
  \varepsilon_{           \gamma\rho}, 
\label{eqn: generalized Hooke's law} 
\end{align} 
where $C^{\alpha\beta\gamma\rho}$ denotes the components of the rank-four stiffness tensor~\cite{Niordson1985_Shell_Theory}, and $\varepsilon_{\gamma\rho}$ the strain components, as previously defined in \Eqref{eqn: defn. of linear strain components}. 
For conventional elastic materials, the stiffness components are symmetric in the first two indices, i.e.: 
$
  C^{\alpha\beta \gamma\rho} 
= C^{\beta \alpha\gamma\rho} 
$; consequently, the stress components are also symmetric: 
\begin{align} 
  \sigma^{\alpha\beta } 
= \sigma^{\beta \alpha}. 
\label{eqn: symmetric stress components} 
\end{align} 
Physically, the symmetric property of the stress components reflects torque balance within a deformed body~\cite{Audoly2019_Elasticity_and_Geometry, LL1986_Elasticity}. 
\par 
In the regime of small deformations, the membrane deformation energy functional $E_\rmm$ is quadratic in the strain components and is therefore given by: 
\begin{align} 
  E_\rmm 
= \frac{1}{2}\int_S d{\calA}\      C^{\alpha\beta\gamma\rho}\, \varepsilon_{\gamma\rho}\, \varepsilon_{\alpha\beta} 
= \frac{1}{2}\int_S d{\calA}\ \sigma^{\alpha\beta          }\,                            \varepsilon_{\alpha\beta}, 
\label{eqn: expression for membrane energy} 
\end{align} 
where $d{\calA}$ is the area two-form [\Eqref{eqn: expression for area two-form}], and the generalized Hooke's law [\Eqref{eqn: generalized Hooke's law}] is used to obtain the second equality. 
\par 
To proceed, recall from \Eqref{eqn: defn. of the deformed metric components} that the linear strain components are defined in terms of the displacement field $\delta{\bfx}$ as: 
\begin{align} 
       \varepsilon_{\alpha\beta} 
\equiv \frac{1}{2}\left( 
       \inProd{ \partial_\alpha{\bfx} }{ \partial_\beta {\delta\bfx} } 
     + \inProd{ \partial_\beta {\bfx} }{ \partial_\alpha{\delta\bfx} } 
       \right). 
\label{eqn: defn. of linear strain components} 
\end{align} 
Substituting \Eqref{eqn: defn. of linear strain components} into \Eqref{eqn: expression for membrane energy} yields, after some algebra, the following expression for the membrane energy associated with a displacement field: 
\begin{align} 
\begin{split} 
    E_\rmm\left[\delta{\bfx}\left(u^1, u^2\right)\right] 
& = \frac{1}{4}\left( 
    \int_S d{\calA}\ \sigma^{\alpha\beta}\inProd{ \partial_\alpha{\bfx} }{ \partial_\beta {\delta\bfx} } 
  + \int_S d{\calA}\ \sigma^{\alpha\beta}\inProd{ \partial_\beta {\bfx} }{ \partial_\alpha{\delta\bfx} } 
    \right) 
    \h 
& = \frac{1}{2}\int_S  d {\calA} \      \sigma^{\alpha\beta}\inProd{ \partial_\beta{\bfx} }{ \partial_\alpha{\delta\bfx} } \h 
& = \frac{1}{2}\int_S \df{u^1  } \wedge 
                      \df{u^2  } \,                         \inProd{ 
                    \sqrt{g    } \,     \sigma^{\alpha\beta}\,       \partial_\beta{\bfx} }{ \partial_\alpha{\delta\bfx} }, 
\end{split} 
\label{eqn: an intermediate step toward deriving the equilibrium equations, 0} 
\end{align} 
where the second line follows from the symmetric property of the stress components [\Eqref{eqn: symmetric stress components}], and the definition of the area two-form [\Eqref{eqn: expression for area two-form}] is applied in the third line. 
\par 
The integrand of the final surface integral in \Eqref{eqn: an intermediate step toward deriving the equilibrium equations, 0} can be rewritten using the product rule as: 
\begin{align} 
                 { \inProd{                      \sqrt{g}\, \sigma^{\alpha\beta}\, \partial_\beta{\bfx}        }{ \partial_\alpha{\delta\bfx} } } 
= \partial_\alpha{ \inProd{                      \sqrt{g}\, \sigma^{\alpha\beta}\, \partial_\beta{\bfx}        }{                {\delta\bfx} } } 
-                { \inProd{ \partial_\alpha\left(\sqrt{g}\, \sigma^{\alpha\beta}\, \partial_\beta{\bfx}\right) }{                {\delta\bfx} } }. 
\label{eqn: an intermediate step toward deriving the equilibrium equations, 1} 
\end{align} 
Substituting \Eqref{eqn: an intermediate step toward deriving the equilibrium equations, 1} into \Eqref{eqn: an intermediate step toward deriving the equilibrium equations, 0} gives: 
\begin{align} 
\begin{split} 
    E_\rmm\left[\delta{\bfx}\left(u^1, u^2\right)\right] 
& = \frac{1}{2}\int_S \df{u^1} \wedge \df{u^2}\, 
    \partial_\alpha { \inProd{                      \sqrt{g}\, \sigma^{\alpha\beta}\, \partial_\beta{\bfx}        }{\delta\bfx} } 
    \h 
& \quad\quad\quad\quad 
  - \frac{1}{2}\int_S \df{u^1} \wedge \df{u^2}\, 
                    { \inProd{ \partial_\alpha\left(\sqrt{g}\, \sigma^{\alpha\beta}\, \partial_\beta{\bfx}\right) }{\delta\bfx} }. 
\end{split} 
\label{eqn: an intermediate step toward deriving the equilibrium equations, 2} 
\end{align} 
\par 
We notice that the integrand of the first surface integral in \Eqref{eqn: an intermediate step toward deriving the equilibrium equations, 2} is a total divergence. 
As a result, that surface integral can be rewritten, using Stokes' theorem [\Eqref{eqn: Stokes' theorem}], as a line integral along the surface boundary $\partial{S}$. 
More specifically, using the identity: 
\begin{align} 
\begin{split} 
    \df{ \left(\epsilon_{\alpha \beta }\,                {v^\alpha}\,                      \df{u^\beta}\right) } 
& =            \epsilon_{\alpha \beta }\,             \df{v^\alpha}                 \wedge \df{u^\beta} 
  =            \epsilon_{\alpha \beta }\, \partial_\gamma{v^\alpha}\, \df{u^\gamma} \wedge \df{u^\beta} 
    \h 
& =            \epsilon^{\gamma \beta }\, 
               \epsilon_{\alpha \beta }\, \partial_\gamma{v^\alpha}\, \df{u^1     } \wedge \df{u^2    } 
  =              \delta_ \alpha^\gamma \, \partial_\gamma{v^\alpha}\, \df{u^1     } \wedge \df{u^2    } 
    \h 
& =                                       \partial_\alpha{v^\alpha}\, \df{u^1     } \wedge \df{u^2    }, 
\end{split} 
\end{align} 
the first surface integral in \Eqref{eqn: an intermediate step toward deriving the equilibrium equations, 2} becomes: 
\begin{align} 
\begin{split} 
    \int_S \df{u^1} \wedge \df{u^2}\, \partial_\alpha{ \inProd{ \sqrt{g}\, \sigma^{\alpha\beta}\, \partial_\beta{\bfx} }{\delta\bfx} } 
& = \int_S \df{ \left( 
    \epsilon_{\alpha\gamma}\,                        { \inProd{ \sqrt{g}\, \sigma^{\alpha\beta}\, \partial_\beta{\bfx} }{\delta\bfx} }\, \df{u^\gamma} 
    \right) } 
    \h 
& = \oint_{\partial S} 
    \epsilon_{\alpha\gamma}\,                        { \inProd{ \sqrt{g}\, \sigma^{\alpha\beta}\, \partial_\beta{\bfx} }{\delta\bfx} }\, \df{u^\gamma} 
    \h 
& \equiv 
    \oint_{\partial S}                                                                                                                \, \df{u^\gamma}\, 
                                                     { \inProd{ 
       \calE_{\alpha\gamma}\,                                              \sigma^{\alpha\beta}\, \partial_\beta{\bfx} }{\delta\bfx} }, 
\end{split} 
\label{eqn: an intermediate step toward deriving the equilibrium equations, 3} 
\end{align} 
where Stokes' theorem is applied in the second line, and $\calE_{\alpha\gamma}$ denotes the covariant components of the area two-form [see \Eqref{eqn: expression for area two-form}]. 
\par 
Finally, by substituting 
\Eqref{eqn: an intermediate step toward deriving the equilibrium equations, 3} into 
\Eqref{eqn: an intermediate step toward deriving the equilibrium equations, 2}, we obtain the following expression for the membrane energy, which is the sum of a boundary term and a bulk term: 
\begin{align} 
\begin{split} 
    E_\rmm\left[\delta{\bfx}\left(u^1, u^2\right)\right] 
& = \frac{1}{2}\oint_{\partial S} \df{u^\gamma}\, 
    \inProd{    \calE_{\alpha\gamma}       \, \sigma^{\alpha\beta}\, \partial_\beta{\bfx}        }{\delta\bfx} 
    \h 
& \quad\quad\quad\quad 
  - \frac{1}{2} \int_{         S} \df{u^1     } \wedge \df{u^2}\, 
    \inProd{ \partial_ \alpha\left(\sqrt{g}\, \sigma^{\alpha\beta}\, \partial_\beta{\bfx}\right) }{\delta\bfx}. 
\end{split} 
\label{eqn: expression for membrane energy, final} 
\end{align} 
\paragraph{Minimizing the membrane energy functional.} To find the equilibrium stress which minimizes the membrane energy functional in the bulk of the surface, we take the functional derivative of the surface integral in \Eqref{eqn: expression for membrane energy, final} with respect to the displacement field $\delta{\bfx}$.\footnote{
In principle, one could also take the functional derivative of the line integral in \Eqref{eqn: expression for membrane energy, final} to obtain boundary conditions associated with the equilibrium equations. 
However, we omit this step, as we focus on the periodic boundary conditions of an interior unit cell.
} 
Since the surface integral in question is linear in $\delta{\bfx}$, we can directly read off its functional derivative: 
\begin{align} 
  \fdfrac{}{ \left(\delta{\bfx}\right) }\left[ 
  \int_S \df{u^1} \wedge \df{u^2}\, 
  \inProd{ \partial_\alpha\left(\sqrt{g}\, \sigma^{\alpha\beta}\, \partial_\beta{\bfx}\right) }{\delta\bfx} 
  \right] 
=        { \partial_\alpha\left(\sqrt{g}\, \sigma^{\alpha\beta}\, \partial_\beta{\bfx}\right) }. 
\label{eqn: an intermediate step toward deriving the equilibrium equations, 4} 
\end{align} 
Setting \Eqref{eqn: an intermediate step toward deriving the equilibrium equations, 4} to zero then yields the desired equilibrium equations: 
\begin{align} 
  \partial_\alpha\left(\sqrt{g}\, \sigma^{\alpha\beta}\, \partial_\beta{\bfx}\right) 
= \zero. 
\label{eqn: equilibrium equations, revisited} 
\end{align} 
\par 
Hence, the membrane energy functional is minimized\footnote{
Strictly speaking, we have only shown above that the membrane energy is extremized when the equilibrium equations are satisfied. 
However, we focus on small deformations from a stable reference state. 
} when the stress components $\sigma^{\alpha\beta}$ satisfy the equilibrium equations [\Eqref{eqn: equilibrium equations, revisited}]. 
For this reason, the corresponding deformation, which results in the equilibrium stress, is referred to as an energy-minimizing deformation. 
The membrane energy cost for the energy-minimizing deformation (denoted by $E_\rmm^\eq$) can be obtained by substituting the equilibrium equations [\Eqref{eqn: equilibrium equations, revisited}] into the expression for the membrane energy functional [\Eqref{eqn: expression for membrane energy, final}], yielding: 
\begin{align} 
  E_\rmm^\eq 
= \frac{1}{2}\oint_{\partial S} \df{u^\gamma}\, 
  \inProd{ \calE_{\alpha\gamma}\, \sigma^{\alpha\beta}\, \partial_\beta{\bfx} }{\delta\bfx}. 
\label{eqn: expression for minimized membrane energy} 
\end{align} 
\par 
As shown in \Appref{appendix: line force density}, the combination $\calE_{\alpha\gamma}\, \sigma^{\alpha\beta}\, \partial_\beta{\bfx}\, \df{u^\gamma}$ in \Eqref{eqn: expression for minimized membrane energy} represents the infinitesimal force exerted on the surface boundary $\partial{S}$. 
The line integral in \Eqref{eqn: expression for minimized membrane energy} therefore corresponds to the net work done by all the infinitesimal forces on $\partial{S}$. 
We thus write: 
\begin{align} 
       E_\rmm^\eq 
     = W 
\equiv \frac{1}{2}\oint_{\partial S} \df{u^\gamma}\, 
       \inProd{ \calE_{\alpha\gamma}\, \sigma^{\alpha\beta}\, \partial_\beta{\bfx} }{\delta\bfx}, 
\label{eqn: work-energy theorem} 
\end{align} 
where $W$ denotes the net work. 
\Eqnref{eqn: work-energy theorem} exemplifies the work-energy theorem, here stating that, in force equilibrium, the bulk membrane energy caused by an imposed deformation equals the work done by the deformation on the surface boundary. 
Physically, the work-energy theorem in our case follows inevitably from the assumptions of energy conservation (e.g., no heat generated during deformation) and that all forces are applied locally, so that the net change in energy within a region only arises from the forces applied to its boundary. 
\subsection{The duality} \label{subsec: the isometry-stress duality, duality} 
To recapitulate, in \Secref{subsec: the isometry-stress duality, isometry}, we showed that any linear isometric deformation corresponds to an infinitesimal rotation, characterized by the corresponding angular velocity field $\bsomega$. 
The derivatives of the angular velocity field, describing its variation across the surface, are called the angular acceleration vectors, which are purely in-plane: 
\begin{align} 
                          \partial_\gamma{\bsomega} 
\equiv {a_\gamma}^\beta\, \partial_\beta {\bfx    }, 
\label{eqn: defn. of the angular acceleration vectors, revised, revisited} 
\end{align} 
where the component matrix $\left( {a_\gamma}^\beta \right)$ is traceless: 
\begin{align} 
  {a_\gamma}^\gamma 
= 0. 
\end{align} 
For the infinitesimal rotation to be geometrically compatible, the angular acceleration components ${a_\gamma}^\beta$ must satisfy the compatibility equations: 
\begin{align} 
  \partial_\alpha\left(\sqrt{g}\, \calE^{\alpha\gamma}\, {a_\gamma}^\beta\, \partial_\beta{\bfx}\right) 
= \zero, 
\label{eqn: compatibility equations} 
\end{align} 
where $\calE^{\alpha\gamma}$ denotes the contravariant components of the area two-form [\Eqref{eqn: contravariant Levi-Civita tensor}]. 
\par 
In \Secref{subsec: the isometry-stress duality, equilibrium stress}, we showed that the equilibrium stress, such as caused by an imposed energy-minimizing deformation,\footnote{Geometrical incompatibilities can also give rise to equilibrium stresses (see, e.g., \Refref{Rocklin2022_Topologically_Protected_Deformations_bookChapter}).} satisfies the equilibrium equations: 
\begin{align} 
  \partial_\alpha\left(\sqrt{g}\, \sigma^{\alpha\beta}\, \partial_\beta{\bfx}\right) 
= \zero. 
\label{eqn: equilibrium equations, revisited 2} 
\end{align} 
\par 
It is worth noting that isometric deformations, as a special case of energy-minimizing deformations, do not cost any membrane energy since they do not strain a surface. 
Nevertheless, as we will see in the following subsubsections, an isometric deformation can still be mapped to a general energy-minimizing deformation, which costs energy, through the duality between angular accelerations and equilibrium stresses. 
\subsubsection{Duality between the angular acceleration components and the equilibrium stress components} 
The duality between the angular acceleration components ${a_\gamma}^\beta$ and the equilibrium stress components $\sigma^{\alpha\beta}$ can be readily seen by comparing the compatibility equations [\Eqref{eqn: compatibility equations}] to the equilibrium equations [\Eqref{eqn: equilibrium equations, revisited 2}]. 
\par 
If a surface can be isometrically deformed with the corresponding angular acceleration components ${a_\gamma}^\beta$, it follows that an energy-minimizing deformation mode of the surface will give rise to the following equilibrium stress pattern: 
\refstepcounter{equation} \label{eqn: relations between equilibrium stress and isometry} 
\begin{align} 
      \sigma^{\alpha\beta } 
= f\,  \calE^{\alpha\gamma}\, {a_\gamma}^\beta 
\label{eqn: relation between equilibrium stress and isometry, i to s} 
\tag{\theequation, iso-str} 
\end{align} 
for some constant of proportionality $f$. 
The dimension of $f$ is determined via dimensional analysis. 
Based on our previous discussions, the stress components $\sigma^{\alpha\beta}$ have the dimension of $\mathrm{ [Force] / [Length] }$ (see the footnote on p.~16); the Levi-Civita tensor $\calE^{\alpha\gamma}$ is dimensionless, since both the metric determinant and the Levi-Civita symbol are dimensionless [see \Eqref{eqn: expression for area two-form}]; and the angular acceleration components ${a_\gamma}^\beta$, as the spatial derivatives of the dimensionless angular velocity field, have the dimension of $\mathrm{ [Length]^{-1} }$. 
Accordingly, the proportionality constant $f$ must have the dimension of $\mathrm{ [Force] }$. 
\par 
Conversely, given an equilibrium stress pattern on a surface, we can obtain the angular acceleration components associated with the corresponding isometry of the surface by contracting both sides of \Eqref{eqn: relation between equilibrium stress and isometry, i to s} with $\calE_{\alpha\rho}$, yielding: 
\begin{align} 
                                         {a_\rho  }^\beta 
=               \delta_ \rho  ^\gamma \, {a_\gamma}^\beta 
=                \calE_{\alpha \rho  }\, 
                 \calE^{\alpha \gamma}\, {a_\gamma}^\beta 
= \frac{1}{f}\,  \calE_{\alpha \rho  }\, 
                \sigma^{\alpha \beta }. 
\label{eqn: relation between equilibrium stress and isometry, s to i} 
\tag{\theequation, str-iso} 
\end{align} 
\par 
Thus, \Eqsref{eqn: relations between equilibrium stress and isometry} establish a duality between isometric and energy-minimizing deformations for a given surface: Any isometric deformation can be mapped to an energy-minimizing deformation, and vice versa. 
We would like to emphasize that isometric deformations themselves never induce (nonzero) equilibrium stresses on a membrane surface. 
Rather, any isometric deformation can be \emph{transformed} into an equilibrium stress, through \Eqref{eqn: relation between equilibrium stress and isometry, i to s}. 
Moreover, we note that, from the antisymmetric property of $\calE_{\alpha\rho}$ in \Eqref{eqn: relation between equilibrium stress and isometry, s to i}, the local position-compatibility condition that the angular acceleration components must be traceless (recall \Secref{subsubsec: local position-compatibility condition}) is tantamount to the symmetric property of the stress components, which arises from torque balance within the deformed surface. 
\subsubsection{Expression for the minimized membrane energy functional in terms of the angular acceleration components} 
As a result of the isometry-stress duality, the equilibrium stress caused by an energy-minimizing deformation maps to the angular acceleration characterizing an isometric deformation. 
The work done by the equilibrium stress can thus be expressed in terms of the angular acceleration corresponding to the mapped isometric deformation. 
By first substituting \Eqref{eqn: relation between equilibrium stress and isometry, s to i} into \Eqref{eqn: work-energy theorem} and then using \Eqref{eqn: defn. of the angular acceleration vectors, revised, revisited}, we obtain: 
\begin{align} 
  E_\rmm^\eq = W 
= \frac{1}{2}\, f\oint_{\partial S} \df{u^\gamma}\, \inProd{ {\wt{a}_\gamma}^\beta\, \partial_\beta {   \bfx    } }{ \delta{\bfx} } 
= \frac{1}{2}\, f\oint_{\partial S} \df{u^\gamma}\, \inProd{                         \partial_\gamma{\wt\bsomega} }{ \delta{\bfx} }, 
\label{eqn: expression for minimized membrane energy, mapped isometry version} 
\end{align} 
where $\delta{\bfx}$ denotes the displacement field corresponding to the imposed energy-minimizing deformation, and $\wt\bsomega$ the angular velocity field characterizing the mapped isometric deformation. 
Hereafter, to clarify distinctions, the notation $\wt{ \left(\cdot\right) }$ will denote quantities associated with the mapped isometric deformation. 
\section{Doubly Periodic surfaces and shape-periodic deformations} \label{sec: DP surfaces and SP deformations} 
\subsection{Doubly Periodic surfaces} \label{subsec: DP surfaces and SP deformations, DP surfaces} 
In the remainder of this work, we focus on doubly periodic surfaces---surfaces invariant under two linearly independent translations. 
Let 
$\bsell_1 \equiv \ell^1\, \unitvec{\bsell}_1$ and 
$\bsell_2 \equiv \ell^2\, \unitvec{\bsell}_2$ denote the two translation vectors. 
Such a surface can always be generated by fixing its unit cell and then translating the unit cell repeatedly by $\bsell_1$ and $\bsell_2$. 
\par 
Mathematically, the geometry of the unit cell can be characterized by a doubly periodic vector-valued function $\bfx^\rmp\left(u^1, u^2\right)$, which satisfies the following periodic conditions: 
\refstepcounter{equation} \label{eqn: function periodic conditions} 
\begin{align} 
\bfx^\rmp\left(u^1, u^2\right) & = \bfx^\rmp\left(u^1 + \ell^1, u^2         \right), \tag{\theequation, a} \h 
\bfx^\rmp\left(u^1, u^2\right) & = \bfx^\rmp\left(u^1         , u^2 + \ell^2\right). \tag{\theequation, b} 
\end{align} 
The corresponding doubly periodic surface can thus be parameterized as: 
\begin{align} 
                                            \bfx     \left(u^1, u^2\right) 
\equiv u^\alpha\, \unitvec{\bsell}_\alpha + \bfx^\rmp\left(u^1, u^2\right). 
\label{eqn: parametrization of doubly periodic surfaces} 
\end{align} 
\par 
It is straightforward to verify that the parametrization in \Eqref{eqn: parametrization of doubly periodic surfaces} is consistent with the aforementioned translation property of doubly periodic surfaces by checking: 
\refstepcounter{equation} \label{eqn: surface periodic conditions} 
\begin{align} 
\bfx\left(u^1 + \ell^1, u^2         \right) & = \bfx\left(u^1, u^2\right) + \bsell_1, \tag{\theequation, a} \h 
\bfx\left(u^1         , u^2 + \ell^2\right) & = \bfx\left(u^1, u^2\right) + \bsell_2. \tag{\theequation, b} 
\end{align} 
It is worth pointing out that the meaning of ``being periodic'' differs for functions [\Eqsref{eqn: function periodic conditions}] and surfaces [\Eqsref{eqn: surface periodic conditions}]. 
Notation-wise, the superscript ``$\rmp$'' will hereafter be used to denote any doubly periodic function with periods $\ell^1$ and $\ell^2$, satisfying periodic conditions like those in \Eqsref{eqn: function periodic conditions}. 
\subsection{Shape-Periodic deformations} \label{subsec: DP surfaces and SP deformations, SP deformations} 
\subsubsection{Defining relations} \label{subsubsec: DP surfaces and SP deformations, SP deformations, defn. of SP deformation} 
The type of deformation we are interested in is termed by us \emph{shape-periodic}. 
Under a shape-periodic deformation, the Euclidean distance between any pair of points on the deformed surface remains invariant when both points are shifted by some number of unit cells, as depicted in the main text. 
\par 
Mathematically, let $(s, t)$ and $(S, T)$ denote the local coordinates of two points on a doubly periodic surface $\bfx$. 
Like before, we denote a deformed state of the surface by $\bfx'$. 
By our definition, a shape-periodic deformation thus has to satisfy: 
\begin{align} 
  \norm{ \bfx'\left(S            , T            \right) - \bfx'\left(s            , t            \right) } 
= \norm{ \bfx'\left(S + n_1\ell^1, T + n_2\ell^2\right) - \bfx'\left(s + n_1\ell^1, t + n_2\ell^2\right) } 
\label{eqn: defn. of shape-periodic deformations} 
\end{align} 
for some integer tuple $(n_1, n_2)$. 
\subsubsection{Rewriting the shape-periodic condition} 
In terms of the displacement field $\delta{\bfx}$, the deformed state is expressed as: 
\begin{align} 
             {\bfx}'\left(u^1, u^2\right) 
\equiv       {\bfx} \left(u^1, u^2\right) 
     + \delta{\bfx} \left(u^1, u^2\right). 
\label{eqn: a deformed state, revisited} 
\end{align} 
When this is substituted into \Eqref{eqn: defn. of shape-periodic deformations}, the shape-periodic condition takes the form: 
\begin{align} 
\begin{split} 
&   \norm{ 
    \left[      {\bfx}\left(S, T\right) -       {\bfx}\left(s, t\right)\right] 
  + \left[\delta{\bfx}\left(S, T\right) - \delta{\bfx}\left(s, t\right)\right] 
    } 
    \h 
& \quad\quad\quad\quad= 
    \left\lVert{ 
    \left[      {\bfx}\left(S + n_1\ell^1, T + n_2\ell^2\right) -       {\bfx}\left(s + n_1\ell^1, t + n_2\ell^2\right)\right] 
    }\right. 
    \h 
& \quad\quad\quad\quad\quad\quad\quad\quad+ 
    \left.{ 
    \left[\delta{\bfx}\left(S + n_1\ell^1, T + n_2\ell^2\right) - \delta{\bfx}\left(s + n_1\ell^1, t + n_2\ell^2\right)\right] 
    }\right\rVert. 
\end{split} 
\label{eqn: an intermediate step toward deriving the nonlinear SP condition, 0} 
\end{align} 
\par 
The periodicity of the surface $\bfx$ [\Eqsref{eqn: surface periodic conditions}] implies that: 
\begin{align} 
  \bfx\left(S + n_1\ell^1, T + n_2\ell^2\right) - \bfx\left(s + n_1\ell^1, t + n_2\ell^2\right) 
= \bfx\left(S            , T            \right) - \bfx\left(s            , t            \right). 
\end{align} 
From this, it follows that \Eqref{eqn: an intermediate step toward deriving the nonlinear SP condition, 0} simplifies to: 
\begin{align} 
\begin{split} 
&   \norm{ 
    \left[      {\bfx}\left(S, T\right) -       {\bfx}\left(s, t\right)\right] 
  + \left[\delta{\bfx}\left(S, T\right) - \delta{\bfx}\left(s, t\right)\right] 
    } 
    \h 
& \quad\quad= 
    \norm{ 
    \left[      {\bfx}\left(S            , T            \right) -       {\bfx}\left(s            , t            \right)\right] 
  + \left[\delta{\bfx}\left(S + n_1\ell^1, T + n_2\ell^2\right) - \delta{\bfx}\left(s + n_1\ell^1, t + n_2\ell^2\right)\right] 
    }. 
\end{split} 
\label{eqn: an intermediate step toward deriving the nonlinear SP condition, 1} 
\end{align} 
\par 
For brevity, we introduce the following notations: 
\begin{align} 
\Delta     {      \bfx} & \equiv       {\bfx}\left(S            , T            \right) -       {\bfx}\left(s            , t            \right), 
                                  \label{eqn: convenient difference notation, a} 
                                  \h 
\Delta_\ini{\delta\bfx} & \equiv \delta{\bfx}\left(S            , T            \right) - \delta{\bfx}\left(s            , t            \right), 
                                  \label{eqn: convenient difference notation, b} 
                                  \h 
\Delta_\fin{\delta\bfx} & \equiv \delta{\bfx}\left(S + n_1\ell^1, T + n_2\ell^2\right) - \delta{\bfx}\left(s + n_1\ell^1, t + n_2\ell^2\right). 
                                  \label{eqn: convenient difference notation, c} 
\end{align} 
With the newly introduced notations, the shape-periodic condition [\Eqref{eqn: an intermediate step toward deriving the nonlinear SP condition, 1}] reads: 
\begin{align} 
  \norm{ \Delta{\bfx} + \Delta_\ini{\delta\bfx} } 
= \norm{ \Delta{\bfx} + \Delta_\fin{\delta\bfx} }. 
\label{eqn: nonlinear shape-periodic condition} 
\end{align} 
\subsubsection{The linearized shape-periodic condition} 
To further simplify the shape-periodic condition [\Eqref{eqn: nonlinear shape-periodic condition}], we square both sides of the equation and expand the resulting expression, yielding: 
\begin{align} 
  \norm{ \Delta{\bfx} }^2 + \norm{ \Delta_\ini{\delta\bfx} }^2 + 2\inProd{ \Delta{\bfx} }{ \Delta_\ini{\delta\bfx} } 
= \norm{ \Delta{\bfx} }^2 + \norm{ \Delta_\fin{\delta\bfx} }^2 + 2\inProd{ \Delta{\bfx} }{ \Delta_\fin{\delta\bfx} }. 
\label{eqn: an intermediate step toward deriving the linearized SP condition} 
\end{align} 
In the regime of small deformations, the quadratic terms in $\delta{\bfx}$ in \Eqref{eqn: an intermediate step toward deriving the linearized SP condition} can be neglected without loss of accuracy. 
As a result, \Eqref{eqn: an intermediate step toward deriving the linearized SP condition} reduces to: 
\begin{align} 
  \inProd{ 
  \Delta     {      \bfx} }{ 
  \Delta_\fin{\delta\bfx} 
- \Delta_\ini{\delta\bfx} 
  } 
= 0. 
\label{eqn: linearized shape-periodic condition} 
\end{align} 
\par 
\Eqnref{eqn: linearized shape-periodic condition} is the linearized shape-periodic condition. 
In the following subsubsection, we will derive the form of the displacement field $\delta{\bfx}$ that satisfies this condition. 
\subsubsection{The form of shape-periodic displacement fields} \label{subsubsec: DP surfaces and SP deformations, SP deformations, form of SP deformation} 
We begin with the following simple observation. 
The zero deformation, or a doubly periodic surface itself, satisfies the definition of shape-periodic deformation (recall \Secref{subsubsec: DP surfaces and SP deformations, SP deformations, defn. of SP deformation}). 
Due to the surface's periodicity [\Eqsref{eqn: surface periodic conditions}], the Euclidean distance between any pair of points on the surface is guaranteed to remain invariant when both points are shifted by some number of unit cells. 
\paragraph{Type-I shape-periodic deformations.} More generally, as this simple observation implies, any deformation that preserves the surface's periodicity, with unchanged translation vectors $\bsell_1$ and $\bsell_2$, is shape-periodic. 
This type of deformation includes: 
\begin{itemize} 
\item[a.] rigid-body translations, 
\item[b.] rigid-body rotations and 
\item[c.] microscopic periodic deformations occurring within each unit cell in the same manner. 
\end{itemize} 
The deformed states corresponding to these three subtypes can be respectively expressed as follows: 
\refstepcounter{equation} \label{eqn: type-I SP displacement fields} 
\begin{align} 
\bfx'_\tran    \left(u^1, u^2\right) & \equiv        {\bfx}     \left(u^1, u^2\right) 
                                            +        {\bfc}, 
                                              \label{eqn: type-I SP displacement fields, translation} 
                                              \tag{\theequation, a} 
                                              \h 
\bfx'_\rot     \left(u^1, u^2\right) & \equiv \bfR\, {\bfx}     \left(u^1, u^2\right), 
                                              \label{eqn: type-I SP displacement fields, rotation} 
                                              \tag{\theequation, b} 
                                              \h 
\bfx'_\periodic\left(u^1, u^2\right) & \equiv        {\bfx}     \left(u^1, u^2\right) 
                                            +  \delta{\bfx}^\rmp\left(u^1, u^2\right), 
                                              \label{eqn: type-I SP displacement fields, periodic} 
                                              \tag{\theequation, c} 
\end{align} 
where $\bfc$ in the first line is a constant vector, $\bfR$ in the second a rotation matrix, and $\delta{\bfx}^\rmp$ in the third a doubly periodic vector-valued function, satisfying \Eqsref{eqn: function periodic conditions}. 
It is straightforward to verify that 
\Eqsref{eqn: type-I SP displacement fields, translation} and 
 \eqref{eqn: type-I SP displacement fields, periodic}    satisfy the linearized shape-periodic condition [\Eqref{eqn: linearized shape-periodic condition}]. 
For 
 \Eqref{eqn: type-I SP displacement fields, rotation}, it is easier to observe that it satisfies the defining relation for shape-periodic deformations [\Eqref{eqn: defn. of shape-periodic deformations}] and is hence guaranteed to fulfill the linearized version [\Eqref{eqn: linearized shape-periodic condition}]. 
\par 
In the following discussions, we exclude the rigid-body deformations, as they only change our global perspective of a surface and therefore amount to coordinate transformations, while leaving the surface locally unaltered and costing zero energy. 
\paragraph{Type-II shape-periodic deformations.} 
Based on the previous case, we can more broadly conclude that a deformation is shape-periodic as long as the deformed surface remains doubly periodic, with its translation vectors potentially altered under the deformation. 
Physically, this type of deformation involves uniform stretching or shearing of the entire surface. 
Since these deformations are confined to the plane spanned by the translation vectors $\bsell_1$ and $\bsell_2$, we refer to them as uniform planar modes. 
\par 
For a flat plane $\bfx_\plane$, whose parametrization is given by: 
\begin{align} 
       \bfx_\plane\left(u^1, u^2\right) 
\equiv         {     u}^\alpha\, 
       \unitvec{\bsell}_\alpha, 
\label{eqn: parametrization of a plane} 
\end{align} 
the displacement field associated with the uniform planar modes takes the following form, as described in \Refref{Audoly2019_Elasticity_and_Geometry}: 
\begin{align} 
\begin{split} 
                               \delta_\rms{\bfx_\plane                       }\left(u^1, u^2\right) 
  \equiv     {E}^{\alpha\beta }    \inProd{\bfx_\plane                       }{\unitvec{\bsell}_\alpha}\unitvec{\bsell}_\beta 
& \equiv     {E}^{\alpha\beta }    \inProd{u^\gamma\, \unitvec{\bsell}_\gamma}{\unitvec{\bsell}_\alpha}\unitvec{\bsell}_\beta 
         \h 
& \equiv     {E}^{\alpha\beta }\, 
         \hat{g}_{\gamma\alpha}\,          u^\gamma\,                                                  \unitvec{\bsell}_\beta 
         \h 
& \equiv     {E}^{\alpha\beta }\,          u_\alpha\,                                                  \unitvec{\bsell}_\beta. 
\end{split} 
\label{eqn: type-II SP displacement fields, plane} 
\end{align} 
Here, $E^{\alpha\beta}$ denotes the symmetric components of the uniform strain tensor, which describes the extent of stretching and shearing of the planar surface. 
Since the vectors $\bsell_1$ and $\bsell_2$ are generally not orthogonal to each other, we need to introduce the planar metric tensor to differentiate between quantities with upper and lower indices. 
As in \Eqref{eqn: defn. of the metric components}, the components of the planar metric tensor are given by: 
\begin{align} 
       \left(\hat{g}_{\alpha\beta}                                                                 \right) 
\equiv \left(\inProd{         \partial_\alpha{\bfx_\plane} }{         \partial_\beta{\bfx_\plane} }\right) 
     = \left(\inProd{ \unitvec{\bsell}_\alpha              }{ \unitvec{\bsell}_\beta              }\right) 
     = \begin{pmatrix} 
       1          & \cos{\phi} \h 
       \cos{\phi} & 1 
       \end{pmatrix}, 
\label{eqn: defn. of the planar metric components} 
\end{align} 
where $\phi$ denotes the angle between $\bsell_1$ and $\bsell_2$. 
In \Eqref{eqn: type-II SP displacement fields, plane}, the quantity $u_\alpha$ in the third line is obtained by lowering the index of the local coordinates $u^\gamma$ with the planar metric tensor: 
\begin{align} 
                                u_\alpha 
\equiv \hat{g}_{\alpha\gamma}\, u^\gamma 
     = \hat{g}_{\gamma\alpha}\, u^\gamma. 
\label{eqn: covariant coordinates} 
\end{align} 
\par 
For completeness, we also define here the components of the inverse planar metric tensor, again as the matrix inverse of $\hat{g}_{\alpha\beta}$: 
\begin{align} 
       \left(\hat{g}^{\alpha\beta}\right) 
\equiv \left(\hat{g}_{\alpha\beta}\right)^{-1} 
     = \frac{1}{ \sin^2{\phi} }\begin{pmatrix} 
       1           & -\cos{\phi} \h 
       -\cos{\phi} & 1 
       \end{pmatrix}. 
\label{eqn: defn. of the inverse planar metric components} 
\end{align} 
As for general curved surfaces, the inverse planar metric tensor can be used to raise indices, e.g.: 
\begin{align} 
                           u^\gamma 
= \hat{g}^{\gamma\alpha}\, u_\alpha. 
\end{align} 
\par 
Motivated by \Eqref{eqn: type-II SP displacement fields, plane}, we can express the displacement fields associated with the uniform planar modes of a doubly periodic surface as: 
\begin{align} 
                  \delta_\rms{\bfx     }\left(u^1, u^2\right) 
\equiv E^{\alpha\beta}\inProd{\bfx     }{\unitvec{\bsell}_\alpha}\unitvec{\bsell}_\beta 
     +            \delta_\rms{\bfx^\rmp}\left(u^1, u^2\right), 
\label{eqn: type-II SP displacement fields} 
\end{align} 
where $\delta_\rms{\bfx^\rmp}$ is a doubly periodic function that represents the microscopic deformation occurring within each unit cell, such as the flattening of the unit cells when the surface is uniformly stretched. 
Here, the strain components $E^{\alpha\beta}$ quantify the changes in the translation vectors and the surface components along them, $\inProd{\bfx}{\unitvec{\bsell}_\alpha}$, thus describing the macroscopic changes of the surface under uniform stretching or shearing. 
One should not confuse this macroscopic strain with the infinitesimal strain defined in \Secref{subsec: the isometry-stress duality, isometry}. 
In some sense, the macroscopic strain components $E^{\alpha\beta}$ characterize changes in the Euclidean distance between any two points on the surface, whereas the infinitesimal strain components $\varepsilon_{\alpha\beta}$ describe the infinitesimal changes in distance between two neighboring points. 
\par 
Substituting the surface parametrization [\Eqref{eqn: parametrization of doubly periodic surfaces}] into \Eqref{eqn: type-II SP displacement fields} yields: 
\begin{align} 
\begin{split} 
                    \delta_\rms{                                     \bfx     } 
& \equiv E^{\alpha\beta}\inProd{u^\gamma\, \unitvec{\bsell}_\gamma + \bfx^\rmp}{\unitvec{\bsell}_\alpha}\unitvec{\bsell}_\beta 
       +            \delta_\rms{                                     \bfx^\rmp} 
         \h 
&      = E^{\alpha\beta}\inProd{u^\gamma\, \unitvec{\bsell}_\gamma            }{\unitvec{\bsell}_\alpha}\unitvec{\bsell}_\beta 
       + E^{\alpha\beta}\inProd{                                     \bfx^\rmp}{\unitvec{\bsell}_\alpha}\unitvec{\bsell}_\beta 
       +            \delta_\rms{                                     \bfx^\rmp} 
         \h 
& \equiv E^{\alpha\beta}\, u     _\alpha\, \unitvec{\bsell}_\beta 
       + E^{\alpha\beta}\, x^\rmp_\alpha\, \unitvec{\bsell}_\beta 
       + \delta_\rms{\bfx^\rmp}, 
\end{split} 
\label{eqn: an intermediate step toward deriving the type-II SP deformed states} 
\end{align} 
where $x^\rmp_\alpha \equiv \inProd{\bfx^\rmp}{\unitvec{\bsell}_\alpha}$ in the third line denotes the in-plane periodic components of the surface. 
The corresponding deformed states can therefore be expressed as: 
\begin{align} 
\begin{split} 
                    {\bfx}'_\rms\left(u^1, u^2\right) 
& \equiv            {\bfx}      \left(u^1, u^2\right) 
       + \delta_\rms{\bfx}      \left(u^1, u^2\right) 
         \h 
& \equiv                   u     ^\beta \, \unitvec{\bsell}_\beta +            {\bfx^\rmp} 
       + E^{\alpha\beta}\, u     _\alpha\, \unitvec{\bsell}_\beta 
       + E^{\alpha\beta}\, x^\rmp_\alpha\, \unitvec{\bsell}_\beta + \delta_\rms{\bfx^\rmp} 
         \h 
& \equiv \left(u^\beta + \hat{g}_{\alpha\gamma}\, E^{\alpha\beta}\, u     ^\gamma\right)\unitvec{\bsell}_\beta 
       + \left(                                                                                                             {\bfx      }^\rmp 
       +                                          E^{\alpha\beta}\, x^\rmp_\alpha\,     \unitvec{\bsell}_\beta 
       +                                                                                                         \delta_\rms{\bfx      }^\rmp 
                                                                                 \right) 
         \h 
& \equiv \left(u^\beta + \hat{g}_{\alpha\gamma}\, E^{\alpha\beta}\, u     ^\gamma\right)\unitvec{\bsell}_\beta +            {\bfx'_\rms}^\rmp, 
\end{split} 
\label{eqn: type-II SP deformed states} 
\end{align} 
where the defining relation for $u_\alpha$ [\Eqref{eqn: covariant coordinates}] is applied in the third line, and ${\bfx'_\rms}^\rmp\left(u^1, u^2\right)$ in the final line combines all the periodic components of $\bfx'_\rms$. 
\par 
Since the function ${\bfx'_\rms}^\rmp$ is doubly periodic, satisfying \Eqsref{eqn: function periodic conditions}, it is straightforward to verify, using \Eqref{eqn: type-II SP deformed states}, that the resulting deformed surfaces remain doubly periodic, satisfying: 
\refstepcounter{equation} \label{eqn: periodic conditions, type-II SP deformed states} 
\begin{align} 
\bfx'_\rms\left(u^1 + \ell^1, u^2         \right) &      = \bfx'_\rms\left(u^1, u^2\right) 
                                                         + \left(\bsell _1 + \hat{g}_{\alpha 1}\, {E}^{\alpha\beta}\, \ell^1\, \unitvec{\bsell}_\beta\right) 
                                                    \equiv \bfx'_\rms\left(u^1, u^2\right) 
                                                         +       \bsell'_1, 
                                                           \tag{\theequation, a} 
                                                           \h 
\bfx'_\rms\left(u^1         , u^2 + \ell^2\right) &      = \bfx'_\rms\left(u^1, u^2\right) 
                                                         + \left(\bsell _2 + \hat{g}_{\alpha 2}\, {E}^{\alpha\beta}\, \ell^2\, \unitvec{\bsell}_\beta\right) 
                                                    \equiv \bfx'_\rms\left(u^1, u^2\right) 
                                                         +       \bsell'_2, 
                                                           \tag{\theequation, b} 
\end{align} 
where $\bsell'_\alpha$ denotes the modified translation vectors [cf.~\Eqsref{eqn: surface periodic conditions}]. 
Thus, from \Eqsref{eqn: periodic conditions, type-II SP deformed states}, it follows that the uniform planar modes are shape-periodic, satisfying the defining relation in \Eqref{eqn: defn. of shape-periodic deformations}. 
\subparagraph{Remark.} In fact, the displacement fields associated with the uniform planar modes [\Eqref{eqn: an intermediate step toward deriving the type-II SP deformed states}] may represent the only \emph{continuous} solution to the following functional equation, which provides a means to solve the linearized shape-periodic condition [\Eqref{eqn: linearized shape-periodic condition}]: 
\begin{align} 
\begin{split} 
         \zero 
&      = \Delta_\fin{\delta\bfx} 
       - \Delta_\ini{\delta\bfx} 
         \h 
& \equiv \left[\delta{\bfx}\left(S + n_1\ell^1, T + n_2\ell^2\right) - \delta{\bfx}\left(s + n_1\ell^1, t + n_2\ell^2\right)\right] 
       - \left[\delta{\bfx}\left(S            , T            \right) - \delta{\bfx}\left(s            , t            \right)\right] 
         \h 
&      = \left[\delta{\bfx}\left(S + n_1\ell^1, T + n_2\ell^2\right) - \delta{\bfx}\left(S, T\right)\right] 
       - \left[\delta{\bfx}\left(s + n_1\ell^1, t + n_2\ell^2\right) - \delta{\bfx}\left(s, t\right)\right], 
\end{split} 
\label{eqn: a functional difference equation} 
\end{align} 
where the definitions in \Eqsref{eqn: convenient difference notation, a}--\eqref{eqn: convenient difference notation, c} are applied in the second line. 
\par 
The reasoning behind our postulate is as follows. 
First, it is evident that doubly periodic functions satisfy \Eqref{eqn: a functional difference equation}, as both terms on the right-hand side of the equation vanish in this case. 
Other than doubly periodic functions, \Eqref{eqn: a functional difference equation} can only be satisfied if both terms on its right-hand side are equal. 
This condition implies that the terms must depend solely on $n_1\ell^1$ and $n_2\ell^2$, i.e.: 
\begin{align} 
  \delta{\bfx}\left(S + n_1\ell^1, T + n_2\ell^2\right) - \delta{\bfx}\left(S, T\right) 
=       {\bff}\left(    n_1\ell^1,     n_2\ell^2\right) 
= \delta{\bfx}\left(s + n_1\ell^1, t + n_2\ell^2\right) - \delta{\bfx}\left(s, t\right) 
\label{eqn: generalized Cauchy's functional equation?} 
\end{align} 
for some continuous vector-valued function $\bff\left(u^1, u^2\right)$.\footnote{The function $\bff$ is continuous, as the displacement fields considered in this work are assumed to be continuous.} 
\par 
\Eqnref{eqn: generalized Cauchy's functional equation?} resembles Cauchy's functional equation, which takes the following form~\cite{Kuczma2009_Functional_Equations}: 
\begin{align} 
  \delta{\bfx}\left(    u,     v\right) 
- \delta{\bfx}\left(    x,     y\right) 
= \delta{\bfx}\left(u - x, v - y\right). 
\end{align} 
Since continuous solutions to Cauchy's functional equation are known to be linear~\cite{Kuczma2009_Functional_Equations}, we hypothesize that the continuous solutions to \Eqref{eqn: generalized Cauchy's functional equation?}, while slightly different from Cauchy's functional equation, must also be linear. 
If this hypothesis holds, the displacement fields that satisfy \Eqref{eqn: generalized Cauchy's functional equation?} will take the form: 
\begin{align} 
       \delta{\bfx}\left(u^1, u^2\right) 
\equiv E^{\alpha\beta}\, u_\alpha\, \unitvec{\bsell}_\beta 
     + \left( 
       M^{\alpha\beta}\, u_\alpha\, \unitvec{\bsell}_\beta 
     + C^{\alpha     }\, u_\alpha\, \unitvec{z     } 
       \right), 
\label{eqn: linear solutions to the generalized Cauchy's functional equation} 
\end{align} 
where 
$\left(E^{\alpha\beta}\right)$ is a      symmetric matrix, 
$\left(M^{\alpha\beta}\right)$    an antisymmetric matrix, and 
$\left(C^{\alpha     }\right)$    an in-plane      vector. 
\par 
Adding doubly periodic functions to \Eqref{eqn: linear solutions to the generalized Cauchy's functional equation} then yields the most general solution to \Eqref{eqn: a functional difference equation}. 
This solution is essentially captured by \Eqref{eqn: an intermediate step toward deriving the type-II SP deformed states}, except that the terms in parentheses from \Eqref{eqn: linear solutions to the generalized Cauchy's functional equation} are excluded, as they correspond to rigid-body rotations.\footnote{
To illustrate this argument, one can use the fact that the displacement vector $\delta_\rot{\bfx}$ corresponding to a small global rotation can be expressed as the cross product of the unrotated vector 
$\bfx     \equiv      u^\alpha\, \unitvec{\bsell}_\alpha + \bfx^\rmp\left(u^1, u^2\right)$ and the \emph{constant} rotation vector 
$\bsomega \equiv \omega^\alpha\, \unitvec{\bsell}_\alpha 
               + \omega_z     \, \unitvec{z     }                                        $, 
where the magnitude and direction of $\bsomega$ represent the rotation angle and axis, respectively. 
Taking the indicated cross product yields, after some algebra: 
\begin{align*} 
  \delta_\rot{\bfx} 
=            {\bfx} \times                             \bsomega 
= \left(-\hat{\calE}^{\alpha          {\beta}      }\,   \omega_z    \right)u_\alpha\, \unitvec{\bsell}_\beta 
+ \left(-\hat{\calE}_ \beta ^{\phantom{\beta}\alpha}\,   \omega^\beta\right)u_\alpha\, \unitvec{z     } 
+ \delta_\rot{\bfx}^\rmp\left(u^1, u^2\right), 
\end{align*} 
which includes the terms in parentheses from \Eqref{eqn: linear solutions to the generalized Cauchy's functional equation}. 
} 
\paragraph{Type-III shape-periodic deformations.} To derive the form of this final type of shape-periodic deformations, we once again consider a planar surface, which is doubly periodic in a trivial sense, as our starting point. 
\par 
The difference vector between any two points on a planar surface always lies within the plane. 
Therefore, the displacement fields satisfying the linearized shape-periodic condition [\Eqref{eqn: linearized shape-periodic condition}] must be strictly out-of-plane, unless they correspond to the uniform planar modes discussed in the previous paragraphs. 
Thus, these deformations are referred to as out-of-plane modes, which typically involve bending or twisting of the planar surface. 
As presented in \Refref{LL1986_Elasticity}, the displacement fields associated with the \emph{uniform} out-of-plane modes of a planar surface can be expressed locally as: 
\begin{align} 
       \delta_\rmb{\bfx_\plane}\left(u^1, u^2\right) 
\equiv \frac{1}{2}\, H^{\alpha\beta}\inProd{\bfx_\plane}{\unitvec{\bsell}_\alpha} 
                                    \inProd{\bfx_\plane}{\unitvec{\bsell}_\beta }\, \unitvec{z}, 
\label{eqn: type-III SP displacement fields, plane} 
\end{align} 
where the constant symmetric matrix $\left(H^{\alpha\beta}\right)$ encodes information about the curvature changes at the origin $\bfx_\plane(0, 0)$, and 
$
       \unitvec{z} 
\equiv      {\bsell_1 \times \bsell_2} / 
       \norm{\bsell_1 \times \bsell_2} 
$ characterizes the out-of-plane direction. 
\par 
For general doubly periodic surfaces, the situation becomes slightly more complex. 
Such a surface inherently features hills and valleys, meaning that the difference vector between two points on the surface typically does not lie within the plane spanned by the translation vectors $\bsell_1$ and $\bsell_2$. 
As a result, unlike in the planar case [\Eqref{eqn: type-III SP displacement fields, plane}], the displacement fields associated with the uniform out-of-plane modes of a doubly periodic surface generally contain an in-plane component. 
In fact, from everyday experience, we know that when bending a surface with hills and valleys upwards, the hills tend to compress while the valleys extend simultaneously. 
The in-plane component captures this behavior precisely. 
\par 
For a doubly periodic surface, the displacement fields associated with the uniform out-of-plane modes may thus be expressed as: 
\begin{align} 
\begin{split} 
         \delta_\rmb{\bfx}\left(u^1, u^2\right) 
& \equiv \frac{1}{2}\, H^{\alpha\beta}\inProd{\bfx}{\unitvec{\bsell}_\alpha}\inProd{\bfx}{\unitvec{\bsell}_\beta}\, \unitvec{z     } 
       -               H^{\alpha\beta}\inProd{\bfx}{\unitvec{\bsell}_\alpha}\inProd{\bfx}{\unitvec{z     }      }\, \unitvec{\bsell}_\beta 
         \h 
& \equiv \frac{1}{2}\, H^{\alpha\beta}\, x_\alpha\, x_\beta\, \unitvec{z     } 
       -               H^{\alpha\beta}\, x_\alpha\, x_z    \, \unitvec{\bsell}_\beta, 
\end{split} 
\label{eqn: type-III SP displacement fields} 
\end{align} 
where 
$x_\alpha\left(u^1, u^2\right) \equiv \inProd{\bfx}{\unitvec{\bsell}_\alpha}$ and 
$x_z     \left(u^1, u^2\right) \equiv \inProd{\bfx}{\unitvec{z     }       }$ describe the in-plane and height profiles of the surface, respectively. 
Here, as in \Eqref{eqn: type-III SP displacement fields, plane}, the symmetric matrix $\left(H^{\alpha\beta}\right)$ quantifies the bending and twisting of the plane spanned by the translation vectors $\bsell_1$ and $\bsell_2$, thereby characterizing the macroscopic out-of-plane deformation of the surface. 
The in-plane component accompanying the macroscopic out-of-plane deformation is represented by the second term on the right-hand side of \Eqref{eqn: type-III SP displacement fields}. 
Notably, this term is linearly related to $-x_z$, indicating that as the surface bends upwards, it undergoes increased compression along the $\unitvec{z}$-direction. 
In the following paragraphs, we will confirm that \Eqref{eqn: type-III SP displacement fields} indeed satisfies the linearized shape-periodic condition [\Eqref{eqn: linearized shape-periodic condition}]. 
\par 
As in \Eqref{eqn: an intermediate step toward deriving the type-II SP deformed states}, substituting the surface parametrization [\Eqref{eqn: parametrization of doubly periodic surfaces}] into \Eqref{eqn: type-III SP displacement fields} yields: 
\begin{align} 
\begin{split} 
         \delta_\rmb{\bfx} 
& \equiv \frac{1}{2}\, H^{\alpha\beta}\left(u_\alpha + x^\rmp_\alpha\right)\left(u_\beta + x^\rmp_\beta\right)   \unitvec{z     } 
       -               H^{\alpha\beta}\left(u_\alpha + x^\rmp_\alpha\right)                x^\rmp_z           \, \unitvec{\bsell}_\beta 
         \h 
&      = \frac{1}{2}\, H^{\alpha\beta}\,      u_\alpha\,      u_\beta\, \unitvec{z     } 
         \h 
& \quad\quad+ 
         \left( 
         \frac{1}{2}\, H^{\alpha\beta}\,      u_\alpha\, x^\rmp_\beta\, \unitvec{z     } 
       + \frac{1}{2}\, H^{\alpha\beta}\, x^\rmp_\alpha\,      u_\beta\, \unitvec{z     } 
       -               H^{\alpha\beta}\,      u_\alpha\, x^\rmp_z    \, \unitvec{\bsell}_\beta 
         \right) 
         \h 
& \quad\quad\quad\quad+ 
         \left( 
         \frac{1}{2}\, H^{\alpha\beta}\, x^\rmp_\alpha\, x^\rmp_\beta\, \unitvec{z     } 
       -               H^{\alpha\beta}\, x^\rmp_\alpha\, x^\rmp_z    \, \unitvec{\bsell}_\beta 
         \right), 
\end{split} 
\label{eqn: type-III SP displacement fields, expanded} 
\end{align} 
where 
$x^\rmp_\alpha\left(u^1, u^2\right) \equiv \inProd{\bfx^\rmp}{\unitvec{\bsell}_\alpha}$ and 
$x^\rmp_z     \left(u^1, u^2\right) \equiv \inProd{\bfx^\rmp}{\unitvec{z     }       } 
                                         = \inProd{\bfx     }{\unitvec{z     }       } 
                                         = x_z 
$ represent the in-plane and out-of-plane periodic components of the surface, respectively. 
Because the matrix components $H^{\alpha\beta}$ are symmetric, we have the following: 
\begin{align} 
  H^{\alpha\beta }\,      u_\alpha\, x^\rmp_\beta 
= H^{\beta \alpha}\,      u_\beta \, x^\rmp_\alpha 
= H^{\alpha\beta }\, x^\rmp_\alpha\,      u_\beta, 
\end{align} 
where the first equality is obtained by interchanging the indices $\alpha \leftrightarrow \beta$, and the second follows from the symmetry of $H^{\alpha\beta}$. 
\Eqnref{eqn: type-III SP displacement fields, expanded} accordingly simplifies to: 
\begin{align} 
       \delta_\rmb{\bfx     } 
\equiv \frac{1}{2}\, H^{\alpha\beta}\, u_\alpha\,      u_\beta\, \unitvec{z     } 
     + \left( 
                     H^{\alpha\beta}\, u_\alpha\, x^\rmp_\beta\, \unitvec{z     } 
     -               H^{\alpha\beta}\, u_\alpha\, x^\rmp_z    \, \unitvec{\bsell}_\beta 
       \right) 
     + \delta_\rmb{\bfx^\rmp}, 
\label{eqn: type-III SP displacement fields, simplified, covariant} 
\end{align} 
where $\delta_\rmb{\bfx^\rmp}$ includes all the periodic components of $\delta_\rmb{\bfx}$, as shown in the last line of \Eqref{eqn: type-III SP displacement fields, expanded}. 
Finally, using the defining relation for $u_\alpha$ [\Eqref{eqn: covariant coordinates}], \Eqref{eqn: type-III SP displacement fields, simplified, covariant} can be rewritten in terms of the local coordinates $u^\alpha$ as: 
\begin{align} 
       \delta_\rmb{\bfx     } 
\equiv \frac{1}{2}\,   H_{\alpha  \beta}\, u^\alpha\,      u^\beta\, \unitvec{z     } 
     + \left( 
                      {H_ \alpha}^\beta \, u^\alpha\, x^\rmp_\beta\, \unitvec{z     } 
     -                {H_ \alpha}^\beta \, u^\alpha\, x^\rmp_z    \, \unitvec{\bsell}_\beta 
       \right) 
     + \delta_\rmb{\bfx^\rmp}, 
\label{eqn: type-III SP displacement fields, simplified, contravariant} 
\end{align} 
where the planar metric tensor [see \Eqref{eqn: defn. of the planar metric components}] is employed to lower the indices of $H^{\alpha\beta}$: 
\refstepcounter{equation} \label{eqn: mixed and covariant components of the H matrix} 
\begin{align} 
{H_ \alpha}^\beta  & \equiv                          \hat{g}_{\alpha\gamma}\, H^{\gamma\beta}, 
                            \label{eqn: mixed components of the H matrix} 
                            \tag{\theequation, a} 
                            \h 
 H_{\alpha  \beta} & \equiv \hat{g}_{\alpha\gamma}\, \hat{g}_{\beta \rho  }\, H^{\gamma\rho }. 
                            \label{eqn: covariant components of the H matrix} 
                            \tag{\theequation, b} 
\end{align} 
Here, since both 
$\left(\hat{g}_{\alpha\gamma}\right)$ and 
$\left(    {H}^{\gamma\rho  }\right)$ are symmetric, the matrix 
$\left(    {H}_{\alpha\beta }\right)$ is also symmetric. 
\par 
We now demonstrate that \Eqref{eqn: type-III SP displacement fields, simplified, contravariant} satisfies the linearized shape-periodic condition [\Eqref{eqn: linearized shape-periodic condition}]. 
To begin, we denote 
$L^1 \equiv n_1\ell^1$ and 
$L^2 \equiv n_2\ell^2$. 
With these convenient notations, we consider the following difference function: 
\begin{align} 
       \Delta{\delta_\rmb\bfx}\left(u^1      , u^2      ; 
                                          L^1,       L^2\right) 
\equiv       {\delta_\rmb\bfx}\left(u^1 + L^1, u^2 + L^2\right) 
     -       {\delta_\rmb\bfx}\left(u^1      , u^2      \right). 
\label{eqn: a difference function of type-III SP displacement fields} 
\end{align} 
\par 
By substituting \Eqref{eqn: type-III SP displacement fields, simplified, contravariant} into \Eqref{eqn: a difference function of type-III SP displacement fields}, we expand the difference function as follows: 
\begin{align} 
\begin{split} 
         \Delta{\delta_\rmb\bfx} 
& \equiv \frac{1}{2}\, H_{\alpha\beta}\left(u^\alpha + L^\alpha\right)\left(u^\beta + L^\beta\right)\unitvec{z} 
       - \frac{1}{2}\, H_{\alpha\beta}\,    u^\alpha           \,           u^\beta          \,     \unitvec{z} 
         \h 
& \quad\quad+ 
         {H_\alpha}^\beta\left(u^\alpha + L^\alpha\right)x^\rmp_\beta\left(u^1 + L^1, u^2 + L^2\right)\, \unitvec{z} 
       - {H_\alpha}^\beta\,    u^\alpha           \,     x^\rmp_\beta\left(u^1      , u^2      \right)\, \unitvec{z} 
         \h 
& \quad\quad\quad\quad- 
         {H_\alpha}^\beta\left(u^\alpha + L^\alpha\right)x^\rmp_z\left(u^1 + L^1, u^2 + L^2\right)\, \unitvec{\bsell}_\beta 
       + {H_\alpha}^\beta\,    u^\alpha           \,     x^\rmp_z\left(u^1      , u^2      \right)\, \unitvec{\bsell}_\beta 
         \h 
& \quad\quad\quad\quad\quad\quad+ 
         \delta_\rmb{\bfx^\rmp}\left(u^1 + L^1, u^2 + L^2\right) 
       - \delta_\rmb{\bfx^\rmp}\left(u^1      , u^2      \right). 
\end{split} 
\label{eqn: an intermediate step for simplifying the difference function, 0} 
\end{align} 
To simplify \Eqref{eqn: an intermediate step for simplifying the difference function, 0}, recall that quantities with the superscript ``$\rmp$'' are doubly periodic and satisfy \Eqsref{eqn: function periodic conditions}; for instance: 
\begin{align} 
  x^\rmp_\beta\left(u^1 + L^1, u^2 + L^2\right) 
= x^\rmp_\beta\left(u^1      , u^2      \right). 
\end{align} 
Thus, after some cancellations, \Eqref{eqn: an intermediate step for simplifying the difference function, 0} reduces to: 
\begin{align} 
  \Delta{\delta_\rmb\bfx} 
= \frac{1}{2}\,  H_{\alpha  \beta}\, L^\alpha\,      L^\beta\, \unitvec{z     } 
+                H_{\alpha  \beta}\, L^\alpha\,      u^\beta\, \unitvec{z     } 
+               {H_ \alpha}^\beta \, L^\alpha\, x^\rmp_\beta\, \unitvec{z     } 
-               {H_ \alpha}^\beta \, L^\alpha\, x^\rmp_z    \, \unitvec{\bsell}_\beta, 
\label{eqn: an intermediate step for simplifying the difference function, 1} 
\end{align} 
where the symmetry of $H_{\alpha\beta}$ is used in the derivation, and we suppress the dependence of all the doubly periodic functions on $u^1$ and $u^2$ to save some writing. 
\par 
To proceed, we combine the second and third terms on the right-hand side of \Eqref{eqn: an intermediate step for simplifying the difference function, 1} as follows: 
\begin{align} 
\begin{split} 
     H_{\alpha  \beta}\, L^\alpha\,      u^\beta\, \unitvec{z} 
  + {H_ \alpha}^\beta \, L^\alpha\, x^\rmp_\beta\, \unitvec{z} 
& = {H_ \alpha}^\beta \, L^\alpha\,      u_\beta\, \unitvec{z} 
  + {H_ \alpha}^\beta \, L^\alpha\, x^\rmp_\beta\, \unitvec{z} 
    \h 
& = {H_ \alpha}^\beta \, L^\alpha\left( 
                                         u_\beta 
  +                                 x^\rmp_\beta 
    \right)                                        \unitvec{z} 
    \h 
& \equiv 
    {H_ \alpha}^\beta \, L^\alpha\,      x_\beta\, \unitvec{z}. 
\end{split} 
\label{eqn: an intermediate step for simplifying the difference function, 2} 
\end{align} 
By substituting 
\Eqref{eqn: an intermediate step for simplifying the difference function, 2} into 
\Eqref{eqn: an intermediate step for simplifying the difference function, 1}, we obtain the final simplified expression for the difference function: 
\begin{align} 
  \Delta{\delta_\rmb\bfx}\left(u^1, u^2; L^1, L^2\right) 
= \frac{1}{2}\,  H_{\alpha  \beta}\, L^\alpha\, L^\beta              \,     \unitvec{z     } 
+               {H_ \alpha}^\beta \, L^\alpha\, x_\beta\left(u^1, u^2\right)\unitvec{z     } 
-               {H_ \alpha}^\beta \, L^\alpha\, x_z    \left(u^1, u^2\right)\unitvec{\bsell}_\beta, 
\label{eqn: a difference function of type-III SP displacement fields, final form} 
\end{align} 
where the identity $x^\rmp_z = x_z^{\phantom\rmp}$ is also used. 
\par 
Now, in terms of the difference function $\Delta{\delta_\rmb\bfx}$, the linearized shape-periodic condition [\Eqref{eqn: linearized shape-periodic condition}] can be rewritten as: 
\begin{align} 
  \inProd{ 
         {           \bfx}\left(S, T          \right) 
-        {           \bfx}\left(s, t          \right) }{ 
   \Delta{\delta_\rmb\bfx}\left(S, T; L^1, L^2\right) 
-  \Delta{\delta_\rmb\bfx}\left(s, t; L^1, L^2\right) } 
= 0. 
\label{eqn: linearized shape-periodic condition in terms of the difference function} 
\end{align} 
[Recall the definitions in \Eqsref{eqn: convenient difference notation, a}--\eqref{eqn: convenient difference notation, c} and \eqref{eqn: a difference function of type-III SP displacement fields}.] 
Since $\Delta{\delta_\rmb\bfx}$ is linear in the surface components [see \Eqref{eqn: a difference function of type-III SP displacement fields, final form}], we have: 
\begin{align} 
  \Delta{\delta_\rmb\bfx}\left(S, T; L^1, L^2\right) 
- \Delta{\delta_\rmb\bfx}\left(s, t; L^1, L^2\right) 
= {H_\alpha}^\beta\, L^\alpha\, \Delta{x}_\beta\, \unitvec{z     } 
- {H_\alpha}^\beta\, L^\alpha\, \Delta{x}_z    \, \unitvec{\bsell}_\beta, 
\label{eqn: the difference of two difference functions} 
\end{align} 
where: 
\begin{align} 
\Delta{x_\beta} \equiv \inProd{ \Delta{\bfx} }{\unitvec{\bsell}_\beta} & \equiv \inProd{ \bfx\left(S, T\right) - \bfx\left(s, t\right) }{\unitvec{\bsell}_\beta}, 
                                                                                \label{eqn: convenient difference notation, d} 
                                                                                \h 
\Delta{x_z    } \equiv \inProd{ \Delta{\bfx} }{\unitvec{z     }      } & \equiv \inProd{ \bfx\left(S, T\right) - \bfx\left(s, t\right) }{\unitvec{z     }      }. 
                                                                                \label{eqn: convenient difference notation, e} 
\end{align} 
By taking the inner product of both sides of \Eqref{eqn: the difference of two difference functions} with $\Delta{\bfx}$, we can ultimately show that the displacement fields associated with the uniform out-of-plane modes of a doubly periodic surface, as given in \Eqref{eqn: type-III SP displacement fields, simplified, contravariant}, satisfy the linearized shape-periodic condition [\Eqref{eqn: linearized shape-periodic condition in terms of the difference function}]: 
\begin{align} 
  \inProd{ 
   \Delta{           \bfx}                            }{ 
   \Delta{\delta_\rmb\bfx}\left(S, T; L^1, L^2\right) 
-  \Delta{\delta_\rmb\bfx}\left(s, t; L^1, L^2\right) 
  } 
= {H_\alpha}^\beta\, L^\alpha\left( 
   \Delta{x}_\beta\, \Delta{x}_z 
-  \Delta{x}_z    \, \Delta{x}_\beta 
  \right) 
= 0, 
\end{align} 
where the definitions in 
\Eqsref{eqn: convenient difference notation, d} and 
 \eqref{eqn: convenient difference notation, e} are utilized. 
\subparagraph{Remark.} Based on the calculations above, we infer that, conversely, for the linearized shape-periodic condition [\Eqref{eqn: linearized shape-periodic condition in terms of the difference function}] to be satisfied, the difference 
$
  \Delta{\delta_\rmb\bfx}\left(S, T; L^1, L^2\right) 
- \Delta{\delta_\rmb\bfx}\left(s, t; L^1, L^2\right) 
$ must be linearly related to the vector 
$
  \Delta{x}_\beta\, \unitvec{z     } 
- \Delta{x}_z    \, \unitvec{\bsell}_\beta 
$, with constant coefficients. 
Consequently, the difference function $\Delta{\delta_\rmb\bfx}$ is necessarily linear in the surface components, as shown in \Eqref{eqn: a difference function of type-III SP displacement fields, final form}. 
This reasoning demonstrates that the displacement fields associated with the uniform out-of-plane modes are the only solutions to the linearized shape-periodic condition, excluding the uniform planar modes. 
\paragraph{Summary.} To summarize, we have shown that there are three types of shape-periodic deformation---deformations that preserve the Euclidean distance between points on a doubly periodic surface under translations by unit cells. 
They include: 
\begin{itemize} 
\item[1.] Type-I shape-periodic deformations, which consist of rigid-body translations, rigid-body rotations and microscopic periodic deformations that occur in each unit cell              in the same manner. 
          Excluding the rigid-body deformations, the displacement field associated with this type of shape-periodic deformation can be represented by a doubly periodic vector-valued function $\delta_\rmI{\bfx^\rmp}\left(u^1, u^2\right)$: 
          \begin{align} 
                 \delta_\periodic{\bfx     }\left(u^1, u^2\right) 
          \equiv \delta_\rmI     {\bfx^\rmp}\left(u^1, u^2\right). 
          \label{eqn: type-I SP displacement fields, revisited} 
          \end{align} 
\item[2.] Type-II shape-periodic deformations, i.e., uniform planar modes, which involve uniform stretching or shearing of a doubly periodic surface. 
          This type of shape-periodic deformation can be characterized by the symmetric macroscopic strain components $E^{\alpha\beta}$, which quantify the stretching or shearing of the plane spanned by the translation vectors of the doubly periodic surface. 
          The corresponding displacement fields are given by: 
          \begin{align} 
                 \delta_\rms {\bfx     }\left(u^1, u^2\right) 
          \equiv         {     E}^{\alpha\beta}\, 
                         {     u}_ \alpha      \, 
                 \unitvec{\bsell}_       \beta 
               + \delta_\rmII{\bfx^\rmp}\left(u^1, u^2\right), 
          \label{eqn: type-II SP displacement fields, revisited} 
          \end{align} 
          where the doubly periodic function $\delta_\rmII{\bfx^\rmp}$ represents the microscopic periodic deformation that occurs when the surface is uniformly stretched or sheared. 
\item[3.] Type-III shape-periodic deformations, or uniform out-of-plane modes, which encompass uniform bending or twisting of a doubly periodic surface. 
          The overall bending and twisting of the surface can be described by the symmetric matrix $\left(H^{\alpha\beta}\right)$, which characterizes the local curvature changes of the plane spanned by the translation vectors of the surface. 
          The corresponding displacement fields can be expressed as: 
          \begin{align} 
                 \delta_\rmb  {\bfx     }\left(u^1, u^2\right) 
          \equiv \frac{1}{2}\, H^{\alpha\beta}\, u_\alpha\,      u_\beta\, \unitvec{z     } 
               + \left( 
                               H^{\alpha\beta}\, u_\alpha\, x^\rmp_\beta\, \unitvec{z     } 
               -               H^{\alpha\beta}\, u_\alpha\, x^\rmp_z    \, \unitvec{\bsell}_\beta 
                 \right) 
               + \delta_\rmIII{\bfx^\rmp}\left(u^1, u^2\right), 
          \label{eqn: type-III SP displacement fields, revisited} 
          \end{align} 
          where the doubly periodic function $\delta_\rmIII{\bfx^\rmp}$ represents the microscopic periodic deformation that accompanies the uniform bending or twisting of the surface. 
          In \Eqref{eqn: type-III SP displacement fields, revisited}, the in-plane component, $-H^{\alpha\beta}\, u_\alpha\, x^\rmp_z\, \unitvec{\bsell}_\beta$, captures the phenomenon that when the doubly periodic surface is bent or twisted, its hills compress while its valleys stretch. 
\end{itemize} 
\par 
A general shape-periodic deformation can involve both planar and out-of-plane deformations. 
Its associated displacement field can therefore be expressed as the sum of \Eqsref{eqn: type-I SP displacement fields, revisited}--\eqref{eqn: type-III SP displacement fields, revisited}: 
\begin{align} 
\begin{split} 
         \delta_\SP      {\bfx}\left(u^1, u^2\right) 
& \equiv \delta_\rms     {\bfx}\left(u^1, u^2\right) 
       + \delta_\rmb     {\bfx}\left(u^1, u^2\right) 
       + \delta_\periodic{\bfx}\left(u^1, u^2\right) 
         \h 
& \equiv \left( 
         E^{\alpha\beta} 
       - H^{\alpha\beta}\, x^\rmp_z 
         \right)        \,      u_\alpha\, \unitvec{\bsell}_\beta 
       + \left( 
         \frac{1}{2}\, H^{\alpha\beta}\, u_\alpha\,      u_\beta 
       +               H^{\alpha\beta}\, u_\alpha\, x^\rmp_\beta 
         \right)                        \, \unitvec{z     } 
       + \delta_\SP{\bfx^\rmp}\left(u^1, u^2\right), 
\end{split} 
\label{eqn: expression for general SP displacement fields} 
\end{align} 
where the doubly periodic function 
$
       \delta_\SP   {\bfx^\rmp} 
\equiv \delta_\rmI  {\bfx^\rmp} 
     + \delta_\rmII {\bfx^\rmp} 
     + \delta_\rmIII{\bfx^\rmp} 
$ sums up all the periodic components of $\delta_\SP{\bfx}$. 
In \Eqref{eqn: expression for general SP displacement fields}, note that the total strain is not constant and has components $E^{\alpha\beta} - H^{\alpha\beta}\, x^\rmp_z\left(u^1, u^2\right)$, which seem to depend on where the zero-height level $x_z = 0$ is chosen. 
This gives rise to the notion of gauge invariance~\cite{Schwichtenberg2019_Gauge_Symmetry} in the system, which will be discussed in \Secref{sec: phase space geometry}. 
\par 
As \Eqref{eqn: expression for general SP displacement fields} demonstrates, any shape-periodic deformation of a given doubly periodic surface can be represented by the six-dimensional vector: 
\begin{align} 
       \left(E^{\alpha\beta}, H^{\alpha\beta}\right) 
\equiv \left(E^{1     1    }, 
             E^{2     2    }, 
             E^{1     2    }, 
                              H^{1     1    }, 
                              H^{2     2    }, 
                              H^{1     2    }\right), 
\label{eqn: expression for macroscopic deformations} 
\end{align} 
which describes the macroscopic deformation of the surface, along with a doubly periodic vector-valued function ($\delta_\SP{\bfx^\rmp}$) that characterizes the microscopic deformation within each unit cell. 
Since each six-dimensional vector corresponds to a unique way of deforming the doubly periodic surface macroscopically, we refer to these vectors as macroscopic deformations, and the space they span is termed the \emph{deformation phase space}. 
Thus, the space of shape-periodic displacement fields can be viewed as a fiber bundle, with the deformation phase space as its base manifold and the space of doubly periodic vector-valued functions as the corresponding fibers. 
\par 
In the following sections, we explore the geometry of the deformation phase space, beginning with the subspace of isometric shape-periodic deformations. 
\section{Isometric shape-periodic deformations} \label{sec: isometric SP deformations} 
\subsection{The dimension of the isometric subspace} \label{subsec: isometric SP deformations, dimension of the isometric subspace} 
First, we demonstrate that the subspace of isometric shape-periodic deformations is three-dimensional. 
This follows as a corollary of the isometry-stress duality established in \Secref{subsec: the isometry-stress duality, duality}, which states that an imposed energy-minimizing deformation can be mapped to an isometric deformation. 
\par 
Let $\bfv \equiv \left(E^{\alpha\beta}, H^{\alpha\beta}\right)$ be changes to the coarse-grained strain and curvature that result from some macroscopic deformation imposed on the system. We assume that in response to this imposed combination of bending and stretching the system relaxes to an energy-minimizing configuration according to some linear, stable constitutive relationship.
According to the established isometry-stress duality, the resultant equilibrium stress can be linearly mapped to the angular acceleration tensor (introduced in Sec.~II of the main text) associated with an \emph{isometric} shape-periodic deformation, which would in turn generate changes to the coarse-grained strain and curvature that we denote by $\wt{\bfv} \equiv \left(\wt{E}^{\alpha\beta}, \wt{H}^{\alpha\beta}\right)$.
Note that this new deformation, which is isometric, is not the same as the imposed deformation, which is in general not isometric. Indeed, when macroscopic deformations consistent with an isometry (such as uniaxially straining a singly corrugated sheet) are imposed, the resulting stress and dual isometric deformation is zero.

The linear isometry-stress duality can thus be represented by a six-by-six matrix, which maps $\bfv$ to $\wt{\bfv}$: 
\begin{align} 
       \wt{\bfv} 
\equiv    {\bfM}_\IS\, 
          {\bfv}. 
\label{eqn: relation between equilibrium stress and isometry, matrix} 
\end{align}

\noindent Since isometric deformations do not induce stresses in a surface, the matrix $\bfM_\IS$ always maps the macroscopic deformations associated with imposed isometric deformations to the zero vector. 
In other words, the nullspace of $\bfM_\IS$ corresponds exactly to the isometric subspace, and hence: 
\begin{align} 
  \nullity\left(\bfM_\IS\right) 
= N_\iso. 
\end{align} 
\par
However, from \Eqref{eqn: relation between equilibrium stress and isometry, matrix}, it follows that every set of macroscopic deformations that does not correspond to an isometry induces a stress that can then be mapped onto an isometry. Hence, the rank of the matrix $\bfM_\IS$ is equal to the dimension of the isometric subspace (denoted by $N_\iso$): 
\begin{align} 
  \rank\left(\bfM_\IS\right) 
= N_\iso. 
\end{align} 
Therefore, by the rank-nullity theorem, we conclude that exactly half of the six-dimensional space of combined coarse-grained strains and curvatures corresponds to isometries. For any given surface, there are three ways to globally deform it isometrically, which also correspond to the low-energy deformations of thin surfaces.
\subsection{Characterizing isometric shape-periodic deformations} \label{subsec: isometric SP deformations, characterization} 
To further examine the isometric subspace, we need a way to characterize isometric shape-periodic deformations. 
Recall from \Secref{subsec: the isometry-stress duality, isometry} that any linear isometric deformation corresponds locally to an infinitesimal rotation of the area elements constituting a surface; these local rotations can be characterized by an angular velocity field. 
\par 
The same concept also applies to doubly periodic surfaces. 
However, for such a surface, the angular velocity field corresponding to a linear isometric shape-periodic deformation must characterize not only the infinitesimal rotation of the area elements within each unit cell but also, on a macroscopic level, how each unit cell rotates relative to its neighbors. 
Since shape-periodic deformations by definition have the same shape in each unit cell (see \Secref{subsubsec: DP surfaces and SP deformations, SP deformations, form of SP deformation}), the relative rotation of the unit cells must also be uniform. 
That is, when a doubly periodic surface undergoes an isometric deformation, each unit cell will rotate identically relative to its neighbors. 
\par 
Mathematically, the above discussion suggests that the angular velocity field associated with an isometric shape-periodic deformation consists of two components: a microscopic part, describing the infinitesimal rotations within each unit cell, and a macroscopic part, describing the relative rotations and displacements of the unit cells. 
Moreover, as dictated by the shape-periodic condition, the microscopic part must be doubly periodic (i.e., identical for each unit cell), while the macroscopic angular velocity field must vary uniformly across the doubly periodic surface. 
Therefore, the angular velocity field can be parameterized as: 
\begin{align} 
       \bsomega_\SP     \left(u^1, u^2\right) 
\equiv                        u^\alpha\, 
                           \bfw_\alpha 
     + \bsomega_\SP^\rmp\left(u^1, u^2\right), 
\label{eqn: parametrization of the SP angular velocity fields} 
\end{align} 
where the doubly periodic vector-valued function $\bsomega_\SP^\rmp$ represents the microscopic angular velocity field, and the constant vectors $\bfw_\alpha$ denote the macroscopic angular acceleration vectors, quantifying the uniform variation of the macroscopic angular velocity field. 
\subsubsection{Relation between the macroscopic angular velocity fields and uniform out-of-plane modes} 
Under an isometric shape-periodic deformation, each unit cell of a doubly periodic surface rotates relative to its neighbors in the same manner, which may cause the surface as a whole to bend or twist out-of-plane. 
In light of this observation, for isometric shape-periodic deformations, we now have two different approaches to quantify the uniform out-of-plane modes of a doubly periodic surface: one using the macroscopic component of the angular velocity field and the other using the symmetric matrix $\left(H^{\alpha\beta}\right)$, which characterizes the curvature changes of the plane spanned by the surface's translation vectors. 
\par 
Certainly, the two approaches must be geometrically equivalent. 
To mathematically relate the two, we recall the definition of the angular velocity field [\Eqref{eqn: defn. of the angular velocity}], which is expressed in terms of the notations for isometric shape-periodic deformations as follows: 
\begin{align} 
       \partial_\alpha{  \delta_\SP\bfx} 
\equiv                 \bsomega_\SP 
\times \partial_\alpha{            \bfx}. 
\label{eqn: defn. of the angular velocity, revisited} 
\end{align} 
\par 
The left-hand side of \Eqref{eqn: defn. of the angular velocity, revisited} can be obtained by taking the derivative of the expression for a general shape-periodic displacement field [\Eqref{eqn: expression for general SP displacement fields}]: 
\begin{align} 
\begin{split} 
    \partial_\alpha{\delta_\SP\bfx} 
& = {E_\alpha}^\beta\,                                        \unitvec{\bsell}_\beta 
  - {H_\alpha}^\beta\,                           {x^\rmp_z}\, \unitvec{\bsell}_\beta 
  - {H_\gamma}^\beta\, u^\gamma\, \partial_\alpha{x^\rmp_z}\, \unitvec{\bsell}_\beta 
    \h 
& \quad\quad+ 
     H_{\alpha  \beta}\, u^\beta \,                                 \unitvec{z} 
  + {H_ \alpha}^\beta \,                           {x^\rmp_\beta}\, \unitvec{z} 
  + {H_ \gamma}^\beta \, u^\gamma\, \partial_\alpha{x^\rmp_\beta}\, \unitvec{z} 
    \h 
& \quad\quad\quad\quad+ 
    \partial_\alpha{\delta_\SP\bfx^\rmp}. 
\end{split} 
\label{eqn: an intermediate step toward computing the derivative of SP displacement fields} 
\end{align} 
Here, as what we did for the matrix $\left(H^{\alpha\beta}\right)$ in \Eqref{eqn: mixed components of the H matrix}, the planar metric tensor [see \Eqref{eqn: defn. of the planar metric components}] is again used to lower the index of the macroscopic strain components: 
\refstepcounter{equation} \label{eqn: mixed and covariant components of the E matrix} 
\begin{align} 
           {E _ \alpha}^\beta 
\equiv \hat{g}_{\alpha  \gamma}\, 
           {E}^{\gamma  \beta }. 
\label{eqn: mixed components of the E matrix} 
\tag{\theequation, a} 
\end{align} 
For future reference, we also define the covariant macroscopic strain components as: 
\begin{align} 
           {E}_{\alpha\beta } 
\equiv \hat{g}_{\alpha\gamma}\, 
       \hat{g}_{\beta \rho  }\, 
           {E}^{\gamma\rho  }. 
\label{eqn: covariant components of the E matrix} 
\tag{\theequation, b} 
\end{align} 
After some rearrangements, \Eqref{eqn: an intermediate step toward computing the derivative of SP displacement fields} becomes: 
\begin{align} 
\begin{split} 
    \partial_\alpha{\delta_\SP\bfx} 
& =  H_{\alpha  \beta}\, u^\beta \, \unitvec{z     } 
    \h 
& \quad\quad- 
    {H_ \gamma}^\beta \, u^\gamma 
    \left( 
    \partial_\alpha{x^\rmp_z    }\, \unitvec{\bsell}_\beta 
  - \partial_\alpha{x^\rmp_\beta}\, \unitvec{z     } 
    \right) 
    \h 
& \quad\quad\quad\quad+ 
    \left( 
    {E_\alpha}^\beta\,                  \unitvec{\bsell}_\beta 
  - {H_\alpha}^\beta\, {x^\rmp_z    }\, \unitvec{\bsell}_\beta 
  + {H_\alpha}^\beta\, {x^\rmp_\beta}\, \unitvec{z     } 
  + \partial_\alpha{\delta_\SP\bfx^\rmp} 
    \right), 
\end{split} 
\label{eqn: expression for the derivative of SP displacement fields, LHS} 
\end{align} 
where the first line of the equation is linear in the local coordinates $u^\alpha$, and the terms in the third line are all doubly periodic. 
\par 
For the right-hand side of \Eqref{eqn: defn. of the angular velocity, revisited}, we substitute the parametrization of doubly periodic surfaces [\Eqref{eqn: parametrization of doubly periodic surfaces}] and the angular velocity fields corresponding to isometric shape-periodic deformations [\Eqref{eqn: parametrization of the SP angular velocity fields}], yielding: 
\begin{align} 
\begin{split} 
         \bsomega_\SP 
  \times \partial_\alpha{\bfx} 
& \equiv                \left(u^\beta \,         {\bfw  }_\beta  + \bsomega_\SP^\rmp\right) 
  \times \partial_\alpha\left(u^\gamma\, \unitvec{\bsell}_\gamma +     \bfx    ^\rmp\right) 
         \h 
&      = \left(u^\beta\,         {\bfw  }_\beta  +                {\bsomega_\SP^\rmp}\right) 
  \times \left(          \unitvec{\bsell}_\alpha + \partial_\alpha{    \bfx    ^\rmp}\right) 
         \h 
&      = u^\beta\, \bfw_\beta \times       \unitvec{\bsell}_\alpha 
       + u^\beta\, \bfw_\beta \times                                 \partial_\alpha{\bfx^\rmp} 
       + \bsomega_\SP^\rmp    \times \left(\unitvec{\bsell}_\alpha + \partial_\alpha{\bfx^\rmp}\right), 
\end{split} 
\label{eqn: expression for the derivative of isometric SP displacement fields, RHS} 
\end{align} 
where the first term on the right-hand side is linear in the coordinates $u^\alpha$, and the last term is doubly periodic. 
\par 
Thus, by equating the terms linear in $u^\alpha$ from \Eqsref{eqn: expression for the derivative of SP displacement fields, LHS} and \eqref{eqn: expression for the derivative of isometric SP displacement fields, RHS}, we obtain the desired relation between the macroscopic angular velocity field and the matrix $\left(H^{\alpha\beta}\right)$: 
\begin{align} 
  H_{\alpha\beta}\, u^\beta\,                   \unitvec{z     } 
=                   u^\beta\, \bfw_\beta \times \unitvec{\bsell}_\alpha, 
\end{align} 
which further implies that: 
\begin{align} 
  H_{\alpha\beta}\,                   \unitvec{z     } 
=                   \bfw_\beta \times \unitvec{\bsell}_\alpha. 
\label{eqn: relation between the macroscopic angular accelerations and the H matrix} 
\end{align} 
\Eqnref{eqn: relation between the macroscopic angular accelerations and the H matrix} is consistent with results established in the literature~\cite{Nassar2024_Number_of_Modes, McInerney2020_Hidden_Symmetries, McInerney2022_Discrete_Symmetries}. 
\par 
From \Eqref{eqn: relation between the macroscopic angular accelerations and the H matrix}, it follows that the macroscopic angular acceleration vectors $\bfw_\alpha$ must lie within the plane spanned by the translation vectors of a doubly periodic surface. 
This result is, in fact, the planar counterpart of the general finding discussed in \Secref{subsec: the isometry-stress duality, isometry}: For any arbitrary smooth surface, the angular acceleration vectors associated with an isometric deformation always lie within the surface's tangent planes [see, e.g., \Eqref{eqn: defn. of the angular acceleration vectors, revised, revisited}]. 
As in \Secref{subsec: the isometry-stress duality, isometry}, we can express the vectors $\bfw_\alpha$ as linear combinations of the unit translation vectors $\unitvec{\bsell}_\alpha$, with the macroscopic angular acceleration components ${W_\alpha}^\beta$ as the corresponding coefficient matrix: 
\begin{align} 
           \bfw_\alpha 
\equiv       {W_\alpha}^\beta\, 
       \unitvec{\bsell}_\beta. 
\label{eqn: defn. of the macroscopic angular acceleration components} 
\end{align} 
\par 
To obtain the exact expression for the macroscopic angular acceleration components ${W_\alpha}^\beta$ in terms of the surface's macroscopic curvature changes $H_{\alpha\beta}$, we substitute \Eqref{eqn: defn. of the macroscopic angular acceleration components} into \Eqref{eqn: relation between the macroscopic angular accelerations and the H matrix}, which yields: 
\begin{align} 
        H_{\alpha  \beta }\,                           \unitvec{z     } 
\equiv {W_ \beta }^\gamma \,                           \unitvec{\bsell}_\gamma \times \unitvec{\bsell}_\alpha 
     = {W_ \beta }^\gamma \, \epsilon_{\gamma\alpha}\, \unitvec{\bsell}_1      \times \unitvec{\bsell}_2, 
\label{eqn: an intermediate step toward deriving the macroscopic angular acceleration components} 
\end{align} 
where $\epsilon_{\gamma\alpha}$ is the Levi-Civita symbol. 
To proceed, we note that the planar unit normal vector $\unitvec{z}$ can be expressed as: 
\begin{align} 
       \unitvec{z} 
\equiv \frac{        {\bsell}_1 \times         {\bsell}_2}{ \norm{\bsell_1 \times \bsell_2} } 
     = \frac{\unitvec{\bsell}_1 \times \unitvec{\bsell}_2}{  \sin{\phi                    } } 
\equiv \frac{\unitvec{\bsell}_1 \times \unitvec{\bsell}_2}{ \sqrt{\hat g                  } }, 
\label{eqn: expression for the planar unit normal vector} 
\end{align} 
where $\phi$ is the angle between the translation vectors of the surface (taken to be between 0 and $\pi$), and $\hat{g}$ denotes the determinant of the planar metric tensor [recall \Eqref{eqn: defn. of the planar metric components}]: 
\begin{align} 
                 \hat{g} 
\equiv \det\left(\hat{g}_{\alpha\beta}\right) 
\equiv \det{ 
       \begin{pmatrix} 
       1          & \cos{\phi} \h 
       \cos{\phi} & 1 
       \end{pmatrix} 
       } 
     = \sin^2{\phi}. 
\end{align} 
By substituting \Eqref{eqn: expression for the planar unit normal vector} into \Eqref{eqn: an intermediate step toward deriving the macroscopic angular acceleration components}, we obtain the following relation between the macroscopic curvature changes $H_{\alpha\beta}$ and the macroscopic angular acceleration components ${W_\alpha}^\beta$: 
\refstepcounter{equation} \label{eqn: relation between the H matrix and W matrix} 
\begin{align} 
                                                    H_{\alpha\beta } 
     = \sqrt{\hat g}\, {W_\beta}^\gamma\,    \epsilon_{\gamma\alpha} 
\equiv                 {W_\beta}^\gamma\, \hat{\calE}_{\gamma\alpha}, 
\label{eqn: expression for the H matrix in terms of the W matrix} 
\tag{\theequation, a} 
\end{align} 
where $\hat{\calE}_{\gamma\alpha} \equiv \sqrt{\hat g}\, \epsilon_{\gamma\alpha}$ denotes the covariant components of the planar area two-form, analogous to the case of curved surfaces [cf.~\Eqref{eqn: expression for area two-form}]. 
Inverting \Eqref{eqn: expression for the H matrix in terms of the W matrix} then yields the desired expression for the macroscopic angular acceleration components ${W_\alpha}^\beta$ in terms of the macroscopic curvature changes $H_{\alpha\beta}$: 
\begin{align} 
                                                           {W_ \beta }^\rho 
     = \frac{1}{ \sqrt{\hat g} }\,    \epsilon^{\rho\alpha} H_{\alpha  \beta} 
\equiv                             \hat{\calE}^{\rho\alpha} H_{\alpha  \beta}, 
\label{eqn: expression for the W matrix in terms of the H matrix} 
\tag{\theequation, b} 
\end{align} 
where $\hat{\calE}^{\rho\alpha}$ denotes the contravariant components of the planar area two-form, defined as [cf.~\Eqref{eqn: contravariant Levi-Civita tensor}]: 
\begin{align} 
                                   \hat{\calE}^{\rho  \alpha} 
\equiv                             \hat{g    }^{\rho  \beta }\, 
                                   \hat{g    }^{\alpha\gamma}\, 
                                   \hat{\calE}_{\beta \gamma} 
     = \frac{1}{ \sqrt{\hat g} }\,    \epsilon^{\rho  \alpha}. 
\label{eqn: contravariant planar Levi-Civita tensor} 
\end{align} 
Here, $\hat{g}^{\rho\beta}$ represents the components of the inverse planar metric tensor, as defined in \Eqref{eqn: defn. of the inverse planar metric components}. 
Finally, we can substitute \Eqref{eqn: expression for the W matrix in terms of the H matrix} into \Eqref{eqn: defn. of the macroscopic angular acceleration components} to obtain the expression for the macroscopic angular acceleration vectors $\bfw_\alpha$ in terms of the macroscopic curvature changes $H_{\alpha\beta}$: 
\begin{align} 
                                                           {\bfw  }_\alpha 
\equiv \hat{\calE}^{\beta\gamma}H_{\gamma\alpha}\, \unitvec{\bsell}_\beta. 
\label{eqn: expression for the macroscopic angular acceleration vectors} 
\end{align} 
\par 
To verify that \Eqref{eqn: expression for the W matrix in terms of the H matrix} is consistent with our framework, we can substitute \Eqref{eqn: expression for the macroscopic angular acceleration vectors} into the second term on the right-hand side of \Eqref{eqn: expression for the derivative of isometric SP displacement fields, RHS}, yielding: 
\begin{align} 
                                               u^\beta\,         {\bfw  }_\beta  \times \partial_\alpha{\bfx^\rmp} 
\equiv \hat{\calE}^{\gamma\rho}H_{\rho\beta}\, u^\beta\, \unitvec{\bsell}_\gamma \times \partial_\alpha{\bfx^\rmp}, 
\label{eqn: an intermediate step toward showing the compatibility of the macroscopic angular acceleration components, 0} 
\end{align} 
which we then compare to the second line of \Eqref{eqn: expression for the derivative of SP displacement fields, LHS}. 
To facilitate the comparison, we decompose the doubly periodic function $\bfx^\rmp$, which characterizes the unit cell geometry, in the planar frame 
$
\left\{ 
\unitvec{\bsell}_1, 
\unitvec{\bsell}_2, 
\unitvec{z     } 
\right\} 
$ as follows: 
\begin{align} 
                            \bfx^\rmp 
\equiv \hat{g}^{\alpha\beta}   x^\rmp_\alpha\, \unitvec{\bsell}_\beta 
     +                         x^\rmp_z     \, \unitvec{z     }. 
\label{eqn: "decomposing" the geometry of unit cells} 
\end{align} 
Substituting \Eqref{eqn: "decomposing" the geometry of unit cells} into \Eqref{eqn: an intermediate step toward showing the compatibility of the macroscopic angular acceleration components, 0} gives: 
\begin{align} 
                                                          u^\beta\,                                       {\bfw  }_\beta  \times \partial_\alpha{\bfx^\rmp} 
= \hat{g}^{\mu\nu}\hat{\calE}^{\gamma\rho}H_{\rho\beta}\, u^\beta\, \partial_\alpha{x^\rmp_\mu}\, \unitvec{\bsell}_\gamma \times \unitvec{\bsell}_\nu 
+                 \hat{\calE}^{\gamma\rho}H_{\rho\beta}\, u^\beta\, \partial_\alpha{x^\rmp_z  }\, \unitvec{\bsell}_\gamma \times \unitvec{z     }. 
\label{eqn: an intermediate step toward showing the compatibility of the macroscopic angular acceleration components, 1} 
\end{align} 
\par 
To proceed, we need the following identities regarding the planar frame, with their derivation provided symbolically in detail: 
\begin{align} 
                                                 \unitvec{\bsell}_\gamma \times \unitvec{\bsell}_\nu 
     =                    \epsilon_{\gamma\nu}\, \unitvec{\bsell}_1      \times \unitvec{\bsell}_2 
     =  \sin{\phi  }\,    \epsilon_{\gamma\nu}\, \unitvec{z     } 
\equiv \sqrt{\hat g}\,    \epsilon_{\gamma\nu}\, \unitvec{z     } 
\equiv                 \hat{\calE}_{\gamma\nu}\, \unitvec{z     }, 
\label{eqn: the first identity regarding the planar frame} 
\end{align} 

\begin{align} 
\begin{split} 
                                   \unitvec{\bsell}_\gamma \times       \unitvec{z     } 
  = \frac{1}{ \sqrt{\hat g} }\,    \unitvec{\bsell}_\gamma \times \left(\unitvec{\bsell}_1 \times \unitvec{\bsell}_2\right) 
& = \frac{1}{ \sqrt{\hat g} }\left( 
    \inProd{\unitvec{\bsell}_\gamma}{\unitvec{\bsell}_2}\unitvec{\bsell}_1 
  - \inProd{\unitvec{\bsell}_\gamma}{\unitvec{\bsell}_1}\unitvec{\bsell}_2 
    \right) 
    \h 
& = \frac{1}{ \sqrt{\hat g} }\left(                       \hat{g}_{\gamma2  }\, \unitvec{\bsell}_1 
  -                                                       \hat{g}_{\gamma1  }\, \unitvec{\bsell}_2  \right) 
  = \frac{1}{ \sqrt{\hat g} }\,       \epsilon^{\mu\nu}\, \hat{g}_{\gamma\nu}\, \unitvec{\bsell}_\mu 
    \h 
& \equiv 
                                   \hat{\calE}^{\mu\nu}\, \hat{g}_{\gamma\nu}\, \unitvec{\bsell}_\mu, 
\end{split} 
\label{eqn: the second identity regarding the planar frame} 
\end{align} 
where the $\mathrm{BAC - CAB}$ rule is used for deriving the second identity. 
With these identities, \Eqref{eqn: an intermediate step toward showing the compatibility of the macroscopic angular acceleration components, 1} becomes: 
\begin{align} 
  u^\beta\, \bfw_\beta \times \partial_\alpha{\bfx^\rmp} 
= \hat{g}^{\mu   \nu}\, \hat{\calE}^{\gamma\rho}\, \hat{\calE}_{\gamma\nu}\, H_{\rho\beta}\, u^\beta\, \partial_\alpha{x^\rmp_\mu}\, \unitvec{z     } 
+ \hat{g}_{\gamma\nu}\, \hat{\calE}^{\gamma\rho}\, \hat{\calE}^{\mu   \nu}\, H_{\rho\beta}\, u^\beta\, \partial_\alpha{x^\rmp_z  }\, \unitvec{\bsell}_\mu. 
\label{eqn: an intermediate step toward showing the compatibility of the macroscopic angular acceleration components, 2} 
\end{align} 
\par 
To further simplify \Eqref{eqn: an intermediate step toward showing the compatibility of the macroscopic angular acceleration components, 2}, we use the following tensor relations: 
\begin{align} 
  \hat{\calE}^{\gamma\rho}\, \hat{\calE}_{\gamma\nu} 
=    \epsilon^{\gamma\rho}\,    \epsilon_{\gamma\nu} 
=      \delta^       \rho               _       \nu, 
\label{eqn: convenient tensor relation, 1} 
\end{align} 

\begin{align} 
                           \hat{\calE}^{\gamma\rho}\, \hat{\calE}^{\mu\nu}\, \hat{g}_{\gamma\nu} 
     = -\frac{1}{\hat g}\,    \epsilon^{\gamma\rho}\,    \epsilon^{\nu\mu}\, \hat{g}_{\gamma\nu} 
\equiv -                                                                     \hat{g}^{\rho  \mu}, 
\label{eqn: convenient tensor relation, 2} 
\end{align} 
where we recognize that the second relation is simply the definition of the matrix inverse of $-\hat{g}_{\gamma\nu}$, expressed in index notation. 
Thus, by substituting 
\Eqsref{eqn: convenient tensor relation, 1} and 
 \eqref{eqn: convenient tensor relation, 2} into \Eqref{eqn: an intermediate step toward showing the compatibility of the macroscopic angular acceleration components, 2}, we can finally see that: 
\begin{align} 
\begin{split} 
    u^\beta\, \bfw_\beta \times \partial_\alpha{\bfx^\rmp} 
& = \hat{g}^{\mu \rho}\, H_{\rho\beta}\, u^\beta\, \partial_\alpha{x^\rmp_\mu}\, \unitvec{z     } 
  - \hat{g}^{\rho\mu }\, H_{\rho\beta}\, u^\beta\, \partial_\alpha{x^\rmp_z  }\, \unitvec{\bsell}_\mu 
    \h 
& = -{H_\beta}^\mu\, u^\beta\left( 
    \partial_\alpha{x^\rmp_z  }\, \unitvec{\bsell}_\mu 
  - \partial_\alpha{x^\rmp_\mu}\, \unitvec{z     } 
    \right), 
\end{split} 
\end{align} 
which matches the second line of \Eqref{eqn: expression for the derivative of SP displacement fields, LHS} exactly. 
\par 
For future reference, the doubly periodic terms in \Eqsref{eqn: expression for the derivative of SP displacement fields, LHS} and \eqref{eqn: expression for the derivative of isometric SP displacement fields, RHS} can also be equated, resulting in the following equality: 
\begin{align} 
  \bsomega_\SP^\rmp \times \partial_\alpha{          \bfx     } 
= {E_\alpha}^\beta\,                  \unitvec{\bsell}_\beta 
- {H_\alpha}^\beta\, {x^\rmp_z    }\, \unitvec{\bsell}_\beta 
+ {H_\alpha}^\beta\, {x^\rmp_\beta}\, \unitvec{z     } 
+                          \partial_\alpha{\delta_\SP\bfx^\rmp}. 
\label{eqn: periodic component of the derivative of SP displacement fields} 
\end{align} 
\subparagraph{Remark.} For general curved surfaces, a relation analogous to \Eqref{eqn: expression for the W matrix in terms of the H matrix} exists between the angular acceleration components ${a_\alpha}^\beta$ associated with an isometric deformation and the corresponding changes in the components of the second fundamental form $\delta{b}_{\alpha\beta}$: 
\begin{align} 
                             {a _ \beta }^\rho 
= \calE^{\rho\alpha}\, \delta{b}_{\alpha  \beta}. 
\label{eqn: expression for the a matrix in terms of the db matrix} 
\end{align} 
A derivation of \Eqref{eqn: expression for the a matrix in terms of the db matrix} using the isometry-stress duality can be found in \Appref{appendix: rewriting the equilibrium equations}. 
\subsection{Geometry of the isometric subspace} \label{subsec: isometric SP deformations, geometry of the isometric subspace} 
\subsubsection{Three useful integral identities} \label{subsubsec: isometric SP deformations, geometry of the isometric subspace, integral identities} 
In this subsubsection, we prove three simple integral identities that play a crucial role in uncovering the geometry of the isometric subspace. 
\paragraph{Integral identity 1.} Let $D$ be a simply connected region on an arbitrary surface, and let $\partial{D}$ denote its boundary, which is a simple closed curve. 
For any two smooth vector fields, 
$\bfa\left(u^1, u^2\right)$ and 
$\bfb\left(u^1, u^2\right)$, the following formula, which resembles integration by parts, holds: 
\begin{align} 
   \oint_{\partial D} \df{u^\alpha}\inProd{ \partial_\alpha{\bfa} }{                {\bfb} } 
= -\oint_{\partial D} \df{u^\alpha}\inProd{                {\bfa} }{ \partial_\alpha{\bfb} }. 
\label{eqn: identity 1, "line integration by parts"} 
\end{align} 

\begin{proof} 
By Stokes' theorem [\Eqref{eqn: Stokes' theorem}], the line integral of the exact one-form $\df{ \inProd{\bfa}{\bfb} }$ over $\partial{D}$ vanishes [see \Eqref{eqn: the final form of the orientation compatibility condition} and the associated footnote]: 
\begin{align} 
  \oint_{\partial D} \df   { \inProd{\bfa}{\bfb} } 
=  \int_{         D} \rmd^2{ \inProd{\bfa}{\bfb} } 
= 0. 
\label{eqn: an intermediate step toward deriving the identity 1, 0} 
\end{align} 
Applying the exterior derivative and the product rule then yields the desired identity: 
\begin{align} 
\begin{split} 
    0 
  = \oint_{\partial D} \df{                             \inProd{                {\bfa} }{                {\bfb} } } 
& = \oint_{\partial D} \df{u^\alpha}\, \partial_\alpha{ \inProd{                {\bfa} }{                {\bfb} } } 
    \h 
& = \oint_{\partial D} \df{u^\alpha}\,                { \inProd{ \partial_\alpha{\bfa} }{                {\bfb} } } 
  + \oint_{\partial D} \df{u^\alpha}\,                { \inProd{                {\bfa} }{ \partial_\alpha{\bfb} } }. 
\end{split} 
\end{align} 
\end{proof} 
\paragraph{Integral identity 2.} Let $\rmC$ be a unit cell of a doubly periodic surface, and let $\partial{\rmC}$ denote its boundary. 
The projected area of $\rmC$ onto the plane spanned by the surface's translation vectors $\bsell_1$ and $\bsell_2$ is known to be $\norm{\bsell_1 \times \bsell_2}$, which can be related to the following line integral: 
\begin{align} 
\oint_{\partial\rmC} u^\alpha\, \df{u^\beta} = \norm{\bsell_1 \times \bsell_2}\, \hat{\calE}^{\alpha\beta}, 
\label{eqn: identity 2, area of unit cells} 
\end{align} 
where $\hat{\calE}^{\alpha\beta}$ denotes the contravariant components of the planar area two-form [\Eqref{eqn: contravariant planar Levi-Civita tensor}]. 

\begin{proof} 
Again, by Stokes' theorem [\Eqref{eqn: Stokes' theorem}], we have: 
\begin{align} 
\begin{split} 
    \oint_{\partial\rmC}             {u^\alpha} \,     \df{u^\beta} 
& =  \int_{        \rmC} \df{ \left( {u^\alpha} \,     \df{u^\beta} \right) } 
  =  \int_{        \rmC} \df         {u^\alpha} \wedge \df{u^\beta} 
    \h 
& = \epsilon^{\alpha\beta}   \int_\rmC                     \df{   u^1} \wedge \df{   u^2} 
  \equiv 
    \epsilon^{\alpha\beta}   \int_{u^1_0}^{u^1_0 + \ell^1} 
                             \int_{u^2_0}^{u^2_0 + \ell^2} \df{   u^1} \,     \df{   u^2} 
  = \epsilon^{\alpha\beta}\,                                  {\ell^1} \,        {\ell^2} 
    \h 
& = \frac{1}{  \sin{\phi  } }\, \epsilon^{\alpha\beta}\,      {  \ell^1 \,       \ell^2}\, \sin{\phi} 
  \equiv 
    \frac{1}{ \sqrt{\hat g} }\, \epsilon^{\alpha\beta}\, \norm{\bsell_1 \times \bsell_2} 
    \h 
& \equiv 
    \norm{\bsell_1 \times \bsell_2}\, \hat{\calE}^{\alpha\beta}, 
\end{split} 
\end{align} 
where the coordinate tuple $\left(u^1_0, u^2_0\right)$ corresponds to the bottom-left corner of the unit cell $\rmC$. 
\end{proof} 
\paragraph{Integral identity 3.} The line integral of any one-form with doubly periodic components over the boundary of a unit cell vanishes identically: 
\begin{align} 
\oint_{\partial\rmC} f^\rmp_\alpha\, \df{u^\alpha} = 0, 
\label{eqn: identity 3, line integral of periodic one-forms} 
\end{align} 
where the doubly periodic functions $f^\rmp_\alpha\left(u^1, u^2\right)$ are the corresponding one-form components, satisfying the following periodic conditions: 
\refstepcounter{equation} \label{eqn: function periodic conditions, revisited} 
\begin{align} 
f^\rmp_\alpha\left(u^1, u^2\right) & = f^\rmp_\alpha\left(u^1 + \ell^1, u^2         \right), 
                                       \label{eqn: function periodic condition 1, revisited} 
                                       \tag{\theequation, a} 
                                       \h 
f^\rmp_\alpha\left(u^1, u^2\right) & = f^\rmp_\alpha\left(u^1         , u^2 + \ell^2\right). 
                                       \label{eqn: function periodic condition 2, revisited} 
                                       \tag{\theequation, b} 
\end{align} 

\begin{proof} 
To demonstrate this identity, we decompose the boundary of the unit cell, $\partial{\rmC}$, into segments corresponding to moving rightward cross the bottom edge of the cell, upward across the right edge, etc. We then express the line integral in question as the sum of the integrals over these segments: 
\begin{align} 
       \oint_{\partial\rmC            } f^\rmp_\alpha\, \df{u^\alpha} 
\equiv  \int_{\partial\rmC_\rightarrow} f^\rmp_\alpha\, \df{u^\alpha} 
     +  \int_{\partial\rmC_   \uparrow} f^\rmp_\alpha\, \df{u^\alpha} 
     +  \int_{\partial\rmC_ \leftarrow} f^\rmp_\alpha\, \df{u^\alpha} 
     +  \int_{\partial\rmC_ \downarrow} f^\rmp_\alpha\, \df{u^\alpha}, 
\label{eqn: an intermediate step toward deriving the identity 3, 0} 
\end{align} 
where the segments correspond to the following sets of coordinate tuples (the bottom, right, top and left edges of the unit cell, respectively): 
\refstepcounter{equation} \label{eqn: "parametrization" of the unit cell boundary} 
\begin{align} 
\partial\rmC_\rightarrow: & \quad \left\{\left.\left(u^1_0 + t\, \ell^1, u^2_0             \right)\right|t \in ( 0, 1]\right\}, 
                                  \tag{\theequation, a} 
                                  \h 
\partial\rmC_   \uparrow: & \quad \left\{\left.\left(u^1_0 +     \ell^1, u^2_0 + t\, \ell^2\right)\right|t \in ( 0, 1]\right\}, 
                                  \tag{\theequation, b} 
                                  \h 
\partial\rmC_ \leftarrow: & \quad \left\{\left.\left(u^1_0 - s\, \ell^1, u^2_0 +     \ell^2\right)\right|s \in (-1, 0]\right\}, 
                                  \tag{\theequation, c} 
                                  \h 
\partial\rmC_ \downarrow: & \quad \left\{\left.\left(u^1_0             , u^2_0 - s\, \ell^2\right)\right|s \in (-1, 0]\right\}. 
                                  \tag{\theequation, d} 
\end{align} 
Using \Eqsref{eqn: "parametrization" of the unit cell boundary}, the integrals on the right-hand side of \Eqref{eqn: an intermediate step toward deriving the identity 3, 0} can be evaluated as follows: 
\refstepcounter{equation} \label{eqn: an intermediate step toward deriving the identity 3, 1} 
\begin{align} 
          \int_{\partial\rmC_\rightarrow} f^\rmp_\alpha                                      \, \df{u^\alpha} 
& =       \int_{\partial\rmC_\rightarrow} f^\rmp_1                                           \, \df{u^1     } 
  = \ell^1\int_0^1                        f^\rmp_1     \left(u^1_0 + t\, \ell^1, u^2_0\right)\, \df{t       }, 
    \label{eqn: an intermediate step toward deriving the identity 3, 1, right} 
    \tag{\theequation, a} 
    \h 
          \int_{\partial\rmC_\uparrow} f^\rmp_\alpha                                               \, \df{u^\alpha} 
& =       \int_{\partial\rmC_\uparrow} f^\rmp_2                                                    \, \df{u^2     } 
  = \ell^2\int_0^1                     f^\rmp_2     \left(u^1_0 + \ell^1, u^2_0 + t\, \ell^2\right)\, \df{t       }, 
    \label{eqn: an intermediate step toward deriving the identity 3, 1, up} 
    \tag{\theequation, b} 
    \h 
           \int_{\partial\rmC_\leftarrow} f^\rmp_\alpha                                               \, \df{u^\alpha} 
& =        \int_{\partial\rmC_\leftarrow} f^\rmp_1                                                    \, \df{u^1     } 
  = -\ell^1\int_{-1}^0                    f^\rmp_1     \left(u^1_0 - s\, \ell^1, u^2_0 + \ell^2\right)\, \df{s       }, 
    \label{eqn: an intermediate step toward deriving the identity 3, 1, left} 
    \tag{\theequation, c} 
    \h 
           \int_{\partial\rmC_\downarrow} f^\rmp_\alpha                                      \, \df{u^\alpha} 
& =        \int_{\partial\rmC_\downarrow} f^\rmp_2                                           \, \df{u^2     } 
  = -\ell^2\int_{-1}^0                    f^\rmp_2     \left(u^1_0, u^2_0 - s\, \ell^2\right)\, \df{s       }. 
    \label{eqn: an intermediate step toward deriving the identity 3, 1, down} 
    \tag{\theequation, d} 
\end{align} 
\par 
To proceed, we utilize the periodicity of the one-form components $f^\rmp_\alpha$. 
In particular, from \Eqref{eqn: function periodic condition 1, revisited}, it follows that the line integrals over the horizontal segments cancel each other: 
\refstepcounter{equation} \label{eqn: an intermediate step toward deriving the identity 3, 2} 
\begin{align} 
\begin{split} 
           \int_{\partial\rmC_ \leftarrow} f^\rmp_\alpha                                       \, \df{u^\alpha} 
& = -\ell^1\int_{-1}^0                     f^\rmp_1     \left(u^1_0 - s \, \ell^1, u^2_0\right)\, \df{s       } 
    \h 
& = -\ell^1\int_{ 0}^1                     f^\rmp_1     \left(u^1_0 + s'\, \ell^1, u^2_0\right)\, \df{s'      } 
  = -      \int_{\partial\rmC_\rightarrow} f^\rmp_\alpha                                       \, \df{u^\alpha}, 
\end{split} 
\label{eqn: an intermediate step toward deriving the identity 3, 2, horizontal} 
\tag{\theequation, a} 
\end{align} 
where the first line follows from the periodicity, and the second line is obtained by performing the change of variables $s \to s' \equiv -s$. 
Similarly, by applying the other periodic condition, \Eqref{eqn: function periodic condition 2, revisited}, we can show that the remaining pair of line integrals also cancel each other: 
\begin{align} 
\begin{split} 
           \int_{\partial\rmC_\downarrow} f^\rmp_\alpha                                                \, \df{u^\alpha} 
& = -\ell^2\int_{-1}^0                    f^\rmp_2     \left(u^1_0 + \ell^1, u^2_0 - s \, \ell^2\right)\, \df{s       } 
    \h 
& = -\ell^2\int_{ 0}^1                    f^\rmp_2     \left(u^1_0 + \ell^1, u^2_0 + s'\, \ell^2\right)\, \df{s'      } 
  = -      \int_{\partial\rmC_  \uparrow} f^\rmp_\alpha                                                \, \df{u^\alpha}. 
\end{split} 
\label{eqn: an intermediate step toward deriving the identity 3, 2, vertical} 
\tag{\theequation, b} 
\end{align} 
\par 
Thus, adding 
\Eqsref{eqn: an intermediate step toward deriving the identity 3, 2, horizontal} and 
 \eqref{eqn: an intermediate step toward deriving the identity 3, 2, vertical} yields the desired identity. 
\end{proof} 
\subsubsection{Coupling of isometric modes} \label{subsubsec: isometric SP deformations, geometry of the isometric subspace, coupling of isometries} 
Finally, with all the groundwork laid, we are now ready to reveal the geometry of the isometric subspace. 
To accomplish this, we examine the interaction of two distinct isometric modes associated with a given doubly periodic surface $\bfx$. 
\par 
Let $\delta_a{\bfx}\left(u^1, u^2\right)$ represent the displacement field corresponding to an isometric shape-periodic deformation. 
As discussed in \Secref{subsec: isometric SP deformations, characterization}, the isometric shape-periodic deformation can be characterized either by the associated angular velocity field $\bsomega_a\left(u^1, u^2\right)$ or by the macroscopic deformation $\left(E^{\alpha\beta}_a, H^{\alpha\beta}_a\right)$, together with the doubly periodic vector-valued function $\delta_a{\bfx^\rmp}\left(u^1, u^2\right)$. 
In a similar manner, we use the subscript ``$b$'' to denote the second isometric mode and its associated quantities, e.g., $\delta_b{\bfx}\left(u^1, u^2\right)$ for the corresponding isometric shape-periodic displacement field and $\bsomega_b\left(u^1, u^2\right)$ for the angular velocity field characterizing the deformation. 
\par 
Our starting point is the following surface integral over the unit cell $\rmC$: 
\begin{align} 
  \int_\rmC \inProd{\bsomega_b}{ 
  \partial_\mu{ 
  \partial_\nu{\delta_a\bfx} 
  } }\df{u^\mu} \wedge 
     \df{u^\nu} 
= 0. 
\label{eqn: the "blessed" surface integral} 
\end{align} 
The surface integral trivially vanishes due to symmetry: The partial derivatives commute, while the wedge product is skew-commutative. However, by manipulating this expression algebraically, we obtain a nontrivial geometric relationship that follows essentially from stating that the displacement vectors come from compatible surface rotations.

Recalling the general definition of angular velocity field [see, e.g., \Eqref{eqn: defn. of the angular velocity, revisited}] for $\bsomega_a$, we rewrite \Eqref{eqn: the "blessed" surface integral} as follows: 
\begin{align} 
\begin{split} 
         0 
& \equiv \int_\rmC \inProd{\bsomega_b}{ \partial_\mu\left( {\bsomega_a} \times             { \partial_\nu{\bfx} } \right) }\df{u^\mu} \wedge \df{u^\nu} 
         \h 
&      = \int_\rmC \inProd{\bsomega_b}{ \partial_\mu       {\bsomega_a} \times             { \partial_\nu{\bfx} }         }\df{u^\mu} \wedge \df{u^\nu} 
       + \int_\rmC \inProd{\bsomega_b}{                    {\bsomega_a} \times \partial_\mu{ \partial_\nu{\bfx} }         }\df{u^\mu} \wedge \df{u^\nu} 
         \h 
&      = \int_\rmC \inProd{\bsomega_b}{ \partial_\mu       {\bsomega_a} \times             { \partial_\nu{\bfx} }         }\df{u^\mu} \wedge \df{u^\nu}, 
\end{split} 
\end{align} 
where the product rule is applied in the second line, and the second surface integral on the right-hand side, containing the term $\partial_\mu{ \partial_\nu{\bfx} }$, again vanishes due to symmetry. 
Using the cyclic property of the scalar triple product, we further obtain: 
\begin{align} 
       0 
     = -\int_\rmC \inProd{ \partial_\mu{\bsomega_a} }{ \bsomega_b \times \partial_\nu{        \bfx} }\df{u^\mu} \wedge \df{u^\nu} 
= -\int_\rmC \inProd{ \partial_\mu{\bsomega_a} }{                   \partial_\nu{\delta_b\bfx} }\df{u^\mu} \wedge \df{u^\nu}, 
\label{eqn: an intermediate step toward rewriting the "blessed" surface integral, 0} 
\end{align} 
where the second equality follows from the definition of the angular velocity field $\bsomega_b$. 
\par 
To proceed, we observe that the integrand of the surface integral in \Eqref{eqn: an intermediate step toward rewriting the "blessed" surface integral, 0} can be expressed using the product rule as: 
\begin{align} 
              { \inProd{             { \partial_\mu{\bsomega_a} } }{ \partial_\nu{\delta_b\bfx} } } 
= \partial_\nu{ \inProd{             { \partial_\mu{\bsomega_a} } }{             {\delta_b\bfx} } } 
-             { \inProd{ \partial_\nu{ \partial_\mu{\bsomega_a} } }{             {\delta_b\bfx} } }. 
\label{eqn: the integrand of the "blessed" surface integral} 
\end{align} 
Substituting \Eqref{eqn: the integrand of the "blessed" surface integral} into \Eqref{eqn: an intermediate step toward rewriting the "blessed" surface integral, 0} yields: 
\begin{align} 
\begin{split} 
    0 
& = -\int_\rmC \partial_\nu{ \inProd{             { \partial_\mu{\bsomega_a} } }{\delta_b\bfx} }\, \df{u^\mu} \wedge \df{u^\nu} 
  +  \int_\rmC             { \inProd{ \partial_\nu{ \partial_\mu{\bsomega_a} } }{\delta_b\bfx} }\, \df{u^\mu} \wedge \df{u^\nu} 
    \h 
& =  \int_\rmC \partial_\nu{ \inProd{             { \partial_\mu{\bsomega_a} } }{\delta_b\bfx} }\, \df{u^\nu} \wedge \df{u^\mu}, 
\end{split} 
\label{eqn: an intermediate step toward rewriting the "blessed" surface integral, 1} 
\end{align} 
where the second surface integral in the first line vanishes due to symmetry, and the order of $\df{u^\mu}$ and $\df{u^\nu}$ is swapped in the second line, introducing an additional negative sign. 
Finally, by applying Stokes' theorem [\Eqref{eqn: Stokes' theorem}] to \Eqref{eqn: an intermediate step toward rewriting the "blessed" surface integral, 1}, we arrive at the following line integral, which demonstrates the key relationship between the two isometric modes: 
\begin{align} 
       0 
     =  \int_{        \rmC} \df{        { \inProd{ \partial_\mu{\bsomega_a} }{\delta_b\bfx} } \wedge \df{u^\mu}         } 
     =  \int_{        \rmC} \df{ \left( { \inProd{ \partial_\mu{\bsomega_a} }{\delta_b\bfx} } \,     \df{u^\mu} \right) } 
     = \oint_{\partial\rmC}             { \inProd{ \partial_\mu{\bsomega_a} }{\delta_b\bfx} } \,     \df{u^\mu} 
\equiv \calI_{a b}. 
\label{eqn: the "blessed" line integral} 
\end{align} 
\par 
Although \Eqref{eqn: the "blessed" line integral} is derived purely from mathematical considerations, it can be physically interpreted as indicating that an imposed isometric deformation on a \emph{prestressed} surface incurs no energy cost. 
To comprehend this interpretation, we recall the expression for the energy cost associated with an imposed energy-minimizing deformation, which takes a form analogous to \Eqref{eqn: the "blessed" line integral} [referencing \Eqref{eqn: expression for minimized membrane energy, mapped isometry version}]: 
\begin{align} 
  E_\rmm^\eq = W 
= \frac{1}{2}\, f\oint_{\partial S} \df{u^\gamma}\, \inProd{ \partial_\gamma{\wt\bsomega} }{ \delta{\bfx} }. 
\label{eqn: expression for minimized membrane energy, mapped isometry version, revisited} 
\end{align} 
Here, $\delta{\bfx}$ represents the displacement field corresponding to the imposed energy-minimizing deformation, and $\wt\bsomega$ denotes the angular velocity field that characterizes the isometric deformation mapped from the equilibrium stress induced by the imposed deformation. 
Thus, by drawing an analogy to \Eqref{eqn: expression for minimized membrane energy, mapped isometry version, revisited}, we can interpret the isometric mode ``$b$'', represented by the displacement field $\delta_b{\bfx}$ in \Eqref{eqn: the "blessed" line integral}, as an imposed isometric deformation. 
Also, despite the fact that imposed isometric deformations are mapped to the zero deformation through the isometry-stress duality, if the surface in question is prestressed (i.e., already under stress before the isometric mode ``$b$'' is imposed), the angular acceleration vectors $\partial_\mu{\bsomega_a}$ in \Eqref{eqn: the "blessed" line integral} can be viewed as being mapped from the equilibrium prestress. 
\par 
We have thus shown that \Eqref{eqn: the "blessed" line integral} can be understood as a special case of \Eqref{eqn: expression for minimized membrane energy, mapped isometry version, revisited}, in which the isometric deformation of even a prestressed surface does no work. 
Moving forward, to uncover the geometry of the isometric subspace, we will evaluate the line integral in \Eqref{eqn: the "blessed" line integral}, $\calI_{a b}$, in terms of the macroscopic deformations corresponding to the two isometric modes. 
\Eqnref{eqn: expression for minimized membrane energy, mapped isometry version, revisited} will serve as our starting point in the next section, where we derive the relationship between rigidity and flexibility of doubly periodic surfaces. 
\par 
To evaluate the line integral $\calI_{a b}$, we first observe that the angular acceleration vectors $\partial_\mu{\bsomega_a}$ are doubly periodic: 
\begin{align} 
       \partial_\mu                                             {\bsomega_a     } 
\equiv \partial_\mu\left(u^\alpha\, \bfw^a_\alpha +             {\bsomega_a^\rmp}\right) 
     =                              \bfw^a_\mu    + \partial_\mu{\bsomega_a^\rmp} 
\equiv \hat{\calE}^{\alpha\beta}  \, 
           {H    }_{\beta \mu  }^a\, \unitvec{\bsell}_\alpha 
     +                                              \partial_\mu{\bsomega_a^\rmp}, 
\label{eqn: expression for the SP angular acceleration vectors} 
\end{align} 
where we have substituted the parametrization of shape-periodic angular velocity fields [\Eqref{eqn: parametrization of the SP angular velocity fields}] and the expression for the macroscopic angular acceleration vectors [\Eqref{eqn: expression for the macroscopic angular acceleration vectors}]. 
Therefore, by the third integral identity [\Eqref{eqn: identity 3, line integral of periodic one-forms}], the part of $\calI_{a b}$ involving the periodic component of $\delta_b{\bfx}$ must vanish. 
\par 
To begin the evaluation, we substitute \Eqref{eqn: expression for the SP angular acceleration vectors} and the general form of the shape-periodic displacement fields [\Eqref{eqn: expression for general SP displacement fields}] for $\delta_b{\bfx}$ into $\calI_{a b}$. 
After discarding the aforementioned part of $\calI_{a b}$ and the terms involving the vanishing inner product $\inProd{\unitvec{\bsell}_\alpha}{\unitvec z}$ in the resulting expression, we obtain: 
\begin{align} 
\begin{split} 
         \calI_{a b} 
& \equiv \calI_{a b}^1 
       + \calI_{a b}^2 
       + \calI_{a b}^3 
         \h 
& \equiv \oint_{\partial\rmC} \df{u^\mu}\, \inProd 
         {\hat{\calE}^{\alpha\beta}\, H_{\beta\mu   }^a\,                     \unitvec{\bsell}_\alpha} 
         {\hat{g    }_{\nu   \rho }\, E^{\nu  \gamma}_b\, u^\rho\,            \unitvec{\bsell}_\gamma} 
       - \oint_{\partial\rmC} \df{u^\mu}\, \inProd 
         {\hat{\calE}^{\alpha\beta}\, H_{\beta\mu   }^a\,                     \unitvec{\bsell}_\alpha} 
         {\hat{g    }_{\nu   \rho }\, H^{\nu  \gamma}_b\, u^\rho\, x^\rmp_z\, \unitvec{\bsell}_\gamma} 
         \h 
& \quad\quad\quad\quad+ 
         \oint_{\partial\rmC} \df{u^\mu}\, \inProd{ \partial_\mu{\bsomega_a^\rmp} }{\delta_b\bfx}. 
\end{split} 
\label{eqn: an intermediate step toward computing I_ab} 
\end{align} 
Next, we compute $\calI_{a b}^1$ and $\calI_{a b}^3$ individually and then combine all three to obtain the final result. 
\paragraph{Computation of \texorpdfstring{$\calI_{a b}^1$}{}.} It is straightforward to compute the line integral $\calI_{a b}^1$. 
We begin by bringing the constant factors outside the integral, as shown below: 
\begin{align} 
\begin{split} 
    \calI_{a b}^1 
& = \hat{g}_{\nu   \rho  }\, \hat{\calE}^{\alpha\beta}\, E^{\nu \gamma}_b\, H_{\beta\mu}^a\, \inProd{\unitvec{\bsell}_\alpha}{\unitvec{\bsell}_\gamma}\, 
                                                                                             \oint_{\partial\rmC} \df{u^\mu}\, u^\rho 
    \h 
& \equiv 
    \hat{g}_{\rho  \nu   }\, 
    \hat{g}_{\alpha\gamma}\, \hat{\calE}^{\alpha\beta}\, E^{\nu \gamma}_b\, H_{\beta\mu}^a\, \oint_{\partial\rmC} \df{u^\mu}\, u^\rho 
  \equiv 
                             \hat{\calE}^{\alpha\beta}\, E_{\rho\alpha}^b\, H_{\beta\mu}^a\, \oint_{\partial\rmC} \df{u^\mu}\, u^\rho, 
\end{split} 
\label{eqn: an intermediate step toward computing I_ab^1} 
\end{align} 
where the definition of the planar metric components $\hat{g}_{\alpha\gamma}$ [see \Eqref{eqn: defn. of the planar metric components}] is applied in the second line, and the planar metric tensors are used to lower the indices of the macroscopic strain components $E^{\nu\gamma}_b$, as in \Eqref{eqn: covariant components of the E matrix}.\footnote{To clarify, the Latin indices $a$ and $b$ are used to differentiate the two isometric modes; their placement (i.e., lower or upper) is therefore inconsequential.} 
The remaining line integral in \Eqref{eqn: an intermediate step toward computing I_ab^1} is directly given by the second integral identity [\Eqref{eqn: identity 2, area of unit cells}]: 
\begin{align} 
\oint_{\partial\rmC} \df{u^\mu}\, u^\rho = \norm{\bsell_1 \times \bsell_2}\, \hat{\calE}^{\rho\mu}. 
\label{eqn: identity 2, revisited} 
\end{align} 
By substituting \Eqref{eqn: identity 2, revisited} into \Eqref{eqn: an intermediate step toward computing I_ab^1}, we obtain the desired result: 
\begin{align} 
  \calI_{a b}^1 
= \norm{\bsell_1 \times \bsell_2}\, 
  \hat{\calE}^{\alpha\beta}\, 
  \hat{\calE}^{\rho  \mu  }\, E_{\alpha\rho}^b\, 
                              H_{\beta \mu }^a. 
\label{eqn: expression for I_ab^1} 
\end{align} 
\paragraph{Computation of \texorpdfstring{$\calI_{a b}^3$}{}.} To compute the line integral $\calI_{a b}^3$, we first integrate by parts using the integral identity \Eqref{eqn: identity 1, "line integration by parts"} and subsequently apply the definition and parametrization of the angular velocity field $\bsomega_b$, as follows: 
\begin{align} 
\begin{split} 
    \calI_{a b}^3 
&      = -\oint_{\partial\rmC} \df{u^\mu}\, \inProd{\bsomega_a^\rmp}{                                     \partial_\mu{\delta_b\bfx} } 
  \equiv -\oint_{\partial\rmC} \df{u^\mu}\, \inProd{\bsomega_a^\rmp}{            \bsomega_b        \times \partial_\mu{        \bfx} } 
         \h 
&      = -\oint_{\partial\rmC} \df{u^\mu}\, \inProd{\bsomega_a^\rmp}{ u^\alpha\,     \bfw^b_\alpha \times \partial_\mu{        \bfx} } 
         -\oint_{\partial\rmC} \df{u^\mu}\, \inProd{\bsomega_a^\rmp}{            \bsomega_b^\rmp   \times \partial_\mu{        \bfx} }. 
\end{split} 
\label{eqn: an intermediate step toward computing I_ab^3, 0} 
\end{align} 
Once again, due to periodicity [\Eqref{eqn: identity 3, line integral of periodic one-forms}], the second line integral on the right-hand side of \Eqref{eqn: an intermediate step toward computing I_ab^3, 0} vanishes, leaving: 
\begin{align} 
       \calI_{a b}^3 
     = \oint_{\partial\rmC} \df{u^\mu}\, \inProd{u^\alpha\,     {\bfw          }^b_      \alpha}{ \bsomega_a^\rmp \times \partial_\mu{\bfx} } 
\equiv \oint_{\partial\rmC} \df{u^\mu}\, \inProd{u^\alpha\, \hat{\calE         }  ^{\nu  \beta }\, 
                                                                {H             }^b_{\beta\alpha}\, 
                                                                {\unitvec\bsell}  _ \nu        }{ \bsomega_a^\rmp \times \partial_\mu{\bfx} }, 
\label{eqn: an intermediate step toward computing I_ab^3, 1} 
\end{align} 
where the cyclic property of the scalar triple product is used to rearrange the resulting line integral, introducing an additional negative sign, and the expression for the macroscopic angular acceleration vectors $\bfw^b_\alpha$ is applied, yielding the second equality. 
\par 
To proceed, we substitute the expression for $\bsomega_a^\rmp \times \partial_\mu{\bfx}$ [see \Eqref{eqn: periodic component of the derivative of SP displacement fields}] into \Eqref{eqn: an intermediate step toward computing I_ab^3, 1}. 
After discarding the term involving the vanishing inner product $\inProd{\unitvec{\bsell}_\nu}{\unitvec z}$, we obtain: 
\begin{align} 
\begin{split} 
         \calI_{a b}^{3   } 
& \equiv \calI_{a b}^{3, 1} 
       + \calI_{a b}^{3, 2} 
       + \calI_{a b}^{3, 3} 
         \h 
& \equiv \oint_{\partial\rmC} \df{u^\mu}\, \inProd 
         {u^\alpha\, \hat{\calE}^{\nu\beta }\, H^b_{\beta \alpha}\,            \unitvec{\bsell}_\nu } 
         {           \hat{g    }_{\mu\gamma}\, E_a^{\gamma\rho  }\,            \unitvec{\bsell}_\rho} 
       - \oint_{\partial\rmC} \df{u^\mu}\, \inProd 
         {u^\alpha\, \hat{\calE}^{\nu\beta }\, H^b_{\beta \alpha}\,            \unitvec{\bsell}_\nu } 
         {           \hat{g    }_{\mu\gamma}\, H_a^{\gamma\rho  }\, x^\rmp_z\, \unitvec{\bsell}_\rho} 
         \h 
& \quad\quad\quad\quad+ 
         \oint_{\partial\rmC} \df{u^\mu}\, \inProd 
         {u^\alpha\, \hat{\calE}^{\nu\beta }\, H^b_{\beta \alpha}\,            \unitvec{\bsell}_\nu }{ \partial_\mu{\delta_a\bfx^\rmp} }. 
\end{split} 
\label{eqn: an intermediate step toward computing I_ab^3, 2} 
\end{align} 
\par 
Among the three newly obtained integrals in \Eqref{eqn: an intermediate step toward computing I_ab^3, 2}, we observe that $\calI_{a b}^{3, 1}$ and $\calI_{a b}^{3, 2}$ resemble $\calI_{a b}^1$ and $\calI_{a b}^2$, respectively [cf.~\Eqref{eqn: an intermediate step toward computing I_ab}]. 
In particular, the integral $\calI_{a b}^{3, 1}$ can be evaluated in exactly the same manner as was done for $\calI_{a b}^1$ [cf.~\Eqsref{eqn: an intermediate step toward computing I_ab^1}--\eqref{eqn: expression for I_ab^1}]: 
\begin{align} 
\begin{split} 
         \calI_{a b}^{3, 1} 
&      = \hat{g}_{\mu\gamma}\, \hat{\calE}^{\nu   \beta }\, E_a^{\gamma\rho}\, H^b_{\beta \alpha}\, \inProd{\unitvec{\bsell}_\nu}{\unitvec{\bsell}_\rho}\, 
         \oint_{\partial\rmC} \df{u^\mu}\, u^\alpha 
         \h 
& \equiv  \norm{\bsell_1 \times \bsell_2}\, 
         \hat{g}_{\mu\gamma}\, 
         \hat{g}_{\nu\rho  }\, \hat{\calE}^{\alpha\mu   }\, 
                               \hat{\calE}^{\nu   \beta }\, E_a^{\gamma\rho}\, H^b_{\beta \alpha} 
         \h 
& \equiv  \norm{\bsell_1 \times \bsell_2}\, 
                               \hat{\calE}^{\alpha\mu   }\, 
                               \hat{\calE}^{\nu   \beta }\, E^a_{\mu   \nu }\, H^b_{\beta \alpha} 
       = -\norm{\bsell_1 \times \bsell_2}\, 
                               \hat{\calE}^{\mu   \alpha}\, 
                               \hat{\calE}^{\nu   \beta }\, E^a_{\mu   \nu }\, H^b_{\alpha\beta }, 
\end{split} 
\label{eqn: expression for I_ab^31} 
\end{align} 
where the indices of $\hat{\calE}^{\alpha\mu}$ are swapped in the final equality, leading to the negative sign. 
\par 
As for the line integral $\calI_{a b}^{3, 3}$, it can be shown to vanish due to periodicity [\Eqref{eqn: identity 3, line integral of periodic one-forms}] after integration by parts [\Eqref{eqn: identity 1, "line integration by parts"}]: 
\begin{align} 
  \calI_{a b}^{3, 3} 
=  \hat{\calE}^{\nu\beta}\, H^b_{\beta\alpha}\, \oint_{\partial\rmC} \df{u^\mu   }\, \inProd{u^\alpha\, \unitvec{\bsell}_\nu}{ \partial_\mu{\delta_a\bfx^\rmp} } 
= -\hat{\calE}^{\nu\beta}\, H^b_{\beta\alpha}\, \oint_{\partial\rmC} \df{u^\alpha}\, \inProd{           \unitvec{\bsell}_\nu}{             {\delta_a\bfx^\rmp} } 
= 0. 
\label{eqn: expression for I_ab^33} 
\end{align} 
\par 
Thus, by substituting \Eqsref{eqn: expression for I_ab^31} and \eqref{eqn: expression for I_ab^33} into \Eqref{eqn: an intermediate step toward computing I_ab^3, 2}, the final expression for $\calI_{a b}^3$ is obtained: 
\begin{align} 
\begin{split} 
    \calI_{a b}^3 
& = -\norm{\bsell_1 \times \bsell_2}\,                    \hat{\calE}^{\mu\alpha}\, 
                                                          \hat{\calE}^{\nu\beta }\, E^a_{\mu   \nu   }\, 
                                                                                    H^b_{\alpha\beta } 
    -\oint_{\partial\rmC} \df{u^\mu}\, \inProd{u^\alpha\, \hat{\calE}^{\nu\beta }\, H^b_{\beta \alpha}\,            \unitvec{\bsell}_\nu } 
                                              {           \hat{g    }_{\mu\gamma}\, H_a^{\gamma\rho  }\, x^\rmp_z\, \unitvec{\bsell}_\rho} 
    \h 
& = -\norm{\bsell_1 \times \bsell_2}\, 
    \hat{\calE}^{\mu\alpha}\, 
    \hat{\calE}^{\nu\beta }\, E^a_{\mu   \nu  }\, 
                              H^b_{\alpha\beta} 
  - \hat{\calE}^{\nu\beta }\, H^a_{\mu   \nu  }\, 
                              H^b_{\alpha\beta}\, 
     \oint_{\partial\rmC} \df{u^\mu   }\, 
                             {u^\alpha}\, x^\rmp_z, 
\end{split} 
\label{eqn: expression for I_ab^3} 
\end{align} 
where the expression for the line integral $\calI_{a b}^{3, 2}$ is simplified in the second line, following the previously discussed steps. 
\paragraph{The final expression for \texorpdfstring{$\calI_{a b}$}{}.} In fact, the line integral $\calI_{a b}^{3, 2}$ is precisely the negative of the line integral $\calI_{a b}^2$, which is defined in \Eqref{eqn: an intermediate step toward computing I_ab}. 
To see this relationship, we simplify the expression for $\calI_{a b}^2$ as in \Eqref{eqn: expression for I_ab^3}, yielding: 
\begin{align} 
       \calI_{a b}^2 
\equiv -\oint_{\partial\rmC} \df{u^\mu}\, \inProd 
       {\hat{\calE}^{\alpha\beta }\, H^a_{\beta\mu   }\,                     \unitvec{\bsell}_\alpha} 
       {\hat{g    }_{\nu   \rho  }\, H_b^{\nu  \gamma}\, u^\rho\, x^\rmp_z\, \unitvec{\bsell}_\gamma} 
     =  \hat{\calE}^{\beta \alpha}\, H^a_{\mu  \beta }\, 
                                     H^b_{\rho \alpha}\, 
        \oint_{\partial\rmC} \df{u^\mu}\,                u^\rho\, x^\rmp_z, 
\label{eqn: expression for I_ab^2} 
\end{align} 
which matches the negative of the second term on the right-hand side of \Eqref{eqn: expression for I_ab^3}, after relabeling the indices. 
\par 
Consequently, by substituting the expressions for $\calI_{a b}^1$, $\calI_{a b}^2$ and $\calI_{a b}^3$ 
[\Eqsref{eqn: expression for I_ab^1}, 
  \eqref{eqn: expression for I_ab^2} and 
  \eqref{eqn: expression for I_ab^3}, respectively] into \Eqsref{eqn: an intermediate step toward computing I_ab} and \eqref{eqn: the "blessed" line integral}, we obtain the final expression for $\calI_{a b}$, after relabeling the indices: 
\begin{align} 
  0 
&= \calI_{a b} 
= \norm{\bsell_1 \times \bsell_2}\, 
  \hat{\calE}^{\alpha\beta}\, 
  \hat{\calE}^{\mu   \nu  }\left( 
  E^b_{\alpha\mu}\, H^a_{\beta\nu} 
- E^a_{\alpha\mu}\, H^b_{\beta\nu} 
  \right) \Rightarrow
  \nonumber
  \\
    0 
&= \epsilon^{\alpha\beta}\, 
  \epsilon^{\mu   \nu  }\left( 
  E^b_{\alpha\mu}\, H^a_{\beta\nu} 
- E^a_{\alpha\mu}\, H^b_{\beta\nu} 
  \right). 
\label{eqn: expression for I_ab, index notation} 
\end{align} 
\noindent 
This constitutes one of our main results. It states that an arbitrary pair of isometric deformations indexed by $a,b$ cannot be programmed into the same surface, since they would violate the above equation. Instead, each of the three isometric deformations of a particular surface imposes constraints on the others, as will be discussed in \Secref{subsec: rigidity-flexibility relation, part I}.
\subsubsection{Isometric subspace as symplectic and Lagrangian} \label{subsubsec: isometric SP deformations, geometry of the isometric subspace, geometry} 
We can express the deformation modes in the above equation as the following six-dimensional vectors:
\refstepcounter{equation} \label{eqn: expression for modified macroscopic deformations} 
\begin{align} 
\bfV_a^\transpose & \equiv \left(E^a_{11}, E^a_{22}, 2E^a_{12}, H^a_{22}, H^a_{11}, -H^a_{12}\right), 
                           \label{eqn: modified macroscopic deformation, mode a} 
                           \tag{\theequation, a} 
                           \h 
\bfV_b^\transpose & \equiv \left(E^b_{11}, E^b_{22}, 2E^b_{12}, H^b_{22}, H^b_{11}, -H^b_{12}\right), 
                           \label{eqn: modified macroscopic deformation, mode b} 
                           \tag{\theequation, b} 
\end{align} 
\noindent where the factors of $2$ and $-1$ are due to conventions. We can then rewrite \Eqref{eqn: expression for I_ab, index notation} as follows: 
\begin{align} 
  0  =
  \bfV_b^\transpose 
  \begin{pmatrix} 
  \phantom{-}\zero_{3 \times 3} &  \one_{3 \times 3} \h 
          {-} \one_{3 \times 3} & \zero_{3 \times 3} 
  \end{pmatrix} 
  \bfV_a 
\equiv 
  \bfV_b^\transpose\, \bfJ\, 
  \bfV_a, 
\label{eqn: expression for I_ab, matrix notation} 
\end{align} 
where $\zero_{3 \times 3}$ and $\one_{3 \times 3}$ denote the three-by-three zero and identity matrices, respectively. 
We recognize that the matrix $\bfJ$ in \Eqref{eqn: expression for I_ab, matrix notation} is the symplectic matrix (with the properties that its transpose is both its inverse and its opposite) in its canonical form~\cite{Morrison1998_Hamiltonian_Fluid}. 
Hence, as \Eqref{eqn: expression for I_ab, matrix notation} reveals, the isometric subspace possesses a natural symplectic structure and is therefore a three-dimensional symplectic vector space (recall \Secref{subsec: isometric SP deformations, dimension of the isometric subspace} for the subspace's dimensionality). 
\par 
In the following sections, we will show that the symplectic structure extends across the entire six-dimensional deformation phase space. 
Accordingly, the isometric subspace, which satisfies the isometry constraint \Eqref{eqn: expression for I_ab, matrix notation}, is a Lagrangian subspace of the deformation phase space (see \Refsref{Arnold1991_Mechanics, da_Silva2008_Symplectic_Geometry, Jeffs2022_Symplectic_Geometry} for definitions of the Lagrangian subspace). 
\section{The relation between rigidity and flexibility of doubly periodic surfaces} \label{sec: rigidity-flexibility relation} 
\subsection{Examples of permitted sets of isometries} \label{subsec: rigidity-flexibility relation, part I} 
In \Eqref{eqn: expression for I_ab, index notation}, the terms $E^a_{\alpha\mu}$ and $E^b_{\alpha\mu}$ represent the macroscopic strain components of two distinct isometries $a$ and $b$, which quantify the uniform stretching and shearing of a given doubly periodic surface. 
Since these macroscopic strain components correspond to isometries, which incur no energy cost, they can effectively quantify the surface's inherent rigidity---that is, its resistance to in-plane deformations. 
Similarly, the terms $H^a_{\beta\nu}$ and $H^b_{\beta\nu}$, which denote the macroscopic local curvature changes of the surface, can effectively capture the surface's inherent flexibility, i.e., its ability to resist out-of-plane deformations, such as bending or twisting. 
In this context, the isometry constraint [\Eqsref{eqn: expression for I_ab, index notation} or \eqref{eqn: expression for I_ab, matrix notation}] can thus be understood as establishing a relationship between the rigidity and flexibility of a given doubly periodic surface. 
\par 
An important consequence of the rigidity-flexibility relationship is the following: If a doubly periodic surface can be isometrically bent in a given direction, then stretching it along a related direction will inevitably result in an energy cost, and vice versa. 
The mathematical basis for this consequence is as follows. 
\par 
Consider a doubly periodic surface that can be isometrically bent along one of its translation vectors, e.g., $\bsell_1$, assuming such a surface exists. 
The corresponding isometric mode can be represented macroscopically by the following six-dimensional vector: 
\begin{align} 
       \bfV_a^\transpose 
\equiv \left(0, 0, 0, 0, H^a_{11}, 0\right). 
\label{eqn: an example modified macroscopic deformation} 
\end{align} 
Here, $H^a_{11}$ denotes the macroscopic curvature of the deformed surface along the $\bsell_1$ direction. 
In other words, under this isometric deformation, the doubly periodic surface curves upward or downward along the $\bsell_1$ direction, macroscopically forming a cylinder with a radius of $1 / \abs{ H^a_{11} }$. 
By substituting \Eqref{eqn: an example modified macroscopic deformation} into the isometry constraint [\Eqsref{eqn: expression for I_ab, index notation} or \eqref{eqn: expression for I_ab, matrix notation}], we obtain a condition that all isometric modes of the surface must satisfy: 
\begin{align} 
  E^b_{22} 
= 0, 
\end{align} 
indirectly indicating that stretching the surface along $\bsell_2$ always costs energy. 
Thus, we have established that if a doubly periodic surface exhibits flexibility along one of its translation vectors, it must necessarily be rigid along the other. 
One simple example is the flat plane: It is flexible in all directions, and stretching it in any direction incurs an energy cost. 
\par 
More generally, when a known isometric mode of a given surface involves both in-plane and out-of-plane deformations, as is often the case, the isometry is macroscopically represented by the vector $\bfV_a$ in \Eqref{eqn: modified macroscopic deformation, mode a}. 
In this case, we conclude that the deformation modes corresponding to vectors of the following form cannot be isometric: 
\begin{align} 
                     \bfV_c 
\equiv \bfJ\, \bfM\, \bfV_a, 
\label{eqn: modified macroscopic deformation, non-isometric modes} 
\end{align} 
for some definite matrix $\bfM$, as they violate the isometry constraint [\Eqsref{eqn: expression for I_ab, index notation} or \eqref{eqn: expression for I_ab, matrix notation}]: 
\begin{align} 
     \bfV_c^\transpose\,                                     \bfJ\, \bfV_a 
   = \bfV_a^\transpose\, \bfM^\transpose\, \bfJ^\transpose\, \bfJ\, \bfV_a 
   = \bfV_a^\transpose\, \bfM^\transpose\,                          \bfV_a 
\neq 0. 
\end{align} 
\par 
As the above discussion shows, given the isometry data of a doubly periodic surface, the isometry constraint is a useful tool for determining whether a proposed deformation is isometric. 
For non-isometric deformation modes, a relation analogous to the isometry constraint can be used to compute the corresponding deformation energy, which we will derive in the following subsection. 
\subsection{Coarse-grained stress-strain relations for periodic membranes} \label{subsec: rigidity-flexibility relation, part II} 
As noted earlier in \Secref{subsubsec: isometric SP deformations, geometry of the isometric subspace, coupling of isometries}, we derive the relation for non-isometric modes by evaluating the following line integral, which is proportional to the energy cost of an imposed energy-minimizing shape-periodic deformation [cf.~\Eqref{eqn: expression for minimized membrane energy, mapped isometry version}]: 
\begin{align} 
       \calW 
\equiv \frac{2W}{f} 
     = \oint_{\partial\rmC} \df{u^\mu}\, \inProd{ \partial_\mu{ {\wt\bsomega}_\SP } }{ \delta_\SP{\bfx} }. 
\label{eqn: defn. of the work integral} 
\end{align} 
Here, $\rmC$ again denotes a unit cell of a doubly periodic surface, and the line integral is taken over its boundary $\partial\rmC$. 
The shape-periodic displacement field $\delta_\SP{\bfx}\left(u^1, u^2\right)$ corresponds to the imposed energy-minimizing deformation, which is generally non-isometric. 
The equilibrium stress resulting from this deformation is mapped to a shape-periodic isometric deformation, characterized by the angular acceleration vectors $\partial_\mu{\wt\bsomega_\SP}\left(u^1, u^2\right)$. 
Hereafter, the notation $\wt{ \left(\cdot\right) }$ will be used to denote quantities related to the mapped isometric deformation. 
\par 
To evaluate the line integral $\calW$, we again substitute the parametrization of the shape-periodic angular acceleration vectors [see \Eqsref{eqn: parametrization of the SP angular velocity fields} and \eqref{eqn: expression for the SP angular acceleration vectors}] and the expression for the shape-periodic displacement fields [\Eqref{eqn: expression for general SP displacement fields}] into \Eqref{eqn: defn. of the work integral}, following the same procedure used for the line integral $\calI_{a b}$ in \Secref{subsubsec: isometric SP deformations, geometry of the isometric subspace, coupling of isometries}. 
After discarding the terms involving either a periodic integrand\footnote{As a reminder, line integrals of the form given in \Eqref{eqn: identity 3, line integral of periodic one-forms} vanish due to periodicity.} or the vanishing inner product $\inProd{\unitvec{\bsell}_\alpha}{\unitvec z}$, we obtain the following expression [cf.~\Eqref{eqn: an intermediate step toward computing I_ab}]: 
\begin{align} 
\begin{split} 
         \calW 
& \equiv \calW^1 
       + \calW^2 
       + \calW^3 
         \h 
& \equiv \oint_{\partial\rmC} \df{u^\mu}\, \inProd 
         {\hat{\calE}^{\alpha\beta}\, \wt{H}_{\beta\mu   }\,                     \unitvec{\bsell}_\alpha} 
         {\hat{g    }_{\nu   \rho }\,    {E}^{\nu  \gamma}\, u^\rho\,            \unitvec{\bsell}_\gamma} 
       - \oint_{\partial\rmC} \df{u^\mu}\, \inProd 
         {\hat{\calE}^{\alpha\beta}\, \wt{H}_{\beta\mu   }\,                     \unitvec{\bsell}_\alpha} 
         {\hat{g    }_{\nu   \rho }\,    {H}^{\nu  \gamma}\, u^\rho\, x^\rmp_z\, \unitvec{\bsell}_\gamma} 
         \h 
& \quad\quad\quad\quad+ 
         \oint_{\partial\rmC} \df{u^\mu}\, \inProd{ \partial_\mu{ {\wt\bsomega}^\rmp_\SP } }{ \delta_\SP{\bfx} }. 
\end{split} 
\label{eqn: an intermediate step toward computing W, 0} 
\end{align} 
\par 
We recognize that the line integrals $\calW^1$ and $\calW^2$ in \Eqref{eqn: an intermediate step toward computing W, 0} take the same forms as $\calI_{a b}^1$ and $\calI_{a b}^2$ in \Eqref{eqn: an intermediate step toward computing I_ab}, respectively. 
Thus, by applying the previously obtained results for $\calI_{a b}^1$ and $\calI_{a b}^2$ [\Eqsref{eqn: expression for I_ab^1} and \eqref{eqn: expression for I_ab^2}, respectively] in \Eqref{eqn: an intermediate step toward computing W, 0}, we arrive at: 
\begin{align} 
  \calW 
= \norm{\bsell_1 \times \bsell_2}\, 
  \hat{\calE}^{\alpha\beta}\, 
  \hat{\calE}^{\rho  \mu  }\,    {E}_{\alpha\rho}\, 
                              \wt{H}_{\beta \mu } 
- \hat{\calE}^{\alpha\beta}\,    {H}_{\alpha\rho}\, 
                              \wt{H}_{\beta \mu } 
  \oint_{\partial\rmC} \df{u^\mu }\, 
                          {u^\rho}\, x^\rmp_z 
+ \calW^3. 
\label{eqn: an intermediate step toward computing W, 1} 
\end{align} 
\par 
To compute the line integral $\calW^3$, we begin by applying the first integral identity [\Eqref{eqn: identity 1, "line integration by parts"}] to perform integration by parts, yielding: 
\begin{align} 
  \calW^3 
= -\oint_{\partial\rmC} \df{u^\mu}\, \inProd{ {\wt\bsomega}^\rmp_\SP }{ \partial_\mu{ \delta_\SP{\bfx} } }. 
\label{eqn: an intermediate step toward computing W^3, 0} 
\end{align} 
Since the imposed deformation in this case is not necessarily isometric, we cannot manipulate \Eqref{eqn: an intermediate step toward computing W^3, 0} in the same way as we did for $\calI_{a b}^3$, where we expressed $\partial_\mu{\delta_b\bfx}$ in terms of the corresponding angular velocity field [see \Eqref{eqn: an intermediate step toward computing I_ab^3, 0}]. 
Instead, we replace $\partial_\mu{ \delta_\SP{\bfx} }$ in \Eqref{eqn: an intermediate step toward computing W^3, 0} with the general expression for the derivative of shape-periodic displacement fields [\Eqref{eqn: expression for the derivative of SP displacement fields, LHS}]. 
After discarding terms with a periodic integrand, we obtain the following expression for $\calW^3$: 
\begin{align} 
  \calW^3 
= -\oint_{\partial\rmC} \df{u^\mu}\, \inProd{ {\wt\bsomega}^\rmp_\SP }{                         H_{\mu   \beta}\, u^\beta\, \unitvec{z} } 
  +\oint_{\partial\rmC} \df{u^\mu}\, \inProd{ {\wt\bsomega}^\rmp_\SP }{ \hat{g}_{\beta\alpha}\, H^{\alpha\nu  }\, u^\beta\left( 
  \partial_\mu{x^\rmp_z  }\, \unitvec{\bsell}_\nu 
- \partial_\mu{x^\rmp_\nu}\, \unitvec{z     } 
  \right) 
  }. 
\label{eqn: an intermediate step toward computing W^3, 1} 
\end{align} 
\par 
To simplify \Eqref{eqn: an intermediate step toward computing W^3, 1}, we express $H_{\mu\beta}$ in the following way: 
\begin{align} 
                          H_{\mu   \beta} 
= \hat{g}_{\mu  \nu   }\, 
  \hat{g}_{\beta\alpha}\, H^{\alpha\nu  }, 
\end{align} 
which enables us to combine the two terms involving $\unitvec{z}$. 
Thus, after some rearrangements, we arrive at: 
\begin{align} 
  \calW^3 
= \hat{g}_{\beta \alpha}\, 
      {H}^{\alpha\nu   } 
  \oint_{\partial\rmC} \df{u^\mu  }\, 
                          {u^\beta}\, \inProd{ {\wt\bsomega}^\rmp_\SP }{ 
                           \partial_\mu{x^\rmp_z  }\,     \unitvec{\bsell}_\nu 
- \left(\hat{g}_{\mu\nu} + \partial_\mu{x^\rmp_\nu}\right)\unitvec{z     } 
  }. 
\label{eqn: an intermediate step toward computing W^3, 2} 
\end{align} 
\par 
To proceed further, we make the following observations based on the parametrization of doubly periodic surfaces [\Eqref{eqn: parametrization of doubly periodic surfaces}]: 
\begin{align} 
x     _z \equiv \inProd{\bfx     }{\unitvec z} 
              = \inProd{\bfx^\rmp}{\unitvec z} 
         \equiv 
x^\rmp_z, 
\label{eqn: convenient observation 1} 
\end{align} 

\begin{align} 
                \partial_\mu{   x     _\nu} 
\equiv \inProd{ \partial_\mu{\bfx         } }{\unitvec{\bsell}_\nu} 
     = \inProd{                              {\unitvec{\bsell}_\mu} 
     +          \partial_\mu{\bfx^\rmp    } }{\unitvec{\bsell}_\nu} 
\equiv \hat{g}_{\mu\nu} 
     +          \partial_\mu{   x^\rmp_\nu}. 
\label{eqn: convenient observation 2} 
\end{align} 
Given these observations, \Eqref{eqn: an intermediate step toward computing W^3, 2} becomes: 
\begin{align} 
\begin{split} 
         \calW^3 
&      = \hat{g}_{\beta \alpha}\, 
             {H}^{\alpha\nu   }\, 
         \oint_{\partial\rmC} \df{u^\mu  }\, 
                                 {u^\beta}\, \inProd{ {\wt\bsomega}^\rmp_\SP }{ 
         \partial_\mu{x_z  }\, \unitvec{\bsell}_\nu 
       - \partial_\mu{x_\nu}\, \unitvec{z     } 
         } 
         \h 
& \equiv \hat{g}_{\beta \alpha}\, 
             {H}^{\alpha\nu   }\, 
         \oint_{\partial\rmC} \df{u^\mu  }\, 
                                 {u^\beta}\left( 
         {\wt\omega}^\rmp_\nu\, \partial_\mu{x_z  } 
       - {\wt\omega}^\rmp_z  \, \partial_\mu{x_\nu} 
         \right), 
\end{split} 
\label{eqn: an intermediate step toward computing W^3, 3} 
\end{align} 
where 
${\wt\omega}^\rmp_\nu$ and 
${\wt\omega}^\rmp_z  $ denote the in-plane and out-of-plane components of ${\wt\bsomega}^\rmp_\SP$, respectively, as defined below: 
\refstepcounter{equation} 
\begin{align} 
{\wt\omega}^\rmp_\nu & \equiv \inProd{ {\wt\bsomega}^\rmp_\SP }{\unitvec{\bsell}_\nu}, 
                              \tag{\theequation, a} 
                              \h 
{\wt\omega}^\rmp_z   & \equiv \inProd{ {\wt\bsomega}^\rmp_\SP }{\unitvec{z     }    }. 
                              \tag{\theequation, b} 
\end{align} 
\par 
In \Eqref{eqn: an intermediate step toward computing W^3, 3}, the term in parentheses is related to the in-plane component of ${\wt\bsomega}^\rmp_\SP \times \partial_\mu{\bfx}$. 
To see this relationship, we begin by decomposing the vectors ${\wt\bsomega}^\rmp_\SP$ and $\partial_\mu{\bfx}$ into the planar frame 
$
\left\{ 
\unitvec{\bsell}_1, 
\unitvec{\bsell}_2, 
\unitvec{z     } 
\right\} 
$, as was done in \Eqref{eqn: "decomposing" the geometry of unit cells}: 
\begin{align} 
{\wt\bsomega}^\rmp_\SP & \equiv \hat{g}^{\nu\rho}\, {\wt\omega}^\rmp_\nu\, \unitvec{\bsell}_\rho 
                              +                     {\wt\omega}^\rmp_z  \, \unitvec{z     }, 
                                \h 
\partial_\mu{\bfx} & \equiv \hat{g}^{\nu\rho}\, \partial_\mu{x_\nu}\, \unitvec{\bsell}_\rho 
                          +                     \partial_\mu{x_z  }\, \unitvec{z     }. 
\end{align} 
The desired relationship can be obtained by computing the following scalar triple product: 
\begin{align} 
\begin{split} 
    \inProd{\unitvec{\bsell}_\gamma}{                    {\wt\bsomega}^\rmp_\SP                         \times \partial_\mu{\bfx    }                        } 
& = \inProd{\unitvec{\bsell}_\gamma}{\hat{g}^{\nu\rho}\, {\wt  \omega}^\rmp_\nu\, \unitvec{\bsell}_\rho \times \partial_\mu{   x_z  }\, \unitvec{z     }     } 
  + \inProd{\unitvec{\bsell}_\gamma}{\hat{g}^{\nu\rho}\, {\wt  \omega}^\rmp_z  \, \unitvec{z     }      \times \partial_\mu{   x_\nu}\, \unitvec{\bsell}_\rho} 
    \h 
& = \hat{g}^{\nu\rho}\left( 
    {\wt\omega}^\rmp_\nu\, \partial_\mu{x_z  } 
  - {\wt\omega}^\rmp_z  \, \partial_\mu{x_\nu} 
    \right)\inProd{\unitvec z}{\unitvec{\bsell}_\gamma \times \unitvec{\bsell}_\rho} 
    \h 
& = \hat{g    }^{\nu   \rho}\, 
    \hat{\calE}_{\gamma\rho}\left( 
    {\wt\omega}^\rmp_\nu\, \partial_\mu{x_z  } 
  - {\wt\omega}^\rmp_z  \, \partial_\mu{x_\nu} 
    \right), 
\end{split} 
\label{eqn: rewriting the integrand of W^3, step 1} 
\end{align} 
where the cyclic property of the scalar triple product is used in the second line to combine the two terms, and the vector identity \Eqref{eqn: the first identity regarding the planar frame} is applied in the last line. 
By utilizing the tensor relations \Eqsref{eqn: defn. of the inverse planar metric components} and \eqref{eqn: convenient tensor relation, 1}, we can further invert \Eqref{eqn: rewriting the integrand of W^3, step 1}, yielding: 
\begin{align} 
  {\wt\omega}^\rmp_\nu\, \partial_\mu{x_z  } 
- {\wt\omega}^\rmp_z  \, \partial_\mu{x_\nu} 
= \hat{g    }_{\nu   \rho}\, 
  \hat{\calE}^{\gamma\rho}\inProd{\unitvec{\bsell}_\gamma}{ {\wt\bsomega}^\rmp_\SP \times \partial_\mu{\bfx} }. 
\label{eqn: rewriting the integrand of W^3, step 2} 
\end{align} 
\par 
Finally, substituting \Eqref{eqn: rewriting the integrand of W^3, step 2} into \Eqref{eqn: an intermediate step toward computing W^3, 3}, we express the line integral $\calW^3$ as follows: 
\begin{align} 
\begin{split} 
    \calW^3 
& = \hat{g    }_{\beta \alpha}\, 
        {H    }^{\alpha\nu   }\, \oint_{\partial\rmC} \df{u^\mu}\, u^\beta\, 
    \hat{g    }_{\nu   \rho  }\, 
    \hat{\calE}^{\gamma\rho  }\, \inProd{\unitvec{\bsell}_\gamma}{ {\wt\bsomega}^\rmp_\SP \times \partial_\mu{\bfx} } 
    \h 
& = \oint_{\partial\rmC} \df{u^\mu  }\, \inProd{ 
                            {u^\beta}\, 
    \hat{\calE}^{\gamma\rho}\, 
        {H    }_{\beta \rho}\, \unitvec{\bsell}_\gamma 
    }{ {\wt\bsomega}^\rmp_\SP \times \partial_\mu{\bfx} }. 
\end{split} 
\label{eqn: an intermediate step toward computing W^3, 4} 
\end{align} 
By comparing \Eqref{eqn: an intermediate step toward computing W^3, 4} to \Eqref{eqn: an intermediate step toward computing I_ab^3, 1}, we observe that $\calW^3$ shares the same form as $\calI_{a b}^3$, where the imposed deformation in \Eqref{eqn: an intermediate step toward computing W^3, 4} corresponds to the isometric mode ``$b$'' in \Eqref{eqn: an intermediate step toward computing I_ab^3, 1}. 
From this observation, we can directly apply the previously derived results for $\calI_{a b}^3$ [\Eqsref{eqn: an intermediate step toward computing I_ab^3, 2}--\eqref{eqn: expression for I_ab^3}] to compute $\calW^3$, yielding: 
\begin{align} 
\begin{split} 
    \calW^3 
& = -\norm{\bsell_1 \times \bsell_2}\, \hat{\calE}^{\mu\beta}\, \hat{\calE}^{\gamma\rho  }\, \wt{E}_{\mu \gamma}\,    {H}_{\beta \rho} 
  -                                                             \hat{\calE}^{\gamma\rho  }\, \wt{H}_{\mu \gamma}\,    {H}_{\beta \rho}\, 
     \oint_{\partial\rmC} \df{u^\mu}\, u^\beta\, x^\rmp_z 
    \h 
& = -\norm{\bsell_1 \times \bsell_2}\, \hat{\calE}^{\mu\beta}\, \hat{\calE}^{\gamma\rho  }\, \wt{E}_{\mu \gamma}\,    {H}_{\beta \rho} 
  +                                                             \hat{\calE}^{\rho  \gamma}\,    {H}_{\rho\beta }\, \wt{H}_{\gamma\mu }\, 
     \oint_{\partial\rmC} \df{u^\mu}\, u^\beta\, x^\rmp_z 
    \h 
& = -\norm{\bsell_1 \times \bsell_2}\, \hat{\calE}^{\mu\beta}\, \hat{\calE}^{\gamma\rho  }\, \wt{E}_{\mu \gamma}\,    {H}_{\beta \rho} 
  - \calW^2, 
\end{split} 
\label{eqn: expression for W^3} 
\end{align} 
where, in the second line, we rearrange the integral on the right-hand side to match the form of $\calW^2$ [see 
\Eqsref{eqn: an intermediate step toward computing W, 0} and 
 \eqref{eqn: an intermediate step toward computing W, 1}]. 
\par 
Thus, by substituting \Eqref{eqn: expression for W^3} into \Eqref{eqn: an intermediate step toward computing W, 1}, we arrive at the final expression for the line integral $\calW$, after relabeling the indices [cf.~\Eqref{eqn: expression for I_ab, index notation}]: 
\begin{align} 
  \calW 
= \norm{\bsell_1 \times \bsell_2}\, 
  \hat{\calE}^{\alpha\beta}\, 
  \hat{\calE}^{\mu   \nu  }\left( 
     {E}_{\alpha\mu}\, \wt{H}_{\beta\nu} 
- \wt{E}_{\alpha\mu}\,    {H}_{\beta\nu} 
  \right). 
\label{eqn: expression for W} 
\end{align} 
From \Eqref{eqn: expression for W}, the expression for the energy cost of the imposed deformation follows directly [recall \Eqref{eqn: defn. of the work integral}]: 
\begin{align} 
  E_\rmm^\eq = W 
= \frac{1}{2}\, f\, \norm{\bsell_1 \times \bsell_2}\, 
  \hat{\calE}^{\alpha\beta}\, 
  \hat{\calE}^{\mu   \nu  }\left( 
     {E}_{\alpha\mu}\, \wt{H}_{\beta\nu} 
- \wt{E}_{\alpha\mu}\,    {H}_{\beta\nu} 
  \right). 
\label{eqn: expression for the energy cost} 
\end{align} 
 \noindent
 These expressions are one of the main results, as discussed in the main text. To better understand \Eqref{eqn: expression for the energy cost} physically, we define the following quantities: 
\refstepcounter{equation} \label{eqn: defn. of the macroscopic stress components} 
\begin{align} 
   {\Sigma}^{\alpha\mu} & \equiv f\, \hat{\calE}^{\alpha\beta}\, \hat{\calE}^{\mu\nu}\,    {H}_{\beta\nu}, 
                          \tag{\theequation, a} 
                          \h 
\wt{\Sigma}^{\alpha\mu} & \equiv f\, \hat{\calE}^{\alpha\beta}\, \hat{\calE}^{\mu\nu}\, \wt{H}_{\beta\nu}, 
                          \tag{\theequation, b} 
\end{align} 
which have the dimensions of stress. 
For the \emph{plane} spanned by the translation vectors of a doubly periodic surface, the out-of-plane deformation modes characterized by $H_{\beta\nu}$ and $\wt{H}_{\beta\nu}$ are isometric. 
Therefore, by \Eqref{eqn: relation between equilibrium stress and isometry, curvature version}, the newly introduced quantities $\Sigma^{\alpha\mu}$ and $\wt{\Sigma}^{\alpha\mu}$ can be interpreted as the macroscopic equilibrium stress components corresponding to these isometric out-of-plane modes of the plane through the isometry-stress duality. 
In this light, the combination $\Sigma^{\alpha\mu}\, E_{\alpha\mu}$ has the meaning of the macroscopic membrane energy density, analogous to its microscopic counterpart $\sigma^{\alpha\mu}\, \varepsilon_{\alpha\mu}$ in \Eqref{eqn: expression for membrane energy}. 
\par 
In terms of $\Sigma^{\alpha\mu}$ and $\wt{\Sigma}^{\alpha\mu}$, the expression for the energy cost can be expressed as [cf.~\Eqref{eqn: expression for membrane energy}]: 
\begin{align} 
  E_\rmm^\eq = W 
= \frac{1}{2}\norm{\bsell_1 \times \bsell_2}\left( 
  \wt{\Sigma}^{\alpha\nu}\,    {E}_{\alpha\mu} 
-    {\Sigma}^{\alpha\nu}\, \wt{E}_{\alpha\mu} 
  \right). 
\label{eqn: expression for the energy cost, Sigma version} 
\end{align} 
As illustrated by \Eqref{eqn: expression for the energy cost, Sigma version}, the cross-couplings of the non-isometric mode and its corresponding isometric mode give rise to two energies, and their difference equals the energy cost associated with the non-isometric mode. 
\par 
To summarize, the isometry constraint [\Eqref{eqn: expression for I_ab, index notation}], resulting from the coupling between isometric modes of a doubly periodic surface, links the surface's rigidity and flexibility (\Secref{subsec: rigidity-flexibility relation, part I}); in contrast, the coupling of an isometry with a non-isometric mode quantifies the energy cost associated with the non-isometric mode (\Secref{subsec: rigidity-flexibility relation, part II}). 
It is worth noting that both types of coupling [\Eqsref{eqn: expression for I_ab, index notation} and \eqref{eqn: expression for the energy cost}] are macroscopic in nature. 
That is, the couplings are independent of the microscopic terms unique to individual doubly periodic surfaces, such as the vector field $\bfx^\rmp\left(u^1, u^2\right)$ that characterizes the unit cell geometry. 
In this sense, our results, \Eqsref{eqn: expression for I_ab, index notation} and \eqref{eqn: expression for the energy cost}, are universal, meaning they apply to any doubly periodic surface.\footnote{As discussed in \Secref{SI subsec: numerical verification, piecewise smooth surfaces}, our results also apply to doubly periodic surfaces that are piecewise smooth.} 
However, this universality does not suggest that all doubly periodic surfaces are macroscopically mechanically equivalent. 
The microscopic effects, though integrated out, remain incorporated into the macroscopic behaviors of doubly periodic surfaces, giving rise to various mechanical properties~\cite{Feng2024_Topological_Origami}. 
As an example, the isometric subspaces associated with two doubly periodic surfaces are generally distinct, as determined by their underlying unit cell geometries. 
\section{Geometry of the deformation phase space} \label{sec: phase space geometry} 
\subsection{Symplectic structure of the deformation phase space} 
The shared form of the two couplings, \Eqsref{eqn: expression for I_ab, index notation} and \eqref{eqn: expression for the energy cost}, indicates that the symplectic structure of the isometric subspace extends across the entire six-dimensional deformation phase space. 
In fact, the deformation phase space can be formulated analogously to the phase space of classical mechanics. 
\par 
To see the similarity, we represent a deformation mode macroscopically using the vector [cf.~\Eqsref{eqn: expression for macroscopic deformations} and \eqref{eqn: expression for modified macroscopic deformations}]: 
\begin{align} 
       \bfv^\transpose 
\equiv \left(\Sigma^{11}, \Sigma^{22}, \Sigma^{12}, E_{11}, E_{22}, 2E_{12}\right) 
\equiv \left(     Q^{A },                           P_{A }                 \right), 
\end{align} 
where $\Sigma^{\alpha\beta}$ denotes the macroscopic stress components, as defined in \Eqsref{eqn: defn. of the macroscopic stress components}. 
The capital Latin index $A$ runs from one to three; for instance, 
$Q^1 \equiv  \Sigma^{11}$, and 
$P_3 \equiv 2     E_{12}$, where the factor of two may seem jarring but is, in fact, consistent with the Voigt notation~\cite{Manik2021_Voigt_Notation}. 
In this way, the macroscopic stress and strain components serve as the coordinates of the deformation phase space, playing a similar role to the generalized coordinates and their conjugate momenta in the phase space of classical mechanics. 
\par 
As in classical mechanics, the symplectic structure of the deformation phase space becomes evident when the symplectic two-form is introduced: 
\begin{align} 
       \bsOmega 
\equiv \df{Q^A} \wedge 
       \df{P_A}. 
\label{eqn: symplectic two-form} 
\end{align} 
In terms of the symplectic two-form $\bsOmega$, \Eqsref{eqn: expression for I_ab, index notation} and \eqref{eqn: expression for the energy cost, Sigma version} can be expressed as: 
\begin{align} 
             0 & =                                           \bsOmega\left(   {\bfv}_a, \bfv_b\right), 
                   \label{eqn: special relation between rigidity and flexibility, form version} 
                   \h 
E_\rmm^\eq = W & = \frac{1}{2}\norm{\bsell_1 \times \bsell_2}\bsOmega\left(\wt{\bfv}  , \bfv  \right). 
                   \label{eqn: general relation between rigidity and flexibility, form version} 
\end{align} 
\Eqnref{eqn: special relation between rigidity and flexibility, form version} demonstrates that the isometric subspace is Lagrangian~\cite{Arnold1991_Mechanics, da_Silva2008_Symplectic_Geometry, Jeffs2022_Symplectic_Geometry}, while \Eqref{eqn: general relation between rigidity and flexibility, form version} suggests that the energy cost of a deformation mode can be interpreted geometrically as the projected phase-space area of the parallelogram formed by the mode and its corresponding isometry. It is unclear what process, if any, plays the same role in this symplectic geometry as does time-evolution in Hamiltonian dynamics.
\subsection{Gauge invariance and canonical transformations} 
Recall from \Secref{subsubsec: DP surfaces and SP deformations, SP deformations, form of SP deformation} [see the text around \Eqref{eqn: expression for general SP displacement fields}] that the total strain of a deformed doubly periodic surface appears to depend on the choice of the zero-height level $x_z = 0$. 
We demonstrate below that our results for the couplings of the deformation modes [\Eqsref{eqn: expression for I_ab, index notation} and \eqref{eqn: expression for the energy cost}] are independent of this gauge choice. 
More remarkably, as we will show, changing the zero-height level corresponds to a canonical transformation of the deformation phase space~\cite{Schutz1999_Geometrical_Methods} that preserves its symplectic structure. 
\par 
The total strain is composed of the uniform macroscopic in-plane strain (with components 
$
      {E}_{\alpha\beta} 
= \hat{g}_{\alpha\mu  }\, 
  \hat{g}_{\beta \nu  }\, 
      {E}^{\mu   \nu  } 
$) and the height-dependent strain arising from the uniform out-of-plane deformation modes. 
The components of the total strain are given by: 
\begin{align} 
       E^\tot_{\alpha\beta}      \left(u^1, u^2\right) 
\equiv E     _{\alpha\beta} 
     - H     _{\alpha\beta}\, x_z\left(u^1, u^2\right), 
\label{eqn: defn. of the total strain components} 
\end{align} 
where $x_z$ describes the height profile of a given doubly periodic surface. 
Given the form of a shape-periodic mode $\delta_\SP{\bfx}\left(u^1, u^2\right)$ [\Eqref{eqn: expression for general SP displacement fields}], the total strain components, which vary from point to point within a unit cell, can be obtained by computing the fractional changes in the lattice vectors via the following lattice derivatives (denoted $\frakd_\alpha$): 
\refstepcounter{equation} \label{eqn: expression for the total strain components} 
\begin{align*} 
         E^\tot_{11}\left(u^1, u^2\right) 
&      = \inProd{ \frac{  { \delta_\SP{\bfx}\left(u^1 + \ell^1, u^2\right) } 
       -                  { \delta_\SP{\bfx}\left(u^1         , u^2\right) } }{\ell^1} }{\unitvec{\bsell}_1} 
  \equiv \inProd{ \frakd_1{ \delta_\SP{\bfx}\left(u^1         , u^2\right) }           }{\unitvec{\bsell}_1}, 
         \tag{\theequation, a} 
         \h 
         E^\tot_{22}\left(u^1, u^2\right) 
&      = \inProd{ \frac{  { \delta_\SP{\bfx}\left(u^1, u^2 + \ell^2\right) } 
       -                  { \delta_\SP{\bfx}\left(u^1, u^2         \right) } }{\ell^2} }{\unitvec{\bsell}_2} 
  \equiv \inProd{ \frakd_2{ \delta_\SP{\bfx}\left(u^1, u^2         \right) }           }{\unitvec{\bsell}_2}, 
         \tag{\theequation, b} 
         \h 
    E^\tot_{12}\left(u^1, u^2\right) 
& = \frac{1}{2}\left( 
    \inProd{ \frakd_1{ \delta_\SP{\bfx}\left(u^1, u^2\right) } }{\unitvec{\bsell}_2} 
  + \inProd{ \frakd_2{ \delta_\SP{\bfx}\left(u^1, u^2\right) } }{\unitvec{\bsell}_1} 
    \right). 
    \tag{\theequation, c} 
\end{align*} 
Combining \Eqsref{eqn: defn. of the total strain components} and \eqref{eqn: expression for the total strain components} yields the following expression for the macroscopic strain components: 
\begin{align} 
  E_{\alpha\beta} 
= \frac{1}{2}\left( 
  \inProd{ \frakd_\alpha{ \delta_\SP{\bfx} } }{\unitvec{\bsell}_\beta } 
+ \inProd{ \frakd_\beta { \delta_\SP{\bfx} } }{\unitvec{\bsell}_\alpha} 
  \right) 
+ H_{\alpha\beta}\, x_z. 
\label{eqn: expression for macro strains in terms of the lattice derivatives, index notation} 
\end{align} 
\par 
Under the passive transformation $x_z \mapsto x_z + z_0$, which can be achieved by choosing a different zero-height level, the surface's deformation modes remain unaffected. 
Accordingly, the measurable quantities associated with a deformation mode---in particular, the total strain components and macroscopic curvature changes---should also remain invariant: 
\refstepcounter{equation} \label{eqn: transformation rules} 
\begin{align} 
E^\tot_{\alpha\beta}\left(u^1, u^2\right) & \mapsto E^\tot_{\alpha\beta}\left(u^1, u^2\right), \label{eqn: transformation rule for the E_tot matrix} \tag{\theequation, a} \h 
H     _{\alpha\beta}                      & \mapsto H     _{\alpha\beta}                     . \label{eqn: transformation rule for the H matrix}     \tag{\theequation, b} 
\end{align} 
Based on these facts, the following transformation rule for $E_{\alpha\beta}$ can be deduced from \Eqref{eqn: defn. of the total strain components}: 
\begin{align} 
        E_{\alpha\beta} 
\mapsto E_{\alpha\beta} 
      + H_{\alpha\beta}\, z_0. 
\label{eqn: transformation rule for the E matrix} 
\tag{\theequation, c} 
\end{align} 
That is, we have shown that the uniform macroscopic in-plane strain depends on the choice of the zero-height level and is, therefore, generally not gauge-invariant on its own.\footnote{As an exception, for pure in-plane modes, the uniform macroscopic in-plane strain is gauge-invariant, since no macroscopic curvature changes occur in this case.} 
It is worth noting a potentially confusing fact: The total strain is gauge-invariant but varies from point to point within a unit cell, whereas the uniform macroscopic strain, though constant, is not gauge-independent. 
\par 
Although the couplings of the deformation modes [\Eqsref{eqn: expression for I_ab, index notation} and \eqref{eqn: expression for the energy cost}] involve the uniform macroscopic in-plane strain, which is not gauge-invariant, the couplings themselves remain independent of the gauge choice of the zero-height level. 
To avoid redundancy, we will show below that \Eqref{eqn: expression for the energy cost} is gauge-invariant; the same reasoning applies to \Eqref{eqn: expression for I_ab, index notation}. 
\par 
We begin by substituting the transformation rules 
\Eqsref{eqn: transformation rule for the H matrix} and 
 \eqref{eqn: transformation rule for the E matrix} into \Eqref{eqn: expression for the energy cost}, yielding: 
\begin{align} 
\begin{split} 
          W 
& \mapsto \frac{1}{2}\, f\,       \norm{\bsell_1 \times \bsell_2}\, 
          \hat{\calE}^{\alpha\beta}\, 
          \hat{\calE}^{\mu   \nu  }\left[ 
          \left(   {E}_{\alpha\mu} +    {H}_{\alpha\mu}\, z_0\right)\wt{H}_{\beta\nu} 
        - \left(\wt{E}_{\alpha\mu} + \wt{H}_{\alpha\mu}\, z_0\right)   {H}_{\beta\nu} 
          \right] 
          \h 
& \quad\quad= 
          W 
        + \frac{1}{2}\, f\, z_0\, \norm{\bsell_1 \times \bsell_2}\, 
          \hat{\calE}^{\alpha\beta}\, 
          \hat{\calE}^{\mu   \nu  }\left( 
             {H}_{\alpha\mu}\, \wt{H}_{\beta\nu} 
        - \wt{H}_{\alpha\mu}\,    {H}_{\beta\nu} 
          \right) 
          \h 
& \quad\quad= 
          W. 
\end{split} 
\label{eqn: transformation rule for the membrane energy} 
\end{align} 
In \Eqref{eqn: transformation rule for the membrane energy}, the term involving the cross-couplings of the macroscopic curvature changes vanishes, as demonstrated below: 
\begin{align} 
  \hat{\calE}^{\alpha\beta }\, \hat{\calE}^{\mu\nu}\,    {H}_{\alpha\mu}\, \wt{H}_{\beta \nu} 
= \hat{\calE}^{\beta \alpha}\, \hat{\calE}^{\nu\mu}\,    {H}_{\beta \nu}\, \wt{H}_{\alpha\mu} 
= \hat{\calE}^{\alpha\beta }\, \hat{\calE}^{\mu\nu}\, \wt{H}_{\alpha\mu}\,    {H}_{\beta \nu}, 
\end{align} 
where the first equality follows from relabeling the indices, 
$\alpha \leftrightarrow \beta$  and 
$\mu    \leftrightarrow \nu  $, and the second is a consequence of the antisymmetric property of both 
$\hat{\calE}^{\beta\alpha}$ and 
$\hat{\calE}^{\nu  \mu   }$. 
We have thus shown that the membrane energy is gauge-invariant, as is expected for a physical observable. 
\par 
In the language of symplectic geometry, the argument for the gauge invariance of the membrane energy can be rephrased as follows. 
First, the transformation rules 
\Eqsref{eqn: transformation rule for the H matrix} and 
 \eqref{eqn: transformation rule for the E matrix} imply the following coordinate transformation in the deformation phase space: 
\refstepcounter{equation} \label{eqn: a canonical transformation} 
\begin{align} 
        \left(Q^A\right) 
 \equiv                          \left(\Sigma^{11}, \Sigma^{22},  \Sigma^{12}\right) 
 \equiv \frac{1}{ \sqrt{\hat g} }\left(     H_{22},      H_{11}, -     H_{12}\right) 
\mapsto \frac{1}{ \sqrt{\hat g} }\left(     H_{22},      H_{11}, -     H_{12}\right) 
 \equiv \left(Q^A\right), 
\tag{\theequation, a} 
\end{align} 

\begin{align} 
\begin{split} 
          \left(P_A                    \right) 
   \equiv \left(E_{11}, E_{22}, 2E_{12}\right) 
& \mapsto \left(E_{11}, E_{22}, 2E_{12}\right) + z_0             \left(     H_{11},      H_{22},  2     H_{12}\right) 
          \h 
& \quad\equiv 
          \left(P_A                    \right) + z_0\sqrt{\hat g}\left(\Sigma^{22}, \Sigma^{11}, -2\Sigma^{12}\right) 
   \equiv \left(P_A + C_{A B}\, Q^B    \right), 
\end{split} 
\tag{\theequation, b} 
\end{align} 
where the \emph{symmetric} matrix $\left(C_{A B}\right)$ is given by: 
\begin{align} 
       \left(C_{A B}\right) 
\equiv z_0\sqrt{\hat g}\begin{pmatrix} 
       0 & 1 & \phantom{-}0 \h 
       1 & 0 & \phantom{-}0 \h 
       0 & 0 &         {-}2 
       \end{pmatrix}. 
\end{align} 
We note that the coordinate transformation \Eqsref{eqn: a canonical transformation} is canonical because it preserves the symplectic two-form, as verified below: 
\begin{align} 
            \df{Q^A} \wedge \df{ \left(P_A + C_{A B}\, Q^B\right) } 
=           \df{Q^A} \wedge \df{       P_A                        } 
+ C_{A B}\, \df{Q^A} \wedge \df{                       Q^B        } 
=           \df{Q^A} \wedge \df{       P_A                        }, 
\end{align} 
where the term $C_{A B}\, \df{Q^A} \wedge \df{Q^B}$ vanishes because the wedge product is skew-commutative, while the matrix $\left(C_{A B}\right)$ is symmetric. 
As a direct consequence, the membrane energy, which is proportional to the projected phase-space area [see \Eqref{eqn: general relation between rigidity and flexibility, form version}], remains invariant under the canonical transformation, since canonical transformations are generally volume-preserving~\cite{Schutz1999_Geometrical_Methods}. 
\par 
As we have just demonstrated, the gauge choice of the zero-height level is essentially a coordinate transformation---a way to parameterize the deformation phase space as well as the real space $\bbR^3$. 
Physical observables in our system, such as the total strain and membrane energy, are all gauge-independent. 
In the next section, when verifying our results numerically, we fix the gauge by setting the surface average height to zero. 
For later reference, the average height of a surface $\bfx\left(u^1, u^2\right)$ is defined as: 
\begin{align} 
       \left<x_z\right> 
\equiv \frac 
       {\displaystyle \int_\rmC \df{u^1}\, \df{u^2}\, \sqrt{     g}\norm{ \unitvec{n} \times \unitvec{z} }x_z} 
       {\displaystyle \int_\rmC \df{u^1}\, \df{u^2}\, \sqrt{\hat g}                                          } 
\equiv \frac{1}{ \norm{\bsell_1 \times \bsell_2} }\int_\rmC d{\calA_\perp}\, x_z; 
\label{eqn: defn. for surface average height} 
\end{align} 
that is, the height function $x_z$ is weighted by the \emph{projected} infinitesimal area 
$
       d{\calA_\perp} 
\equiv \norm{ 
       \left( 
       \partial_1{\bfx}\, \df{u^1} \times 
       \partial_2{\bfx}\, \df{u^2} 
       \right)                     \times \unitvec{z} 
       } 
$ onto the lattice plane. 
The chosen gauge is consistent with the fact that, for a thin plate without transverse shear, the mid-surface remains stress-free under bending deformations~\cite{LL1986_Elasticity}. 
\section{Numerical Analysis} \label{SI sec: numerical verification} 
In this section, we present our approach for numerically verifying the results obtained in the previous sections. 
The section is organized into three parts. 
First, we argue that, although derived under the assumption of smoothness, our analytical results extend to periodic surfaces that are piecewise smooth (but can contain sharp creases, as in origami). 
In the second subsection, we describe the procedure used to compute the isometric modes of triangulated surfaces and provide an explanation of the underlying theory. 
Finally, we conclude the section by introducing our method for quantifying the distance between subspaces of a symplectic vector space. 
\subsection{Extension of our results to piecewise smooth surfaces} \label{SI subsec: numerical verification, piecewise smooth surfaces} 
Although our analytical results---namely, the isometric constraint [\Eqref{eqn: expression for I_ab, index notation}] and the symplectic energy-cost formula [\Eqref{eqn: expression for the energy cost}]---are derived under the assumption that the periodic surfaces considered thus far are smooth, they generalize straightforwardly to piecewise smooth periodic surfaces, such as origami tessellations. 
We briefly sketch a proof of this statement below. 
\par 
To begin with, by a piecewise smooth periodic surface, we mean that each unit cell of the surface consists of smooth pieces whose boundaries are joined together. 
The tangent planes are well-defined on each of the adjacent pieces, but not at their shared boundary. 
For such surfaces, the duality between equilibrium stresses and isometric deformations holds on each piece. 
Mathematically, this means that the equilibrium stress components ($\sigma^{\alpha\beta}$) and the angular acceleration components (${a_\gamma}^\beta$) in \Eqsref{eqn: relation between equilibrium stress and isometry, i to s} and \eqref{eqn: relation between equilibrium stress and isometry, s to i} are piecewise smooth functions across the surface. 
In light of this, it can be shown that the expression for the boundary work [\Eqref{eqn: defn. of the work integral}] remains valid for piecewise smooth surfaces by first breaking the corresponding surface integral [\Eqref{eqn: expression for membrane energy}] into pieces, then applying Stokes' theorem [\Eqref{eqn: Stokes' theorem}] for each piece and finally summing the resulting line integrals. 
The identity used to derive the isometric constraint, \Eqref{eqn: the "blessed" line integral}, holds for the same underlying reason, as discussed in a similar context in~\Refref{Nassar2024_Rigidity_and_Flexibility}. 
Therefore, from \Eqsref{eqn: the "blessed" line integral} and \eqref{eqn: defn. of the work integral}, we obtain the desired results [\Eqsref{eqn: expression for I_ab, index notation} and \eqref{eqn: expression for the energy cost}] by breaking the curve $\partial{\rmC}$ into pieces and evaluating the resulting line integrals. 
\subsection{Numerical method for obtaining isometry data associated with triangulated surfaces} 
In this subsection, we first characterize the discrete geometry of a triangulated surface, then outline the computational framework for determining its isometries and conclude by presenting the expressions used to extract the associated macroscopic strains and curvature changes from the resulting isometry data. 
\subsubsection{Discrete geometry of triangulated surfaces} 
Triangulated surfaces, a type of piecewise smooth surface, are commonly used in mechanics simulations (e.g., the finite-element method) as approximations for true smooth surfaces (or shells), particularly when they consist of a large number of triangular faces, the overall dimensions of which are much smaller than the deformation length scale of interest. 
Below, we characterize the geometry of a unit cell of an arbitrary triangulated periodic surface. 
\par 
One unit cell of our doubly periodic surface corresponds to a rectangular region $\left(u^1, u^2\right) \in \left[0, \ell^1\right] \times \left[0, \ell^2\right]$ in the parameter space. 
We evenly divide this region into an $N_1 \times N_2$ grid, with the index of the $(i, j)$-th grid point given by: 
\begin{align} 
  \left(         {   u^1}     ,          {   u^2}     \right) 
= \left(i\, \frac{\ell^1}{N_1}, j\, \frac{\ell^2}{N_2}\right) 
\end{align} 
for $i \in \left\{0, 1, \ldots, N_1 - 1\right\}$ and $j \in \left\{0, 1, \ldots, N_2 - 1\right\}$. 
These grid points in the parameter space produce a mesh on the surface with the grid points on the surface given by: 
\begin{align} 
  \bfx\left(         {   u^1}     ,          {   u^2}     \right) 
= \bfx\left(i\, \frac{\ell^1}{N_1}, j\, \frac{\ell^2}{N_2}\right) 
\end{align} 
for $i \in \left\{0, 1, \ldots, N_1 - 1\right\}$ and $j \in \left\{0, 1, \ldots, N_2 - 1\right\}$. 
In this subsection, we simplify our notation by using the subscripts $\bfu_{i, j}$ and $\bfx_{i, j}$ to represent, respectively, the grid points in the parameter space and on the surface. 
\par
\begin{figure}[H] 
\centering 
\includegraphics[width = 0.95\textwidth]{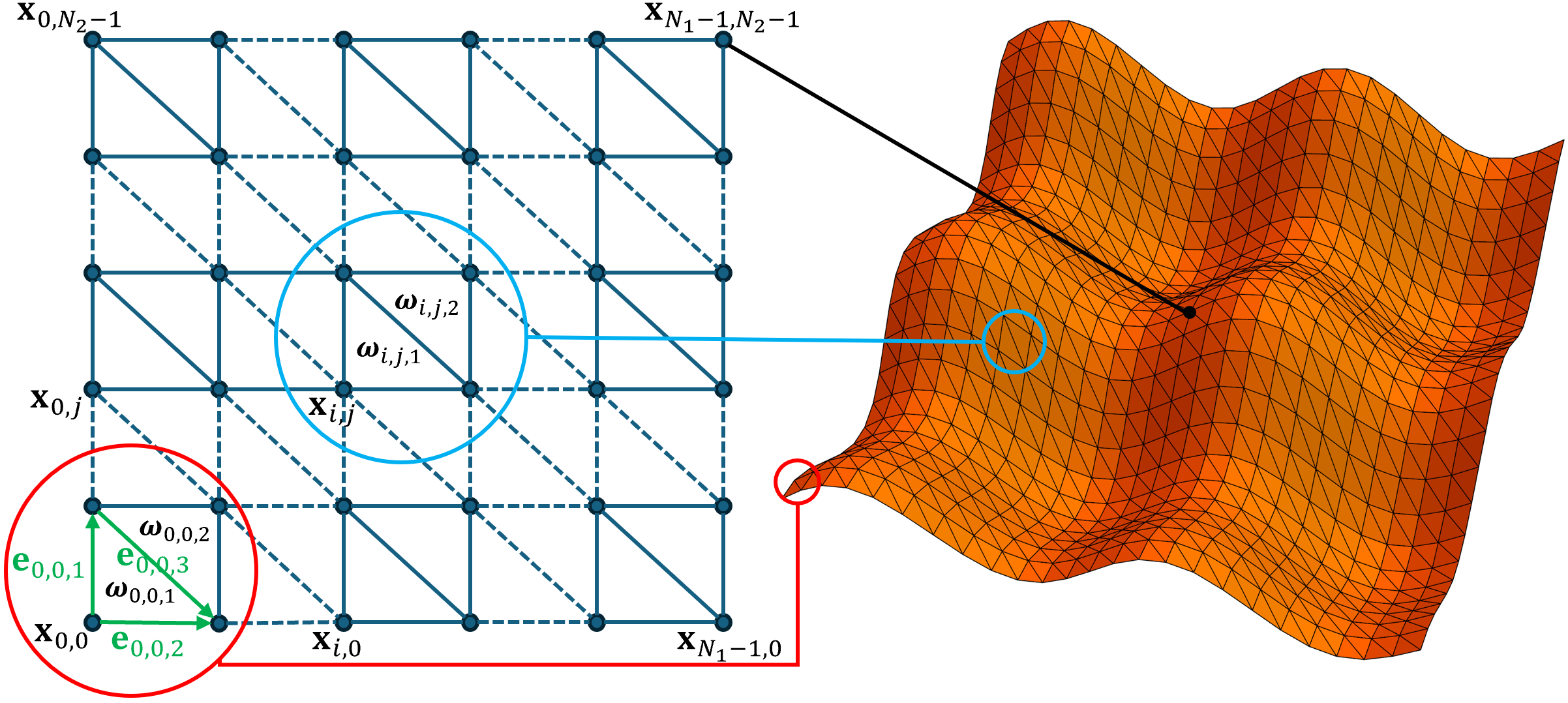} 
\caption{
A triangulated sheet with four unit cells (two-by-two) is shown on the right. 
How different quantities are indexed inside a unit cell, including the vertices, edges and angular velocities associated with each triangular face, are shown in the diagram to the left. 
} 
\label{SI fig: unit cell of a triangulated surface} 
\end{figure}

 
For a surface with an arbitrary triangulation, the expression for the surface average height [\Eqref{eqn: defn. for surface average height}] reduces to: 
\begin{align} 
\begin{split} 
    \left<x_z\right> 
& = \frac{1}{ \norm{\bsell_1 \times \bsell_2} }\sum_{ \text{all }\Delta\text{'s} } \int_\Delta d{\calA_\perp}\, x_z \h 
& = \frac{1}{ \norm{\bsell_1 \times \bsell_2} }\sum_{ \text{all }\Delta\text{'s} } 
    \calA_\perp\left(\Delta\right) 
    \frac{ 
    x_z\left(\Delta_1\right) 
  + x_z\left(\Delta_2\right) 
  + x_z\left(\Delta_3\right) 
    }{3}, 
\end{split} 
\label{eqn: expression for surface average height, arbitrary triangulation} 
\end{align} 
where $\calA_\perp\left(\Delta\right)$ denotes the projected area of a triangle, and $x_z\left(\Delta_i\right)$ the height at the triangle's $i^\textrm{th}$ vertex. 
In our case, since the triangular grid is set up uniformly (see \Figref{SI fig: unit cell of a triangulated surface}), every triangular face has the same projected area, and, after some counting, the expression for the surface average height therefore simplifies further to: 
\begin{align} 
  \left<x_z\right> 
= \frac{1}{N_1 N_2}\sum_{i, j} \inProd{ \bfx_{i, j} }{ \unitvec{z} }. 
\label{eqn: expression for surface average height, uniform triangulation} 
\end{align} 
Moving forward, when analyzing the isometries of a chosen triangulated surface in the following subsubsections, we apply a translation along the $z$ direction to set its average height (computed using \Eqref{eqn: expression for surface average height, uniform triangulation}) to zero if it is not already.

We use horizontal, vertical and diagonal edges to connect neighboring grid points on the surface to generate a triangulation of the surface. 
We denote these edge vectors using: 
\begin{align} 
\bfe_{i, j, 1} \equiv \bfx_{i + 1, j    } - \bfx_{i, j    } , \quad 
\bfe_{i, j, 2} \equiv \bfx_{i    , j + 1} - \bfx_{i, j    } \textAnd 
\bfe_{i, j, 3} \equiv \bfx_{i + 1, j    } - \bfx_{i, j + 1} 
\label{eqn: edge_vectors} 
\end{align} 
for each grid point on the surface indexed by $\left(i, j\right)$. 
There are $3\, N_1\, N_2$ edges per unit cell. (By periodicity, the grid point $\bfx_{N_1, j}$ corresponds to $\bfx_{0, j}$ within each unit cell, and similarly for $\bfx_{i, N_2}$.) 
Denoting using $e_{i, j, k}$ the length of the edge vector $\bfe_{i, j, k}$, we concatenate all the $e_{i, j, k}$ to form a new vector: 
\begin{align} 
             \bfe^\transpose 
\equiv \left(   e_{0      ,       0, 1}, 
                e_{0      ,       0, 2}, 
                e_{0      ,       0, 3}, \ldots, 
                e_{N_1 - 1, N_2 - 1, 1}, 
                e_{N_1 - 1, N_2 - 1, 2}, 
                e_{N_1 - 1, N_2 - 1, 3} 
       \right). 
\end{align} 
\subsubsection{Characterization and numerical computation of linear isometries of triangulated surfaces} 
We allow each grid point on the surface to displace freely in the three-dimensional Euclidean space in which the surface embeds. 
To represent a generic deformation of the triangulated surface, we use the following vector with $3\, N_1\, N_2$ components: 
\begin{align} 
             \delta{\bfx}^\transpose 
\equiv \left( 
       \left(\delta{\bfx}_{0      , 0      }\right)_x, 
       \left(\delta{\bfx}_{0      , 0      }\right)_y, 
       \left(\delta{\bfx}_{0      , 0      }\right)_z, \ldots, 
       \left(\delta{\bfx}_{N_1 - 1, N_2 - 1}\right)_x, 
       \left(\delta{\bfx}_{N_1 - 1, N_2 - 1}\right)_y, 
       \left(\delta{\bfx}_{N_1 - 1, N_2 - 1}\right)_z 
       \right), 
\end{align} 
which is simply a concatenation of the displacement of each grid point on the surface, denoted as: 
$
        \delta{\bfx}_{i, j} 
= \left( 
  \left(\delta{\bfx}_{i, j}\right)_x, 
  \left(\delta{\bfx}_{i, j}\right)_y, 
  \left(\delta{\bfx}_{i, j}\right)_z 
  \right) 
$. 
\par 
If we impose a generic displacement field $\delta{\bfx}$, i.e., arbitrarily displacing the grid points on a surface, we cause all the edges to change their lengths. 
To linear order, the change in the edge length is given by: 
\begin{align} 
         {   \delta{   e}_{i, j, k} } 
= \inProd{   \delta{\bfe}_{i, j, k} }{ 
           \unitvec{   e}_{i, j, k} }, 
\end{align} 
and the resulting change in $\bfe$ is related to $\delta{\bfx}$ via a square matrix $\bfC$, known as the compatibility matrix~\cite{Rocklin2022_Topologically_Protected_Deformations_bookChapter}: 
\begin{align} 
         \delta{\bfe} 
= \bfC\, \delta{\bfx}, 
\end{align} 
where $\delta{\bfe}$ denotes the change in the concatenated edge length vector $\bfe$. 
If, to linear order, the lengths of all edges in this mesh do not change as we impose the infinitesimal displacement $\delta{\bfx}$, we call the corresponding $\delta{\bfx}$ a linear isometry. 
We see that linear isometries correspond to vectors in the nullspace of the compatibility matrix $\bfC$. 
For a generic surface, the nullspace of $\bfC$ consists of three linear independent vectors, corresponding to three uniform translations of the surface. 
\par 
In order to obtain nontrivial linear isometries (not uniform translation or global rotation), we need to take a less direct but very effective approach of describing linear isometries, namely via changes in the dihedral angles between adjacent triangular faces. 
If the infinitesimal displacement $\delta{\bfx}$ corresponds to an isometry, it should not change the length of any edge on the mesh, but it should in general cause neighboring triangular panels to rotate with respect to each other, causing the dihedral angles along the edges to change. 
Therefore, we may use a vector $\bsphi$ to describe the dihedral angle along every edge and $\delta{\bsphi}$ its infinitesimal changes. 
An arbitrary choice of $\delta{\bsphi}$ would tear the triangulated surface somewhere, and it is shown in \Refref{belcastro2002_BH_Condition} that $\delta{\bsphi}$ has to satisfy the belcastro-Hull condition at each vertex. 
To linear order, this condition has the following form at a given vertex $\bfx_{i, j}$: 
\begin{align} 
\sum_{ \alpha \in I_{i, j} } (\pm)\, \delta{\phi}_\alpha\, \unitvec{e}_\alpha = \zero, 
\label{eqn: belcastro_hull} 
\end{align} 
where the summation is over all edges emanating out from that vertex. 
If written out explicitly, the set $I_{i, j}$ contains the following indices for the edge vectors: 
\begin{align} 
  I_{i, j} 
= \left\{ 
  (i    , j    , 1), 
  (i    , j    , 2), 
  (i - 1, j    , 1), 
  (i    , j - 1, 2), 
  (i - 1, j + 1, 3), 
  (i + 1, j - 1, 3) 
  \right\}. 
\end{align} 
The plus or minus sign in \Eqref{eqn: belcastro_hull} is used to adjust the direction of $\unitvec{e}_\alpha$ because the edge vectors may take the opposite direction (instead of emanating out from the vertex $\alpha$) when defined in \Eqref{eqn: edge_vectors}. 
\par 
The two ways of representing linear isometries, respectively via the infinitesimal vertex displacement $\delta{\bfx}$ and infinitesimal folding $\delta{\bsphi}$ are equivalent (up to global translation and rotation, which are isometries that do not generate any folding), and they are related by the following observation. 
Assume the edges are massless springs with spring constant set to unity. 
The energy required to realize an infinitesimal deformation $\delta{\bfx}$ is given by: 
\begin{align} 
  \scrE 
= \frac{1}{2}             \delta{\bfe}       ^\transpose\, \delta{\bfe} 
= \frac{1}{2}\left(\bfC\, \delta{\bfx}\right)^\transpose\, \delta{\bfe} 
= \frac{1}{2}             \delta{\bfx}       ^\transpose\, 
                   \bfC                      ^\transpose\, \delta{\bfe}. 
\label{eqn: energy_ex} 
\end{align} 
Notice that $\delta{\bfe}$ is exactly the tension $\bft$ on each edge, given the spring constant is set to unity. 
Therefore, \Eqref{eqn: energy_ex} can be rewritten as: 
\begin{align} 
\scrE = \frac{1}{2}\delta{\bfx}^\transpose\, \bfC^\transpose\, \bft. 
\end{align} 
By conservation of energy, $\scrE$ is equal to the amount of work required to overcome the stress at each vertex as we displace them. 
Therefore, if we use the vector $\bfs$ to represent the resulting stress (a scalar tension) on each vertex due to the infinitesimal displacement $\delta{\bfx}$, we have: 
\begin{align} 
  \frac{1}{2}\delta{\bfx}^\transpose\,                   \bfs = \scrE 
= \frac{1}{2}\delta{\bfx}^\transpose\, \bfC^\transpose\, \bft, 
\label{eqn: energy_sx} 
\end{align} 
which has to hold for any given infinitesimal $\delta{\bfx}$, and so we have the following relation: 
\begin{align} 
                    \bfs 
= \bfC^\transpose\, \bft, 
\end{align} 
which shows the transpose of $\bfC$, known as the equilibrium matrix, maps the vector $\bft$ representing tension in each edge to the vector $\bfs$ representing the stress at each vertex. 
In analogy to the duality between isometry and equilibrium stress for continuous surfaces established in \Secref{subsec: the isometry-stress duality, duality}, we have the exact duality for the discrete case. 
If the grid points are in a state of equilibrium stress, then for each vertex, we have: 
\begin{align} 
\sum_{ \alpha \in I_{i, j} } (\pm)\, t_\alpha\, \unitvec{e}_\alpha = \zero, 
\label{eqn: duality} 
\end{align} 
which has the same form as the linearized belcastro-Hull condition [\Eqref{eqn: belcastro_hull}]. 
It follows that if we take a tension vector $\bft$ in the nullspace of the equilibrium matrix $\bfC^\transpose$, it should correspond to a state of equilibrium stress and satisfy \Eqref{eqn: duality}. 
If we treat this $\bft$ as $\delta{\bsphi}$, it should also satisfy \Eqref{eqn: belcastro_hull} at every vertex because the two equations have exactly the same form. 
\par 
Once we obtain an isometry in terms of change in dihedral angles $\delta{\bsphi}$ by solving a system of linear equations $\bfC^\transpose\, \delta{\bsphi} = \zero$, we can define an angular velocity field $\bsomega$ correspondingly. 
This field assigns an angular velocity vector $\bsomega_{i, j, k}$ to each triangular panel, where the indices $i, j$ range from $0$ to respectively $N_1$, $N_2$, and $k$ can be $1$ or $2$. (See \Figref{SI fig: unit cell of a triangulated surface}.) 
More specifically, a triangular panel with the third index $k = 1$ has $\bfx_{i, j}$, $\bfx_{i + 1, j}$ and $\bfx_{i, j + 1}$ as its three vertices, while for $k = 2$ the corresponding vertices are $\bfx_{i + 1, j + 1}$, $\bfx_{i + 1, j}$ and $\bfx_{i, j + 1}$. 
Each angular velocity describes how the associated panel rotates in the three-dimensional embedding space. 
Since $\delta{\bsphi}$ describes the change in dihedral angles, it follows directly that for two neighboring panels with indices $\alpha$ and $\beta$, the difference between their angular velocities can be related to the angular velocity field, as follows: 
\begin{align} 
  \bsomega_\alpha 
- \bsomega_\beta 
= (\pm)\,   \delta{\phi}_{ \gamma(\alpha, \beta) }\, 
          \unitvec{   e}_{ \gamma(\alpha, \beta) }, 
\label{eqn: angular_velocity} 
\end{align} 
where the mapping $\gamma(\alpha, \beta)$ gives the index of the edge shared by the two panels indexed by $\alpha$ and $\beta$. 
It is worth noting that we should follow a consistent sign convention to define a proper angular velocity field. 
Also, since we can always add a global rotation to every angular velocity, it is convenient to set the angular velocity of the starting panel---the panel indexed by $(0, 0, 1)$---to zero: $\bsomega_{0, 0, 1} \equiv \zero$ and calculate other $\bsomega_{i, j, k}$ recursively using \Eqref{eqn: angular_velocity} and our isometry $\delta{\bsphi}$. 
Another thing to note is that the result is indeed independent of the path we choose to evaluate $\bsomega_{i, j, k}$ because $\delta{\bsphi}$ satisfies the linearized belcastro-Hull condition [\Eqref{eqn: belcastro_hull}]. 
\par 
Once we get the angular velocity field, we may calculate the displacement field it induces. 
Again, due to a possible uniform translation, we may choose one of the vertices not to move. 
Without loss of generality, we take $\delta{\bfx}_{0,0} = \zero$ and define recursively that for two neighboring vertices with indices $\alpha', \beta'$, we have: 
\begin{align} 
  \delta{\bfx}_{\alpha'} 
- \delta{\bfx}_{\beta '} 
= \left( 
        {\bfx}_{\alpha'} 
-       {\bfx}_{\beta '} 
  \right) \times 
  \bsomega_{ \gamma'(\alpha', \beta') }, 
\label{eqn: displacement_field} 
\end{align} 
where the mapping $\gamma'(\alpha', \beta')$ gives the index of the triangular panel of which $\bfx_{\alpha'}$ and $\bfx_{\beta'}$ are vertices. 
With all those primed indices, we are emphasizing the way $\gamma'$ maps indices should be different from that of $\gamma$ in \Eqref{eqn: angular_velocity}. 
One obvious difference is that the indices $\alpha, \beta$ each has three numbers, while $\alpha', \beta'$ each only has two. 
Again, the final result should be path-independent as long as $\bsomega$ is path-independent, which is truly given $\delta{\bsphi}$ is an isometry and satisfies \Eqref{eqn: belcastro_hull}. 
\subsubsection{Extraction of macroscopic strains and curvature changes from isometry data} 
Given the isometry data obtained from \Eqref{eqn: displacement_field}, one can readily compute the associated macroscopic curvature changes and strains, $H_{\alpha\beta}$ and $E_{\alpha\beta}$, using \Eqsref{eqn: relation between the macroscopic angular accelerations and the H matrix} and \eqref{eqn: expression for macro strains in terms of the lattice derivatives, index notation}, respectively. 
\par 
Within those two expressions, the translation vectors $\bsell_\alpha$ are obtained as the differences between the position vectors corresponding to the corners of a unit cell: 
\begin{align} 
\bsell_1 \equiv \bfx_{N_1, 0  } - \bfx_{0, 0} = \bfx_{N_1, 0  } \textAnd 
\bsell_2 \equiv \bfx_{0  , N_2} - \bfx_{0, 0} = \bfx_{0  , N_2}; 
\end{align} 
recall that our gauge choice in simulations is $\bfx_{0, 0} = \zero$. 
It follows that the macroscopic unit normal vector, $\unitvec{z}$, is given by: 
\begin{align} 
       \unitvec{z} 
\equiv \frac{ \bsell_{  1   } \times \bsell_{     2} }{ \norm{ \bsell_{  1   } \times \bsell_{     2} } } 
     = \frac{ \bfx  _{N_1, 0} \times \bfx  _{0, N_2} }{ \norm{ \bfx  _{N_1, 0} \times \bfx  _{0, N_2} } }. 
\end{align} 
\par 
For the macroscopic curvature changes, the macroscopic angular acceleration vectors $\bfw_\alpha$ that appear in \Eqref{eqn: relation between the macroscopic angular accelerations and the H matrix} are computed from the discrete angular velocity vectors as: 
\refstepcounter{equation} 
\begin{align*} 
\bfw_1 \equiv \frakd_1{\bsomega_\SP}\left(0, 0\right) & \equiv \frac{ \bsomega_\SP\left(\ell^1, 0\right) 
                                                             -        \bsomega_\SP\left(0     , 0\right) }{        \ell^{  1   }   } 
                                                             = \frac{ \bsomega_        {   N_1, 0, 1   } }{ \norm{ \bfx_{N_1, 0} } }, \tag{\theequation, a} \h 
\bfw_2 \equiv \frakd_2{\bsomega_\SP}\left(0, 0\right) & \equiv \frac{ \bsomega_\SP\left(0, \ell^2\right) 
                                                             -        \bsomega_\SP\left(0, 0     \right) }{        \ell^{     2}   } 
                                                             = \frac{ \bsomega_        {0,    N_2, 1   } }{ \norm{ \bfx_{0, N_2} } }, \tag{\theequation, b} 
\end{align*} 
where, again, $\bsomega_\SP\left(u^1, u^2\right)$ denotes the angular velocity field [\Eqref{eqn: parametrization of the SP angular velocity fields}] corresponding to an isometric shape-periodic deformation. 
Recall that, in our simulations, we set $\bsomega_\SP\left(0, 0\right) \equiv \bsomega_{0, 0, 1} = \zero$ (i.e., the origin does not get rotated). 
\par 
For the macroscopic strains, the displacement vectors 
$\delta_\SP{\bfx}\left(u^1         , u^2         \right)$, 
$\delta_\SP{\bfx}\left(u^1 + \ell^1, u^2         \right)$ and 
$\delta_\SP{\bfx}\left(u^1         , u^2 + \ell^2\right)$---which are involved in computing the lattice derivatives in \Eqref{eqn: expression for the total strain components}---are expressed in the discrete language as: 
\begin{align} 
\delta_\SP{\bfx}\left(u^1         , u^2         \right) \equiv \delta{\bfx}_{i      , j      },       \halfQuad 
\delta_\SP{\bfx}\left(u^1 + \ell^1, u^2         \right) \equiv \delta{\bfx}_{i + N_1, j      }  \mbox{\halfQuad{and}\halfQuad} 
\delta_\SP{\bfx}\left(u^1         , u^2 + \ell^2\right) \equiv \delta{\bfx}_{i      , j + N_2}. 
\end{align} 
\par 
To summarize, by substituting the above expressions into \Eqsref{eqn: relation between the macroscopic angular accelerations and the H matrix} and \eqref{eqn: expression for macro strains in terms of the lattice derivatives, index notation}, one can compute the macroscopic curvature changes and strains from the isometry data as: 
\refstepcounter{equation} 
\begin{gather*} 
  \left(H_{\alpha\beta}\right) 
= \begin{pmatrix} 
  \displaystyle \inProd{ \frac{ \bsomega_{N_1, 0  , 1} }{ \norm{ \bfx_{N_1, 0  } } } \times \unitvec{\bsell}_1  } 
                       { \frac{ \bfx_{N_1, 0} \times \bfx_{0, N_2} }{ \norm{ \bfx_{N_1, 0} \times \bfx_{0, N_2} } } } & 
  \displaystyle \inProd{ \frac{ \bsomega_{N_1, 0  , 1} }{ \norm{ \bfx_{N_1, 0  } } } \times \unitvec{\bsell}_2  } 
                       { \frac{ \bfx_{N_1, 0} \times \bfx_{0, N_2} }{ \norm{ \bfx_{N_1, 0} \times \bfx_{0, N_2} } } } \hh 
  \displaystyle \inProd{ \frac{ \bsomega_{N_1, 0  , 1} }{ \norm{ \bfx_{N_1, 0  } } } \times \unitvec{\bsell}_2  } 
                       { \frac{ \bfx_{N_1, 0} \times \bfx_{0, N_2} }{ \norm{ \bfx_{N_1, 0} \times \bfx_{0, N_2} } } } & 
  \displaystyle \inProd{ \frac{ \bsomega_{0  , N_2, 1} }{ \norm{ \bfx_{0  , N_2} } } \times \unitvec{\bsell}_2  } 
                       { \frac{ \bfx_{N_1, 0} \times \bfx_{0, N_2} }{ \norm{ \bfx_{N_1, 0} \times \bfx_{0, N_2} } } } 
  \end{pmatrix}, 
\label{eqn: expression for the macroscopic curvature changes, full glory} 
\tag{\theequation, a} 
\hh 
\begin{split} 
    \left(E_{\alpha\beta}\right) 
& = \left(H_{\alpha\beta}\right) 
    \inProd{ \bfx_{i  , j} }{ 
    \frac{ { \bfx_{N_1, 0} \times \bfx_{0, N_2} } } 
    { \norm{ \bfx_{N_1, 0} \times \bfx_{0, N_2} } } 
    } 
    \h 
& \quad\quad+ 
    \frac{1}{2}\left[ 
    \begin{pmatrix} 
    \displaystyle \inProd{ \frac{ \delta{\bfx}_{i + N_1, j      } - \delta{\bfx}_{i, j} }{ \norm{ \bfx_{N_1, 0  } } } }{\unitvec{\bsell}_1} & 
    \displaystyle \inProd{ \frac{ \delta{\bfx}_{i + N_1, j      } - \delta{\bfx}_{i, j} }{ \norm{ \bfx_{N_1, 0  } } } }{\unitvec{\bsell}_2} \hh 
    \displaystyle \inProd{ \frac{ \delta{\bfx}_{i      , j + N_2} - \delta{\bfx}_{i, j} }{ \norm{ \bfx_{0  , N_2} } } }{\unitvec{\bsell}_1} & 
    \displaystyle \inProd{ \frac{ \delta{\bfx}_{i      , j + N_2} - \delta{\bfx}_{i, j} }{ \norm{ \bfx_{0  , N_2} } } }{\unitvec{\bsell}_2} 
    \end{pmatrix} 
    \right. 
    \h 
& \quad\quad\quad\quad\quad\quad\quad\quad\quad\quad+ 
    \left. 
    \begin{pmatrix} 
    \displaystyle \inProd{ \frac{ \delta{\bfx}_{i + N_1, j      } - \delta{\bfx}_{i, j} }{ \norm{ \bfx_{N_1, 0  } } } }{\unitvec{\bsell}_1} & 
    \displaystyle \inProd{ \frac{ \delta{\bfx}_{i      , j + N_2} - \delta{\bfx}_{i, j} }{ \norm{ \bfx_{0  , N_2} } } }{\unitvec{\bsell}_1} \hh 
    \displaystyle \inProd{ \frac{ \delta{\bfx}_{i + N_1, j      } - \delta{\bfx}_{i, j} }{ \norm{ \bfx_{N_1, 0  } } } }{\unitvec{\bsell}_2} & 
    \displaystyle \inProd{ \frac{ \delta{\bfx}_{i      , j + N_2} - \delta{\bfx}_{i, j} }{ \norm{ \bfx_{0  , N_2} } } }{\unitvec{\bsell}_2} 
    \end{pmatrix} 
    \right]. 
\end{split} 
\tag{\theequation, b} 
\end{gather*} 
We would like to remind our readers that the average heights of the chosen surfaces are set to zero when performing the above computations. 
We have also numerically verified that the macroscopic curvature changes computed using \Eqref{eqn: expression for the macroscopic curvature changes, full glory} are identical to those obtained using Eq.~(3.5,~a) of the \MT, which involves taking the lattice derivatives twice. 
\subsection{Quantifying the distance between subspaces of the deformation phase space} 
As discussed in the \MT, our simulations---carried out according to the procedure outlined in the preceding subsections---show that the nullspace of the compatibility matrix $\bfC$ associated with a generic triangulated surface is three-dimensional, indicating that a generic triangulated periodic surface admits three linearly independent isometric modes. 
This finding is consistent with literature results established through the Maxwell-Calladine index theorem (see, e.g., \Refref{McInerney2020_Hidden_Symmetries}). 
\par 
As in the case of smooth surfaces, each of the three isometric modes associated with a triangulated periodic surface can be represented macroscopically by a six-dimensional vector [see \Eqsref{eqn: expression for modified macroscopic deformations}], which combines the macroscopic strains and curvature changes. 
Together, these three six-dimensional vectors span an isometric Lagrangian subspace. 
By concatenating these vectors, we obtain the following three-by-six matrix, which represents the isometric subspace associated with the surface: 
\begin{align} 
                       \bfG 
\equiv \begin{pmatrix} \bfV_1^\transpose                                            \h 
                       \bfV_2^\transpose                                            \h 
                       \bfV_3^\transpose                                            \end{pmatrix} 
\equiv \begin{pmatrix} E^1_{11}, E^1_{22}, 2E^1_{12}, H^1_{22}, H^1_{11}, -H^1_{12} \h 
                       E^2_{11}, E^2_{22}, 2E^2_{12}, H^2_{22}, H^2_{11}, -H^2_{12} \h 
                       E^3_{11}, E^3_{22}, 2E^3_{12}, H^3_{22}, H^3_{11}, -H^3_{12} \end{pmatrix} 
\equiv \begin{pmatrix} \bfE                                                         & 
                       \bfH                                                         \end{pmatrix}, 
\label{eqn: matrix representing the isometric subspace} 
\end{align} 
where the three-by-three matrices $\bfE$ and $\bfH$ collect the macroscopic strains and curvature changes, respectively, associated with all three isometries. 
\par 
In principle, we expect that the isometric subspaces associated with different triangulations of the same periodic surface will converge as the number of triangular faces in the unit cell increases. 
To quantify this convergence, we have considered the following metrics to measure the ``distances'' or ``angles'' between the isometric subspaces. 
All the metrics considered have the following properties: 
\begin{itemize} 
\item The metric is zero if the two isometric subspaces are the same subspace. 
\item The metric is one  if the two isometric subspaces have no nontrivial intersection. 
\end{itemize} 
\subsubsection{The distance measure based on the Euclidean metric} 
The first distance measure, presented in the \MT, utilizes the Euclidean inner product. 
For this distance measure to be well-defined, we set the characteristic length scale of the surface to unity, ensuring that the macroscopic strains and curvature changes are both dimensionless and on equal footing. 
After doing so, we can treat the isometric subspaces corresponding to triangulated periodic surfaces with $\calN$ and $\calN - 1$ grid squares per row and per column in a unit cell as two ordinary subspaces of $\bbR^6$ and compute the ``distance'' between them using the following formula: 
\begin{align} 
                    {d^{\bbR^6}_{\calN,\, \calN - 1}      } 
\equiv 1 - \cos\left(    \theta_{\calN,\, \calN - 1}\right) 
\equiv 1 - \frac{ \det\left\{\bfG_{\calN    }^\transpose\, \bfG_{\calN - 1}^{\phantom\transpose}\right\} }{ \sqrt{ 
                { \det\left\{\bfG_{\calN    }^\transpose\, \bfG_{\calN    }^{\phantom\transpose}\right\} }\, 
                { \det\left\{\bfG_{\calN - 1}^\transpose\, \bfG_{\calN - 1}^{\phantom\transpose}\right\} } } }, 
\label{eqn: distance measure 1, Euclid} 
\end{align} 
where the matrices $\bfG_\calN$ and $\bfG_{\calN - 1}$ represent the isometric subspaces in question, and $\theta_{\calN,\, \calN - 1}$ is the angle between the subspaces. 
\par 
It may seem that this metric depends on the choice of the characteristic length scale, but it can be verified that the convergence of the isometric subspaces does not depend on it. 
\subsubsection{The distance measures based on the matrix inner product} 
The following two metrics, independent of the characteristic length scale (and thus the Euclidean metric), are defined using the matrix inner product. 
\par 
First, we apply elementary row operations to bring the matrix $\bfG$ in \Eqref{eqn: matrix representing the isometric subspace} into its reduced row echelon form, yielding: 
\begin{align} 
                                            \bfG' 
\equiv \begin{pmatrix} \one_{3 \times 3} &  \bfE^\inverse\, \bfH \end{pmatrix} 
\equiv \begin{pmatrix} \one_{3 \times 3} & \calG                 \end{pmatrix}. 
\label{eqn: reduced row echelon form of the matrix encoding the isometry data} 
\end{align} 
From \Eqref{eqn: reduced row echelon form of the matrix encoding the isometry data}, it follows that the three-by-three matrix $\calG \equiv \bfE^{-1}\, \bfH$ serves as an alternative representation of the isometric subspace associated with a given surface. 
\par 
Thus, inspired by the notion of distance or angle between two vectors in real space, we define the two metrics below to measure the ``closeness'' between the isometric subspaces associated with triangulated periodic surfaces with $\calN$ and $\calN - 1$ grid squares per row and per column in a unit cell, by computing the following matrix inner products~\cite{Itskov2019_Linear_Algebra}, which involve the representing matrices $\calG_\calN$ and $\calG_{\calN - 1}$: 
\begin{align} 
d^{\, \GL, 1}_{\calN,\, \calN - 1} & \equiv \frac{\norm{\calG_\calN -        \calG_{\calN - 1} } }{ 
                                                  \norm{\calG_\calN}\, \norm{\calG_{\calN - 1} } } 
                                     \equiv \sqrt{ 
                                            \frac{\tr\left\{\left(\calG_{\calN    }           -  \calG_{\calN - 1}\right)^{ \transpose} 
                                                            \left(\calG_{\calN    }           -  \calG_{\calN - 1}\right)              \right\} } 
                                                 {\tr\left\{      \calG_{\calN    }^\transpose\, \calG_{\calN    }^{\phantom\transpose}\right\}\, 
                                                  \tr\left\{      \calG_{\calN - 1}^\transpose\, \calG_{\calN - 1}^{\phantom\transpose}\right\} } }, 
\label{eqn: distance measure 2, Rocklin} 
\hh 
d^{\, \GL, 2}_{\calN,\, \calN - 1} & \equiv 1 - \frac{  \inProd{\calG_{\calN    } }                 { \calG_{\calN - 1}                    }         } 
                                                     {    \norm{\calG_{\calN    } }         \, \norm{ \calG_{\calN - 1}                    }         } 
                                     \equiv 1 - \frac{\tr\left\{\calG_{\calN    }^\transpose\,        \calG_{\calN - 1}^{\phantom\transpose}\right\} }{\sqrt{ 
                                                      \tr\left\{\calG_{\calN    }^\transpose\,        \calG_{\calN    }^{\phantom\transpose}\right\} \, 
                                                      \tr\left\{\calG_{\calN - 1}^\transpose\,        \calG_{\calN - 1}^{\phantom\transpose}\right\} } }. 
\label{eqn: distance measure 3, Sun} 
\end{align} 
\subsubsection{Other types of distance measures} 
In addition, as demonstrated in \Refref{Ye2016_Distance_Measures_Between_Subspaces}, several well-established metrics for quantifying the distance between subspaces of a vector space prove to be well-suited to our purposes. 
Among these, we choose the Grassmann distance, as it appears to be among the more commonly known options: 
\begin{align} 
       d^{\, \Gr}_{\calN,\, \calN - 1} 
\equiv            \norm{                                     \bfG_{\calN    }^{        \transpose}\, 
                        \bfJ                     \,          \bfG_{\calN - 1}^{\phantom\transpose}                                              } 
\equiv \sqrt{ \tr\left\{\bfJ^{\phantom\transpose}\!\!\!\left(\bfG_{\calN    }^{\phantom\transpose}\, \bfG_{\calN    }^\transpose\right) 
                        \bfJ^{        \transpose}      \left(\bfG_{\calN - 1}^{\phantom\transpose}\, \bfG_{\calN - 1}^\transpose\right)\right\} }, 
\label{eqn: distance measure 4, Grassmann} 
\end{align} 
where $\bfJ$ is the symplectic matrix, as defined in \Eqref{eqn: expression for I_ab, matrix notation}. 
\par 
The figure below illustrates the convergence behavior of the various metrics discussed above for the triangulated surfaces presented in Fig.~4 of the \MT.

\begin{figure}[H] 
\centering 
\includegraphics[width = 0.75\textwidth]{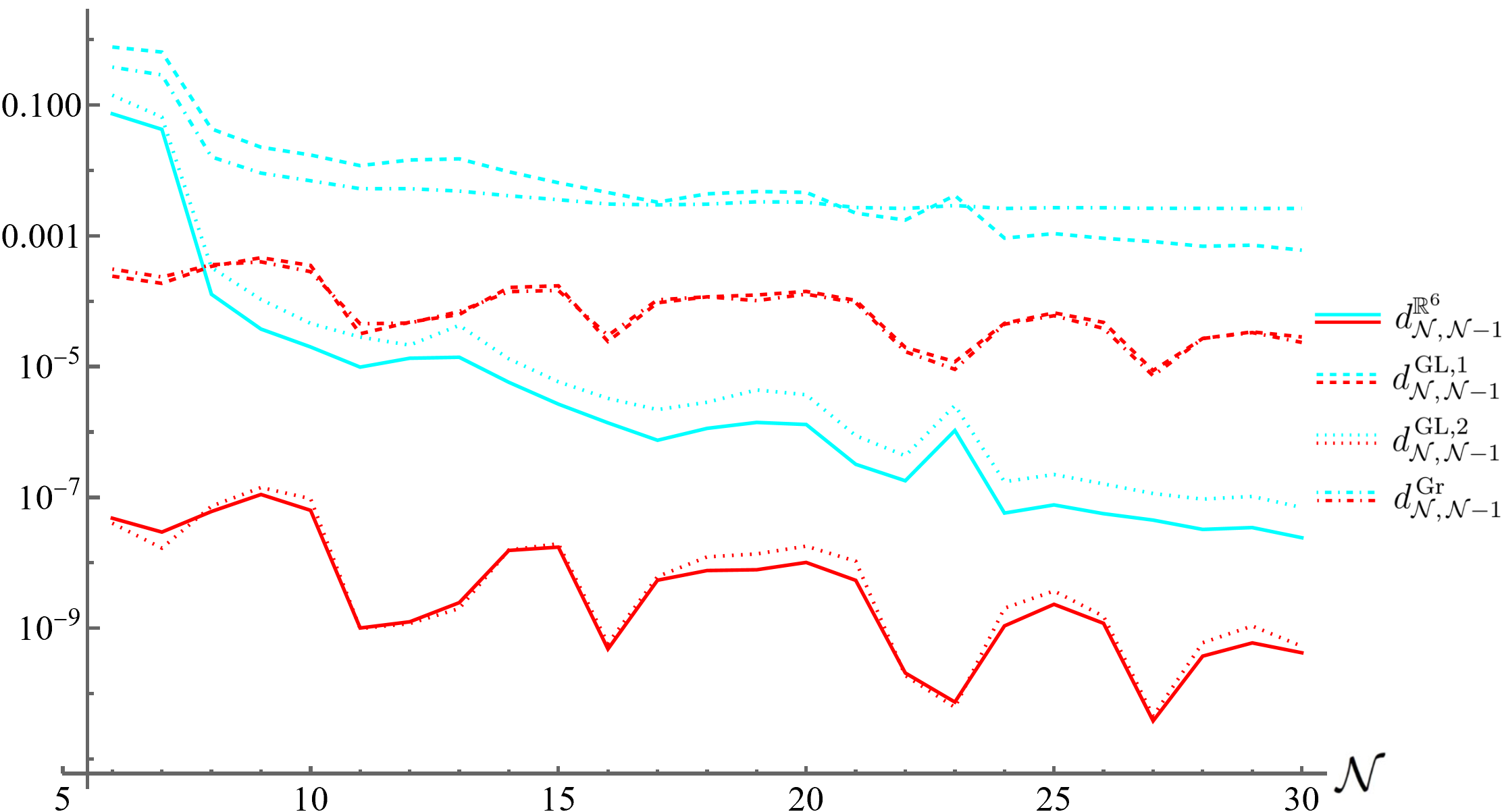} 
\caption{
Convergence of the isometric subspaces corresponding to triangulated surfaces with $\calN$ grid squares per row and per column in a unit cell. 
Each type of curve corresponds to data points computed using a different distance measure [\Eqsref{eqn: distance measure 1, Euclid}, \eqref{eqn: distance measure 2, Rocklin}, \eqref{eqn: distance measure 3, Sun} and \eqref{eqn: distance measure 4, Grassmann}], as indicated in the legend. 
Each color corresponds to a different smooth surface that the triangulated surfaces approximate. 
Both surfaces are graphs of a height function $h\left(u^1, u^2\right)$, that is, of the form $\bfx_\graph\left(u^1, u^2\right) \equiv \left(u^1, u^2, h\left(u^1, u^2\right)\right)$. 
Specifically, the red curves correspond to a translation surface defined by the height function 
$
       h_\rmr\left(u^1, u^2\right) 
\equiv   \sin\left(u^1     \right) 
     +   \cos\left(     u^2\right) 
     - 1 
$, while the cyan curves correspond to a more general height function 
$
       h_\rmc\left(  u^1, u^2\right) 
\equiv 0.8\left[ 
         \sin\left(  u^1     \right) 
     -   \sin\left(2 u^1     \right) 
     +   \sin\left(3 u^1     \right) 
     +   \cos\left(       u^2\right) 
     + 2 \cos\left(2      u^2\right) 
     - 3 
       \right] 
     + 0.2\left[ 
       \sin\left(u^1 - \sin\left(u^2\right)\right) 
     - \cos\left(u^2 - \cos\left(u^1\right)\right) + \cos{1}\right] 
$. 
} 
\label{SI fig: convergence plot} 
\end{figure}

\newpage 
\bibliography{Refs_SI} 
\appendix 

\renewcommand{\theequation}{\thesection.\arabic{equation}} 
\newpage 
\section{Lines of principal stress} \label{appendix: stress lines} 
In this appendix, we present the calculations relevant to generating Fig.~2 in the \MT{}. 
In particular, we demonstrate how the isometry-stress duality applies to a specific surface and describe the procedure used to plot the lines of principal stress on the surface. 
\par 
The surface chosen for Fig.~2 in the \MT{} is a translation surface~\cite{Struik1988_Classical_DG}, parameterized as follows: 
\begin{align} 
       \bfx_\trans\left(u, v\right) 
\equiv \bfU       \left(u   \right) 
     + \bfV       \left(   v\right) 
\equiv \left(u, 0, -\frac{1}{2}             u^2       \right) 
     + \left(0, v, -\frac{1}{2}\left(v^3 - 2v^2\right)\right). 
\label{eqn: parametrization of the chosen surface} 
\end{align} 
\par 
For translation surfaces, it has been established that the displacement field associated with one of their isometric modes takes the following form~\cite{Nassar2023_Isometries_of_Translation_Surfaces}: 
\begin{align} 
       \delta_\iso{\bfx_\trans}\left(u, v\right) 
\equiv                        \bfU(u ) \times        { \bfV(v ) } 
     + \int_{u_0}^u \df{u'}\, \bfU(u') \times \odfrac{ \bfU(u') }{u'} 
     - \int_{v_0}^v \df{v'}\, \bfV(v') \times \odfrac{ \bfV(v') }{v'}. 
\end{align} 
And the corresponding angular velocity field is given by~\cite{Nassar2023_Isometries_of_Translation_Surfaces}: 
\begin{align} 
       \bsomega_\trans\left(u, v\right) 
\equiv \bfU           \left(u   \right) 
     - \bfV           \left(   v\right). 
\end{align} 
\par 
Accordingly, for our chosen surface, the angular velocity field characterizing its known isometric mode can be expressed as: 
\begin{align} 
  \bsomega_\trans\left(u, v\right) 
= \left(u, 0, -\frac{1}{2}             u^2       \right) 
- \left(0, v, -\frac{1}{2}\left(v^3 - 2v^2\right)\right). 
\label{eqn: parametrization of the angular velocity, the chosen surface} 
\end{align} 
\par 
Recall that the angular acceleration components are related to the derivative of the angular velocity field via \Eqref{eqn: defn. of the angular acceleration vectors, revised}: 
\begin{align} 
                      \partial_\alpha{\bsomega_\trans} 
= {a_\alpha}^\gamma\, \partial_\gamma{\bfx    _\trans}. 
\label{eqn: defn. of the angular acceleration vectors, revised, revisited 2} 
\end{align} 
Thus, to obtain the expression for the angular acceleration components, we take the inner product of both sides of \Eqref{eqn: defn. of the angular acceleration vectors, revised, revisited 2} with $\partial_\beta{\bfx_\trans}$ and then invert the resulting relation using the inverse metric components, as shown below: 
\begin{align} 
\begin{split} 
\inProd{ {a_\alpha}^\gamma\, \partial_ \gamma{\bfx_\trans} }{ \partial_\beta{\bfx_\trans} } & = \inProd{ \partial_ \alpha{\bsomega_\trans} }{ \partial_\beta {\bfx_\trans} } 
                                                                                                \h 
         {a_\alpha}^\gamma\,        g_{\gamma                          \beta}               & = \inProd{ \partial_ \alpha{\bsomega_\trans} }{ \partial_\beta {\bfx_\trans} } 
                                                                                                \h 
         {a_\alpha}^\gamma                                                                  & = \inProd{ \partial_ \alpha{\bsomega_\trans} }{ \partial_\beta {\bfx_\trans} } 
                                                                                                                g^{\beta                               \gamma}. 
\end{split} 
\label{eqn: expression for the angular acceleration components, index notation} 
\end{align} 
\par 
In matrix notation, which is more convenient for practical calculations, \Eqref{eqn: expression for the angular acceleration components, index notation} can be written as: 
\begin{align} 
\begin{split} 
    \left( {a_\alpha}^\gamma \right) 
& = \frac{1}{ 
    \norm{ 
    \partial_u{\bfx_\trans} \times 
    \partial_v{\bfx_\trans} 
    }^2 
    } 
    \h 
& \ \ \times 
    \begin{pmatrix} 
    \inProd{ \partial_u{\bsomega_\trans} }{ \partial_u{\bfx_\trans} } & \inProd{ \partial_u{\bsomega_\trans} }{ \partial_v{\bfx_\trans} } \h 
    \inProd{ \partial_v{\bsomega_\trans} }{ \partial_u{\bfx_\trans} } & \inProd{ \partial_v{\bsomega_\trans} }{ \partial_v{\bfx_\trans} } 
    \end{pmatrix} 
    \begin{pmatrix} 
    \phantom{-}\inProd{ \partial_v{\bfx_\trans} }{ \partial_v{\bfx_\trans} } &         {-}\inProd{ \partial_u{\bfx_\trans} }{ \partial_v{\bfx_\trans} } \h 
            {-}\inProd{ \partial_v{\bfx_\trans} }{ \partial_u{\bfx_\trans} } & \phantom{-}\inProd{ \partial_u{\bfx_\trans} }{ \partial_u{\bfx_\trans} } 
    \end{pmatrix}. 
\end{split} 
\label{eqn: expression for the angular acceleration components, matrix notation} 
\end{align} 
Substituting 
\Eqsref{eqn: parametrization of the chosen surface} and 
 \eqref{eqn: parametrization of the angular velocity, the chosen surface} into \Eqref{eqn: expression for the angular acceleration components, matrix notation} yields: 
\begin{align} 
\begin{split} 
    \left( {a_\alpha}^\gamma \right) 
& = \frac{1}{ 
    \norm{ 
    \bfU' \times 
    \bfV' 
    }^2 
    } 
    \begin{pmatrix} 
    \phantom{-}  \norm       {\bfU'}^2 & \phantom{-}\inProd{\bfU'}{\bfV'}   \h 
            {-}\inProd{\bfV'}{\bfU'}   &         {-}  \norm       {\bfV'}^2 
    \end{pmatrix} 
    \begin{pmatrix} 
    \phantom{-}  \norm{\bfV'}^2      &         {-}\inProd{\bfU'}{\bfV'} \h 
            {-}\inProd{\bfV'}{\bfU'} & \phantom{-}  \norm{\bfU'}^2 
    \end{pmatrix} 
    \h 
& = \frac 
    {\norm{\bfU'}^2 \norm{\bfV'}^2 - \inProd{\bfU'}{\bfV'}^2} 
    {\norm{\bfU'   \times \bfV'}^2                          } 
    \begin{pmatrix} 
    1 & \phantom{-}0 \h 
    0 &         {-}1 
    \end{pmatrix} 
    \h 
& = \begin{pmatrix} 
    1 & \phantom{-}0 \h 
    0 &         {-}1 
    \end{pmatrix}. 
\end{split} 
\label{eqn: expression for the angular acceleration components, the chosen surface} 
\end{align} 
It is worth emphasizing that \Eqref{eqn: expression for the angular acceleration components, the chosen surface} applies to the isometric mode considered in this appendix for any translation surface. 
\par 
Now, we apply the isometry-stress duality to determine the equilibrium stress components corresponding to the given mode of isometry. 
Substituting \Eqref{eqn: expression for the angular acceleration components, the chosen surface} into \Eqref{eqn: relation between equilibrium stress and isometry, i to s} and setting $f = 1$ in the resulting expression yields: 
\begin{align} 
\begin{split} 
    \left(\sigma^{\alpha\beta}\right) 
& = \frac{1}{ \norm{\bfU' \times \bfV'} } 
    \begin{pmatrix} \phantom{-}0 &            1 \h -1 &  0 \end{pmatrix} 
    \begin{pmatrix}            1 & \phantom{-}0 \h  0 & -1 \end{pmatrix} 
    \h 
& = -\left[1 + u^2 + \left(\frac{3}{2}v^2 - 2v\right)^2\right]^{-1/2} 
    \begin{pmatrix}            0 &            1 \h  1 &  0 \end{pmatrix}, 
\end{split} 
\end{align} 
which is symmetric, as expected. 
\par 
Just as the principal curvature directions on a surface are given by the eigenvectors of the shape operator~\cite{Shifrin2021_Classical_Differential_Geometry}, rather than the surface's second fundamental form (recall \Secref{subsec: DG, the second fundamental form}), the principal stress directions on the surface are associated with the following mixed stress matrix\footnote{Among all types of rank-two tensors, only mixed tensors map vectors to vectors, making it meaningful to discuss their eigenvectors, as other types of rank-two tensors, strictly speaking, do not have eigenvectors. 
One may also wonder why the other mixed stress matrix, $\left( {\sigma_\gamma}^\alpha \right)$---the transpose of $\left( {\sigma^\alpha}_\gamma \right)$---is not considered in this context. 
The reason is that, between the two, it is the stress components ${\sigma^\alpha}_\gamma$ that contract on their right with vector components, which have an upper index, producing a new set of vector components, akin to matrix multiplication. 
}: 
\begin{align} 
\begin{split} 
    \left( {\sigma^ \alpha}_\gamma                   \right) 
  = \left(  \sigma^{\alpha  \beta }\, g_{\beta\gamma}\right) 
& = -\frac{1}{ \norm{\bfU' \times \bfV'} } 
    \begin{pmatrix}              0 &                     1 \h                     1 &              0 \end{pmatrix} 
    \begin{pmatrix} \norm{\bfU'}^2 & \inProd{\bfU'}{\bfV'} \h \inProd{\bfV'}{\bfU'} & \norm{\bfV'}^2 \end{pmatrix} 
    \h 
& = -\frac{1}{ \norm{\bfU' \times \bfV'} } 
    \begin{pmatrix} 
    \inProd{\bfU'}{\bfV'}   &   \norm       {\bfV'}^2 \h 
      \norm{\bfU'}       ^2 & \inProd{\bfU'}{\bfV'} 
    \end{pmatrix} 
    \h 
& = -\left[ 
    \displaystyle 1 + u^2 + \left(\frac{3}{2}v^2 - 2v\right)^2 
    \right]^{-1/2}\begin{bmatrix} 
    \displaystyle     u     \left(\frac{3}{2}v^2 - 2v\right)   & 
    \displaystyle 1 +       \left(\frac{3}{2}v^2 - 2v\right)^2 \hh 
    \displaystyle 1 + u^2                                      & 
    \displaystyle     u     \left(\frac{3}{2}v^2 - 2v\right) 
    \end{bmatrix}. 
\end{split} 
\end{align} 
By diagonalizing the matrix $\left( {\sigma^\alpha}_\gamma \right)$, we obtain its eigenvalues and corresponding eigenvectors: 
\refstepcounter{equation} 
\begin{align} 
\sigma^+ \equiv -\frac{ \inProd{\bfU'}{\bfV'} + \norm{\bfU'}\norm{\bfV'} }{ \norm{\bfU' \times \bfV'} }, \quad\quad\quad \bfv^+ \equiv \norm{\bfV'}\, \partial_u{\bfx_\trans} 
                                                                                                                                     + \norm{\bfU'}\, \partial_v{\bfx_\trans}, 
\tag{\theequation, a} 
\hh 
\sigma^- \equiv -\frac{ \inProd{\bfU'}{\bfV'} - \norm{\bfU'}\norm{\bfV'} }{ \norm{\bfU' \times \bfV'} }, \quad\quad\quad \bfv^- \equiv \norm{\bfV'}\, \partial_u{\bfx_\trans} 
                                                                                                                                     - \norm{\bfU'}\, \partial_v{\bfx_\trans}. 
\tag{\theequation, b} 
\end{align} 
\par 
The principal stresses---the maximum and minimum stresses at a point on the chosen surface---are characterized by the eigenvalues $\sigma^+$ and $\sigma^-$, with the corresponding principal stress directions given by the eigenvectors $\bfv^+$ and $\bfv^-$. 
It is worth noting that the principal stress directions are orthogonal to each other, as confirmed by the relation $\inProd{\bfv^+}{\bfv^-} = 0$. 
To obtain the expressions for the lines of principal stress, we first determine the relations between the local coordinates $u$ and $v$ along these lines and subsequently substitute the derived relations into the surface parametrization. 
The relations between the local coordinates can be established from the eigenvectors $\bfv^+$ and $\bfv^-$ in the following way. 
\par 
To motivate the approach, we consider a curve on an arbitrary surface $\bfx(u, v)$. 
The curve can be parameterized as $\bfx\big(u(t), v(t)\big)$ for some real parameter $t$. 
The tangent vectors of the curve can be derived by differentiating $\bfx\big(u(t), v(t)\big)$ with respect to $t$, yielding: 
\begin{align} 
  \odfrac{\bfx}{t} 
= \odfrac{u   }{t}\, \partial_u{\bfx} 
+ \odfrac{v   }{t}\, \partial_v{\bfx}. 
\end{align} 
Thus, if the expression for the tangent vector field of a curve embedded in a surface is known, we can determine the relation between the local coordinates along the curve by solving the following differential equation: 
\begin{align} 
  \odfrac{u}{v} 
=   \frac 
  { \df{u} / \df{t} } 
  { \df{v} / \df{t} }. 
\label{eqn: ODE relating the local coordinates along an embedded curve} 
\end{align} 
In our case, \Eqref{eqn: ODE relating the local coordinates along an embedded curve} implies the following for the lines of principal stress: 
\begin{align} 
  \left(\odfrac{u}{v}\right)_\pm 
= \pm\frac{ \norm{\bfV'} }{ \norm{\bfU'} } 
= \pm\frac 
  { \sqrt{1 + \left(3v^2 / 2 - 2v\right)^2} } 
  { \sqrt{1 +        u^2                  } }. 
\label{eqn: ODEs relating the local coordinates along the lines of principal stress} 
\end{align} 
\par 
From \Eqref{eqn: ODEs relating the local coordinates along the lines of principal stress}, one can derive the relations between the local coordinates for the lines of principal stress, which can take the form of either $u = u(v)$ or $v = v(u)$. 
Substituting the relations into the surface parametrization then yields the desired expressions for the lines of principal stress: 
$\bfx_\trans\big(u(v), v   \big)$ or 
$\bfx_\trans\big(u   , v(u)\big)$. 
\newpage 
\section{Recasting the equilibrium equations in a form commonly found in the literature} \label{appendix: rewriting the equilibrium equations} 
In this appendix, we demonstrate that the equilibrium equations derived in \Secref{subsec: the isometry-stress duality, equilibrium stress} are equivalent to the more commonly used form in the literature. 
Using this form, we derive a relation between the equilibrium stress tensor and the second fundamental form of a surface. 
\par 
We begin by briefly reviewing the concept of covariant derivatives. 
\subsection{A brief review of covariant derivatives} 
When acting on scalar functions, covariant derivatives reduce to ordinary partial derivatives. 
For tangent vectors, covariant derivatives account for both the changes in their components (as ordinary partial derivatives do) and the in-plane variations of the coordinate basis vectors (see \Secref{subsec: DG, tangent vectors and one-forms}), which are typically non-constant for curved surfaces. 
\par 
The in-plane variations of the coordinate basis vectors are captured by the Christoffel symbols of the second kind $\Chs{\alpha\beta}{\gamma}$, defined by: 
\begin{align} 
                                   \partial_\alpha{ \partial_\beta{\bfx} } 
\equiv \Chs{\alpha\beta}{\gamma}\, \partial_\gamma{               {\bfx} } 
     +   b_{\alpha\beta}        \, \unitvec{n}, 
\label{eqn: variations of the coordinate basis vectors} 
\end{align} 
where $b_{\alpha\beta}$ denotes the components of the second fundamental form of a surface, and $\unitvec{n}$ the surface unit normal vector (see \Secref{subsec: DG, the second fundamental form}). 
Since the partial derivatives commute, the Christoffel symbols of the second kind are symmetric in their lower indices: 
\begin{align} 
  \Chs{\alpha\beta }{\gamma} 
= \Chs{\beta \alpha}{\gamma}. 
\end{align} 
\par 
For coordinate basis vectors, whose components are constant, their covariant derivatives are given solely in terms of the Christoffel symbols: 
\begin{align} 
                                   D_\alpha{ \partial_\beta {\bfx} } 
\equiv \Chs{\alpha\beta}{\gamma}\,         { \partial_\gamma{\bfx} }. 
\label{eqn: defn. of connection coefficients} 
\end{align} 
More generally, the covariant derivatives of a tangent vector field $\bfv \equiv v^\alpha\, \partial_\alpha{\bfx}$, with spatially varying components $v^\alpha$, are given by: 
\begin{align} 
\begin{split} 
                D_\beta                                                                       {\bfv} 
  \equiv        D_\beta\left( {v^\alpha}\,                                   { \partial_\alpha{\bfx} }\right) 
& \equiv \partial_\beta       {v^\alpha}\,                                   { \partial_\alpha{\bfx} } 
       +                      {v^\alpha}\,                            D_\beta{ \partial_\alpha{\bfx} } 
         \h 
& \equiv \partial_\beta       {v^\alpha}\,                                   { \partial_\alpha{\bfx} } 
       +                      {v^\alpha}\, \Chs{\beta\alpha}{\gamma}\,       { \partial_\gamma{\bfx} } 
       = \partial_\beta       {v^\alpha}\,                                   { \partial_\alpha{\bfx} } 
       +                      {v^\gamma}\, \Chs{\beta\gamma}{\alpha}\,       { \partial_\alpha{\bfx} } 
         \h 
&      = \left( 
         \partial_\beta{v^\alpha} + \Chs{\beta\gamma}{\alpha}\, v^\gamma 
         \right)\, \partial_\alpha{\bfx}, 
\end{split} 
\end{align} 
where the indices $\alpha$ and $\gamma$ are interchanged in the second line. 
Conventionally, the covariant derivatives of a tangent vector are written solely in terms of its components: 
\begin{align} 
              D_\beta{v^\alpha} 
\equiv \partial_\beta{v^\alpha} + \Chs{\beta\gamma}{\alpha}\, v^\gamma. 
\end{align} 
\par 
The covariant derivatives of a $\binom{2}{0}$ tensor are computed in a similar manner to the treatment of tangent vector fields. 
Taking the stress tensor [see \Eqref{eqn: introducing the stress tensor}] as an example, we first apply the Leibniz product rule: 
\begin{align} 
         D_\gamma\left( { \sigma^{\alpha\beta} }\,         { \partial_\alpha{\bfx} } \otimes         { \partial_\beta{\bfx} } \right) 
= \partial_\gamma       { \sigma^{\alpha\beta} }\,         { \partial_\alpha{\bfx} } \otimes         { \partial_\beta{\bfx} } 
+                       { \sigma^{\alpha\beta} }\, D_\gamma{ \partial_\alpha{\bfx} } \otimes         { \partial_\beta{\bfx} } 
+                       { \sigma^{\alpha\beta} }\,         { \partial_\alpha{\bfx} } \otimes D_\gamma{ \partial_\beta{\bfx} }. 
\end{align} 
Then, using \Eqref{eqn: defn. of connection coefficients} and relabeling the indices accordingly, we obtain the desired expression: 
\begin{align} 
\begin{split} 
           D_\gamma\left( { \sigma^{\alpha\beta} }\,                              { \partial_\alpha{\bfx} } \otimes                            { \partial_\beta{\bfx} } \right) 
& = \partial_\gamma       { \sigma^{\alpha\beta} }\,                              { \partial_\alpha{\bfx} } \otimes                            { \partial_\beta{\bfx} } 
  +                       { \sigma^{\alpha\beta} }\, \Chs{\gamma\alpha}{\rho  }\, { \partial_\rho  {\bfx} } \otimes                            { \partial_\beta{\bfx} } 
  +                       { \sigma^{\alpha\beta} }\,                              { \partial_\alpha{\bfx} } \otimes \Chs{\gamma\beta}{\rho }\, { \partial_\rho {\bfx} } 
    \h 
& = \partial_\gamma       { \sigma^{\alpha\beta} }\,                              { \partial_\alpha{\bfx} } \otimes                            { \partial_\beta{\bfx} } 
  +                       { \sigma^{\rho  \beta} }\, \Chs{\gamma\rho  }{\alpha}\, { \partial_\alpha{\bfx} } \otimes                            { \partial_\beta{\bfx} } 
  +                       { \sigma^{\alpha\rho } }\,                              { \partial_\alpha{\bfx} } \otimes \Chs{\gamma\rho }{\beta}\, { \partial_\beta{\bfx} } 
    \h 
& = \left( 
    \partial_\gamma{ \sigma^{\alpha\beta} } + \Chs{\gamma\rho}{\alpha}\, \sigma^{\rho  \beta} 
                                            + \Chs{\gamma\rho}{\beta }\, \sigma^{\alpha\rho } 
    \right)\, \partial_\alpha{\bfx} \otimes 
              \partial_\beta {\bfx} 
\end{split} 
\end{align} 
or, more conventionally: 
\begin{align} 
              D_\gamma{ \sigma^{\alpha\beta} } 
\equiv \partial_\gamma{ \sigma^{\alpha\beta} } + \Chs{\gamma\rho}{\alpha}\, \sigma^{\rho  \beta} 
                                               + \Chs{\gamma\rho}{\beta }\, \sigma^{\alpha\rho }. 
\label{eqn: covariant derivatives of the stress tensor, conventional} 
\end{align} 
In particular, by contracting the indices $\gamma$ and $\alpha$ in \Eqref{eqn: covariant derivatives of the stress tensor, conventional}, we obtain the divergence of the stress tensor: 
\begin{align} 
         D_\alpha{ \sigma^{\alpha\beta} } 
= \partial_\alpha{ \sigma^{\alpha\beta} } + \Chs{\alpha\rho}{\alpha}\, \sigma^{\rho  \beta} 
                                          + \Chs{\alpha\rho}{\beta }\, \sigma^{\alpha\rho }, 
\label{eqn: divergence of the stress tensor, conventional} 
\end{align} 
in terms of which the commonly used form of the equilibrium equations is expressed. 
\subsection{The commonly used form of the equilibrium equations} 
In the literature, the membrane equilibrium equations are often written in the following form, as presented in \Refref{Niordson1985_Shell_Theory}: 
\refstepcounter{equation} \label{eqn: equilibrium equations, common form} 
\begin{align} 
D_\alpha{ \sigma^{\alpha\beta} } = 0, 
\label{eqn: in-plane equilibrium equation} 
\tag{\theequation, a} 
\h 
        {      b_{\alpha\beta} }\, 
        { \sigma^{\alpha\beta} } = 0. 
\label{eqn: out-of-plane equilibrium equation} 
\tag{\theequation, b} 
\end{align} 
We now demonstrate that this commonly used form is equivalent to the equilibrium equations derived earlier in \Secref{subsec: the isometry-stress duality, equilibrium stress}: 
\begin{align} 
  \partial_\alpha\left(\sqrt{g}\, \sigma^{\alpha\beta}\, \partial_\beta{\bfx}\right) 
= \zero. 
\label{eqn: equilibrium equations, appendix} 
\end{align} 
\par 
To begin, we divide both sides of \Eqref{eqn: equilibrium equations, appendix} by $\sqrt{g}$, yielding the following expression: 
\begin{align} 
  \frac{1}{\sqrt g}\, \partial_\alpha\left(\sqrt{g}\, \sigma^{\alpha\beta}\, \partial_\beta{\bfx}\right) 
= \zero. 
\label{eqn: an intermediate step toward deriving the equivalence relationship, 0} 
\end{align} 
Expanding \Eqref{eqn: an intermediate step toward deriving the equivalence relationship, 0} using the product rule and substituting \Eqref{eqn: variations of the coordinate basis vectors} into the resulting expression gives: 
\begin{align} 
\begin{split} 
    \zero 
& = \left(\frac{1}{\sqrt g}\, \partial_\alpha{ \sqrt g              }\right)\sigma^{\alpha\beta}\,                             \partial_\beta {\bfx} 
  + \left(                    \partial_\alpha{ \sigma^{\alpha\beta} }\right)                                                   \partial_\beta {\bfx} 
  +                                          { \sigma^{\alpha\beta} }                           \,                             \partial_\alpha{ 
                                                                                                                               \partial_\beta {\bfx} } 
    \h 
& = \left(\frac{1}{\sqrt g}\, \partial_\alpha{ \sqrt g              }\right)\sigma^{\alpha\beta}\,                             \partial_\beta {\bfx} 
  + \left(                    \partial_\alpha{ \sigma^{\alpha\beta} }\right)                                                   \partial_\beta {\bfx} 
  +                                          { \sigma^{\alpha\beta} }                           \, \Chs{\alpha\beta}{\gamma}\, \partial_\gamma{\bfx} 
  +                                          { \sigma^{\alpha\beta} }                           \,   b_{\alpha\beta}        \, \unitvec{n}. 
\end{split} 
\label{eqn: an intermediate step toward deriving the equivalence relationship, 1} 
\end{align} 
\par 
To proceed, we need the following identity involving the Christoffel symbols~\cite{Schutz2018_General_Relativity}: 
\begin{align} 
\frac{1}{\sqrt g}\, \partial_\alpha{\sqrt g} = \Chs{\alpha\gamma}{\gamma}. 
\end{align} 
Using this identity, \Eqref{eqn: an intermediate step toward deriving the equivalence relationship, 1} becomes: 
\begin{align} 
  \zero 
= \Chs{\alpha\gamma}{\gamma}\,                      { \sigma^{\alpha\beta} }       \, \partial_\beta {\bfx} 
+                              \left(\partial_\alpha{ \sigma^{\alpha\beta} }\right)   \partial_\beta {\bfx} 
+ \Chs{\alpha\beta }{\gamma}\,                      { \sigma^{\alpha\beta} }       \, \partial_\gamma{\bfx} 
+   b_{\alpha\beta }        \,                      { \sigma^{\alpha\beta} }       \, \unitvec{n}. 
\label{eqn: an intermediate step toward deriving the equivalence relationship, 2} 
\end{align} 
Interchanging the indices $\beta$ and $\gamma$ in the third term on the right-hand side of \Eqref{eqn: an intermediate step toward deriving the equivalence relationship, 2} simplifies the equation to, after some rearrangements: 
\begin{align} 
  \zero 
= \left( 
                               \partial_\alpha{ \sigma^{\alpha\beta } } 
+ \Chs{\alpha\gamma}{\gamma}\,                { \sigma^{\alpha\beta } } 
+ \Chs{\alpha\gamma}{\beta }\,                { \sigma^{\alpha\gamma} } 
  \right)                                                                 \partial_\beta{\bfx} 
+   b_{\alpha\beta }        \,                { \sigma^{\alpha\beta } }\, \unitvec{n}. 
\label{eqn: an intermediate step toward deriving the equivalence relationship, 3} 
\end{align} 
\par 
In \Eqref{eqn: an intermediate step toward deriving the equivalence relationship, 3}, we recognize that the quantity in parentheses is nothing but the divergence of the stress tensor [\Eqref{eqn: divergence of the stress tensor, conventional}]. 
Thus, the equilibrium equations commonly used in the literature [\Eqsref{eqn: equilibrium equations, common form}] follow from the requirement that both the in-plane and out-of-plane components of \Eqref{eqn: an intermediate step toward deriving the equivalence relationship, 3} must vanish. 
\subsection{Relation between the components of the equilibrium stress tensor and the second fundamental form} 
Using the common form of the equilibrium equations, specifically \Eqref{eqn: out-of-plane equilibrium equation}, we can establish a relation between the equilibrium stress tensor and the change of the second fundamental form of a surface under an isometric deformation---a relation equivalent to the isometry-stress duality discussed in \Secref{subsec: the isometry-stress duality, duality}. 
\par 
First, recall Gauss' Theorema Egregium, which states that the Gaussian curvature [\Eqref{eqn: mean curvature and Gaussian curvature}] of a surface depends only on the corresponding metric components and their derivatives. 
As a result, the surface Gaussian curvature is invariant under isometric deformations, which preserve the metric. 
\par 
Mathematically, this means that the variation of the surface Gaussian curvature with respect to an isometric deformation is zero: 
\begin{align} 
\begin{split} 
    0 
  = \delta{K} 
& = \delta\left(\frac{1}{2}\, \calE^{\alpha\gamma}\, \calE^{\beta\rho}\, b_{\alpha\beta}\,       {b}_{\gamma\rho }\right) 
    \h 
& =             \frac{1}{2}\, \calE^{\alpha\gamma}\, \calE^{\beta\rho}\, b_{\alpha\beta}\, \delta{b}_{\gamma\rho } 
  +             \frac{1}{2}\, \calE^{\alpha\gamma}\, \calE^{\beta\rho}\, b_{\gamma\rho }\, \delta{b}_{\alpha\beta}, 
\end{split} 
\label{eqn: variation of the Gaussian curvature} 
\end{align} 
where $\calE^{\alpha\gamma} \equiv \epsilon^{\alpha\gamma} / \sqrt{g}$ denotes the contravariant components of the area two-form [\Eqref{eqn: contravariant Levi-Civita tensor}]. 
The variation of $\calE^{\alpha\gamma}$ is zero because the metric determinant $g$ is invariant under the isometric deformation. 
\par 
In \Eqref{eqn: variation of the Gaussian curvature}, by interchanging the indices 
$\alpha \leftrightarrow \gamma$ and 
$\beta  \leftrightarrow \rho  $, we find that the two terms on the right-hand side are equal: 
\begin{align} 
  \frac{1}{2}\, \calE^{\alpha\gamma}\, \calE^{\beta\rho }\, b_{\alpha\beta}\, \delta{b}_{\gamma\rho } 
= \frac{1}{2}\, \calE^{\gamma\alpha}\, \calE^{\rho \beta}\, b_{\gamma\rho }\, \delta{b}_{\alpha\beta} 
= \frac{1}{2}\, \calE^{\alpha\gamma}\, \calE^{\beta\rho }\, b_{\gamma\rho }\, \delta{b}_{\alpha\beta}, 
\end{align} 
where the second equality follows from the antisymmetric property of $\calE^{\alpha\gamma}$. 
Accordingly, \Eqref{eqn: variation of the Gaussian curvature} implies that: 
\begin{align} 
      \calE^{\alpha\gamma}\, 
      \calE^{\beta \rho  }\, 
        {b}_{\alpha\beta }\, 
  \delta{b}_{\gamma\rho } 
= 0. 
\label{eqn: an implication of the invariance of the Gaussian curvature} 
\end{align} 
\par 
By comparing \Eqref{eqn: an implication of the invariance of the Gaussian curvature} to \Eqref{eqn: out-of-plane equilibrium equation}, we deduce that: 
\begin{align} 
         \sigma^{\alpha\beta } 
= f\,     \calE^{\alpha\gamma}\, 
          \calE^{\beta \rho  }\, 
      \delta{b}_{\gamma\rho  }, 
\label{eqn: relation between equilibrium stress and isometry, curvature version} 
\end{align} 
where $f$ is a constant with the dimension of force, as determined from dimensional analysis: 
\begin{align} 
\left[   \sigma^{\alpha\beta }\right] = \mathrm{ \frac{ {Force}^{\phantom 2} }{meter} }, \quad\halfQuad 
\left[      {f}               \right] = \mathrm{        {Force}                       }, \quad\halfQuad 
\left[    \calE^{\alpha\gamma}\right] = 
\left[ \sqrt{g}               \right] = 1                                                \mbox{\quad\halfQuad{and}\quad\halfQuad} 
\left[\delta{b}_{\gamma\rho  }\right] = \mathrm{ \frac{1                     }{meter} }. 
\end{align} 
\par 
As demonstrated by \Eqref{eqn: relation between equilibrium stress and isometry, curvature version}, the equilibrium stress caused by an energy-minimizing deformation can be related to the curvature changes under an isometric deformation, characterized by the changes in the components of the second fundamental form $\delta{b}_{\alpha\beta}$. 
Now, recall the isometry-stress duality [\Eqsref{eqn: relations between equilibrium stress and isometry}] in \Secref{subsec: the isometry-stress duality, duality}, which establishes a relation between the aforementioned equilibrium stress and the angular acceleration components associated with the same mapped isometric deformation. 
Thus, by combining 
\Eqsref{eqn: relation between equilibrium stress and isometry, curvature version} and 
 \eqref{eqn: relation between equilibrium stress and isometry, i to s}, we can obtain the following geometric relation between the angular acceleration components ${a_\alpha}^\beta$, which characterize an isometric deformation, and the resulting curvature changes $\delta{b}_{\alpha\beta}$: 
\begin{align} 
                            {a _ \gamma}^\beta 
= \calE^{\beta\rho}\, \delta{b}_{\rho    \gamma}. 
\label{eqn: expression for the a matrix in terms of the db matrix, appendix} 
\end{align} 
\newpage 
\section{Expression for the line force density along the boundary of a surface} \label{appendix: line force density} 
In this appendix, we demonstrate that the term $\calE_{\alpha\gamma}\, \sigma^{\alpha\beta}\, \partial_\beta{\bfx}\, d{u^\gamma}$ indeed represents the infinitesimal force exerted on the boundary of a surface. 
To avoid potential confusion, we interpret the symbol $d{u^\gamma}$ here as the infinitesimal change in the local coordinate $u^\gamma$, rather than a differential one-form.\footnote{In fact, to be pedantic, we reserve the symbol $\df{u^\gamma}$---with the upright ``$\df{}$''---specifically for the corresponding differential one-form.} 
\par 
We begin by rewriting the term in the following way: 
\begin{align} 
\begin{split} 
                     \calE_{\alpha  \gamma}\, \sigma^{\alpha\beta}\, \partial_\beta{\bfx}\, d{u^\gamma} 
& = g_{\alpha\mu}\, {\calE^ \mu   }_\gamma \, \sigma^{\alpha\beta}\, \partial_\beta{\bfx}\, d{u^\gamma} 
    \h 
& = \sigma^{\alpha\beta}\inProd 
    {                                    \partial_\alpha{\bfx} } 
    { {\calE^\mu}_\gamma\, d{u^\gamma}\, \partial_\mu   {\bfx} } 
    {                                    \partial_\beta {\bfx} }, 
\end{split} 
\label{eqn: an intermediate step toward explaining the meaning of line force density} 
\end{align} 
where the definition of the metric components $g_{\alpha\mu}$ is applied to obtain the second line. 
Here, the vector ${\calE^\mu}_\gamma\, d{u^\gamma}\, \partial_\mu{\bfx} \equiv \bfN$, which lies within the tangent plane of the surface, is orthogonal to the surface boundary. 
To see this orthogonal relationship, we compute the inner product of $\bfN$ with the tangent vector to the surface boundary, $d{\bfx} \equiv d{u^\nu}\, \partial_\nu{\bfx}$, yielding: 
\begin{align} 
  \inProd{\bfN}{ d{\bfx} } 
= \inProd 
  {            {\calE^ \mu}_\gamma \, d{u^\gamma}\, \partial_\mu{\bfx} } 
  {                                   d{u^\nu   }\, \partial_\nu{\bfx} } 
= g_{\nu\mu}\, {\calE^ \mu}_\gamma \, d{u^\nu   }\, 
                                      d{u^\gamma} 
=               \calE_{\nu  \gamma}\, d{u^\nu   }\, 
                                      d{u^\gamma} 
= 0. 
\end{align} 
Moreover, the magnitude of $\bfN$ can be shown to equal the infinitesimal arc length of the surface boundary, $d{s} \equiv \norm{ d{\bfx} }$: 
\begin{align} 
\begin{split} 
    \inProd{\bfN}{\bfN} 
& = \inProd{ {\calE^ \mu}_\gamma \, d{u^\gamma}\, \partial_\mu{\bfx} }{ {\calE^ \nu}_\alpha \, d{u^\alpha}\, \partial_\nu{\bfx} } 
  = \inProd{  \calE^{\mu  \gamma}\, d{u_\gamma}\, \partial_\mu{\bfx} }{  \calE^{\nu  \alpha}\, d{u_\alpha}\, \partial_\nu{\bfx} } 
    \h 
& = \calE^{\mu\gamma}\, 
    \calE^{\nu\alpha}\, g_{\mu   \nu   }\, d{u_\gamma}\, d{u_\alpha} 
  =                     g^{\gamma\alpha}\, d{u_\gamma}\, d{u_\alpha} 
  =                     g_{\gamma\alpha}\, d{u^\gamma}\, d{u^\alpha} 
    \h 
& = \inProd 
    { d{u^\gamma}\, \partial_\gamma{\bfx} } 
    { d{u^\alpha}\, \partial_\alpha{\bfx} } 
  = \inProd{ d{\bfx} }{ d{\bfx} } 
  \equiv \norm{ d{\bfx} }^2 
  \equiv      { d{s   } }^2, 
\end{split} 
\end{align} 
where, in the second line, we recognize that the combination $\calE^{\mu\gamma}\, \calE^{\nu\alpha}\, g_{\mu\nu}$ is nothing but the matrix inverse of $\left(g_{\mu\nu}\right)$ expressed in index notation. 
\par 
Using the properties of the tensor product, \Eqref{eqn: an intermediate step toward explaining the meaning of line force density} can be further expressed as: 
\begin{align} 
  \calE_{\alpha\gamma}\,  \sigma^{\alpha\beta }\, \partial_\beta {\bfx}                    \, d{u^\gamma} 
=               \inProd{  \sigma^{\beta \alpha}\, \partial_\beta {\bfx} \otimes 
                                                  \partial_\alpha{\bfx} }{ \unitvec{\bfN} }\, d{s       } 
=               \inProd{\bssigma                                        }{ \unitvec{\bfN} }\, d{s       }. 
\label{eqn: explaining the meaning of line force density} 
\end{align} 
From \Eqref{eqn: explaining the meaning of line force density}, it is evident that the term in question is proportional to the projection of the stress tensor along the $\unitvec{\bfN}$ direction, thus representing the infinitesimal force exerted perpendicular to the surface boundary. 